\documentclass[a4paper]{article}%
\usepackage{cite}%
\usepackage{amsmath,amssymb,amsfonts}%
\usepackage{algorithmic}%
\usepackage{graphicx}%
\usepackage{textcomp}%
\usepackage{authblk}%
\usepackage[square,numbers]{natbib}%

\usepackage{url}%
\usepackage{doi}%
\usepackage{multirow}%
\usepackage{amsthm}%
\usepackage{mathrsfs}%
\usepackage{manyfoot}%
\usepackage{booktabs}%
\usepackage{listings}%
\usepackage{footmisc}%
\usepackage{bm}%
\usepackage{graphicx}%

\usepackage{enumitem}%
\setlist{nolistsep}

\usepackage[a4paper, left=3cm, right=3cm, top=3cm, bottom=3cm]{geometry}%

\usepackage{tikz}%
\usetikzlibrary{arrows, arrows.meta, calc, positioning, decorations.pathreplacing, shapes.geometric}%

\tikzstyle{nn}=[circle,thick,draw=black!75,minimum size=6mm,fill=white]
\tikzstyle{rr}=[rectangle,rounded corners,thick,draw=black!75,minimum size=6mm,fill=white]
\usepackage{hyperref}%
\hypersetup{
    bookmarksnumbered=true, bookmarksopen=true,
	unicode=true, colorlinks=true, linktoc=all, 
	linkcolor=blue, citecolor=blue, filecolor=blue, urlcolor=blue,
	pdfstartview=FitH
}

\RequirePackage{tabularx} 						    
\newcolumntype{L}{>{$}l<{$}}                        
\newcolumntype{R}{>{$}r<{$}}                        
\newcolumntype{Y}{>{\centering\arraybackslash}X}    
\newcolumntype{Z}{>{\raggedleft\arraybackslash}X}   

\usepackage{placeins}%

\providecommand{\keywords}[1]{\textbf{\textit{Keywords---}} #1}

\newcommand{\dotp}{
    \mathop{
        \mathchoice{\vcenter{\hbox{\LARGE$\cdot$}}}
                   {\vcenter{\hbox{\LARGE$\cdot$}}}
                   {\vcenter{\hbox{\normalsize$\cdot$}}}
                   {\vcenter{\hbox{\small$\cdot$}}}
    }
}

\usepackage{lastpage}%
\usepackage{fancyhdr}%
\fancypagestyle{plain}{%
  \fancyhf{}%
  \cfoot{\thepage\ / \pageref*{LastPage}} 
}
\begin{document}

\title{Approximating Spatial Distance Through \textit{Confront} Networks: Application to the Segmentation of Medieval Avignon}
\author[1]{Margot Ferrand}
\author[2]{Vincent Labatut}
\affil[1]{CIHAM -- UMR 5648, 74 rue Louis-Pasteur, Avignon, F-84029, France\hspace{1.4cm} Ville d'Avignon, Maison du Patrimoine, 20 rue du Roi René, Avignon, F-84045, France. email: \href{mailto:margot.ferrand@alumni.univ-avignon.fr}{\texttt{margot.ferrand@alumni.univ-avignon.fr}}}
\affil[2]{LIA -- UPR 4128, Avignon Université, 339 chemin des Meinajariès, Avignon, F-84911, France. email: \href{mailto:vincent.labatut@univ-avignon.fr}{\texttt{vincent.labatut@univ-avignon.fr}}}

\maketitle

\begin{abstract}
In historical studies, the older the sources, the more common it is to have access to data that are only partial, and/or unreliable or imprecise. This can make it difficult, or even impossible, to perform certain tasks of interest, such as the segmentation of some urban space based on the location of its constituting elements. Indeed, traditional approaches to tackle this specific task require knowing the position of all these elements before clustering them. Yet, alternative information is sometimes available, which can be leveraged to address this challenge. For instance, in the Middle Ages, land registries typically do not provide exact addresses, but rather locate spatial objects relative to each other, e.g. $x$ being to the North of $y$. Spatial graphs are particularly adapted to model such spatial relationships, called \textit{confronts}, which is why we propose their use over standard tabular databases. 
However, historical data are rich and allow extracting confront networks in many ways, making the process non-trivial. In this article, we propose several extraction methods and compare them to identify the most appropriate. We postulate that the best candidate must constitute an optimal trade-off between covering as much of the original data as possible, and providing the best graph-based approximation of spatial distance. Leveraging a dataset that describes Avignon during its papal period, we show empirically that the best results require ignoring some of the information present in the original historical sources, and that including additional information from secondary sources significantly improves the confront network. We illustrate the relevance of our method by partitioning the best graph that we extracted, and discussing its community structure in terms of urban space organization, from a historical perspective. Our data and source code are both publicly available online.  
\end{abstract}


\medskip
\textcolor{red}{\textbf{Cite as:} Margot Ferrand \& Vincent Labatut, Approximating Spatial Distance Through \textit{Confront} Networks: Application to the Segmentation of Medieval Avignon, \textit{Journal of Complex Networks}, 13(1):cnae046, 2025. DOI: \href{http://doi.org/10.1093/comnet/cnae046}{10.1093/comnet/cnae046}}

\section{Introduction} 
\label{sec:Intro}
To study urban spatiality in medieval history, it is necessary to cross-reference different types of sources related to the morphology and use of space: written, planimetric, iconographic, and archaeological sources~\cite{Arnaud2008}. The relevant written sources are particularly numerous and can be of various kinds: legal, tax or religious records, census data, chronicles and annals, correspondence, testaments... Among them, land documentation, which encompasses inventories of properties, administrative surveys, terriers, censiers, and recognition books, is particularly useful when focusing on the Middle Age, due to the state of medieval archives. Indeed, from the 13\textsuperscript{th} century onwards, this documentation increases considerably and constitutes a major part of the written records~\cite{Bertrand2015}. Regardless of the language in which they are written, these documents have common characteristics. Each one of them contains a series of declarations in which tenants declare the properties they hold under the direct domain of a lord and for which they must pay an annual fee. 

However, historians encounter serious  difficulties in dealing with land documentation, which is too often confined to strictly accounting treatment, thereby overlooking what these sources can reveal in terms of the uses and representation of urban space. The thorny issue with this documentation is the spatialization of information, to which its incompleteness must often be added. Indeed, historians who have studied this documentation have often tried to reconstruct medieval parcels, i.e. to locate them as precisely as possible on a map~\cite{Hautefeuille2016}. But this task is almost impossible: before the modern era, it is very rare for these sources to be accompanied by maps, and the provided locations do not yet refer to precise addresses. They are only relative, and difficult to interpret. In this interdisciplinary work combining History, Geography, Computer Science, and Network Science, we show that this step is not mandatory to analyze the distribution of properties recorded in the land documentation and to fully exploit these sources. Historical analysis at the city scale does not need the land holdings to be precisely located but rather requires that relative locations be possible, in order to interpret the distribution of masses, the filled and empty spaces, proximities and distances, attractions, and separations.

In this article, we propose a graph-based approach to model the urban space based on historical sources. Its main advantage is to focus on the spatial \textit{relationships} between the represented objects, e.g. $x$ being to the North of $y$. By using such a \textit{relative} way of locating these objects, our method does not require estimating their \textit{absolute} location on a map, which is a major issue of the standard approach. As a consequence, our method can handle incomplete data that would provide the absolute location of only a part of the described objects. The resulting graph is designed to encode the notion of spatial proximity, and can be used as an approximate representation of the urban space to conduct tasks such as spatial segmentation. 

Our contributions are two-fold. On the methodological side, we propose several methods to extract so-called \textit{confront} networks from historical data taking the form of land registries. We also propose two objective criteria to compare these extraction methods and identify the most appropriate one. On the applications side, we present a dataset describing the city of Avignon during the papacy, and use it as a benchmark for our methods. We extract and compare 12 different versions of the confront network, and select the most suitable to a spatial analysis. We then show empirically its relevance by segmenting the urban space through community detection, and discussing the historical value of the resulting subdivisions of the city.

The rest of this document is organized as follows. In Section~\ref{sec:Histo}, we provide further historical background, especially regarding medieval Avignon. Section~\ref{sec:Data} describes how we built a geographic database by combining the information coming from five primary historical sources. We then turn to graphs, by highlighting the interest of such modeling approach in Section~\ref{sec:GraphJustif}, and describing in detail the proposed graph extraction methods in Section~\ref{sec:Extraction}. We compare experimentally these methods by applying them to our Avignon database in Section~\ref{sec:MethComp}, a process that results in the identification of the optimal confront network. We detect the communities of this graph, which we discuss from a historical perspective in Section~\ref{sec:ComStruct}. Finally, Section~\ref{sec:Conclusion} summarizes our main contributions and findings, and identifies our most promissing research perspectives.








\section{Historical Background}
\label{sec:Histo}
After the division of Provence in 1125~\cite{Mazel2011}, the city of Avignon was subject to a complex seigniorial fragmentation. Different powers manifested themselves in the city and many of them gradually acquired rights over the land and urban property. Before the middle of the 14\textsuperscript{th} century, these rights were not clearly defined by all the land lords. Only the greatest of them, and more particularly the Counts of Provence, had surveys carried out to identify the rights and property on their land. It was with the arrival of the papacy in Avignon that all the land lords, whatever their importance in the city, expressed the need to precisely delimit the rights they possessed over the land. The phenomenon became even more pronounced when Pope Clement VI bought the city from the Countess of Provence, Queen Jeanne, in 1348. The local lords had to protect and guarantee their rights, as they were an important source of economic income. It was therefore essential to be able to clearly identify the plots of land that felt under their lordship. 

Most of these plots were granted to third parties, called \textit{tenures}. The landowner could indeed transfer the beneficial domain of the property while retaining the eminent domain. The person benefiting from the beneficial domain, whom we will call the \textit{tenant}, could then sell, transfer, or lease the property. The landowner received an annual fee, called \textit{cens}, from the tenant, as well as \textit{transfer fees} in the event of the sale of the beneficial domain.

From the second half of the 14\textsuperscript{th} century on, inventories of tenures multiplied, and census practices were perfected. An increasingly complex listing of the rights of the landed lords, cataloging all the leased tenures under their lordship, was put in place. It took the form of various land registers, the most successful of which, particularly from a legal standpoint, is known today as \textit{terrier}~\cite{Fossier1978}.

In the rest of this section, we describe the main characteristics of these land registers (Section~\ref{sec:HistoTerrier}), and the challenges related to their exploitation (Section~\ref{sec:HistoIssues}).

\subsection{Notion of \textit{Terrier}}
\label{sec:HistoTerrier}
\begin{figure}[!h]
    \centering
    \includegraphics[width=0.8\textwidth]{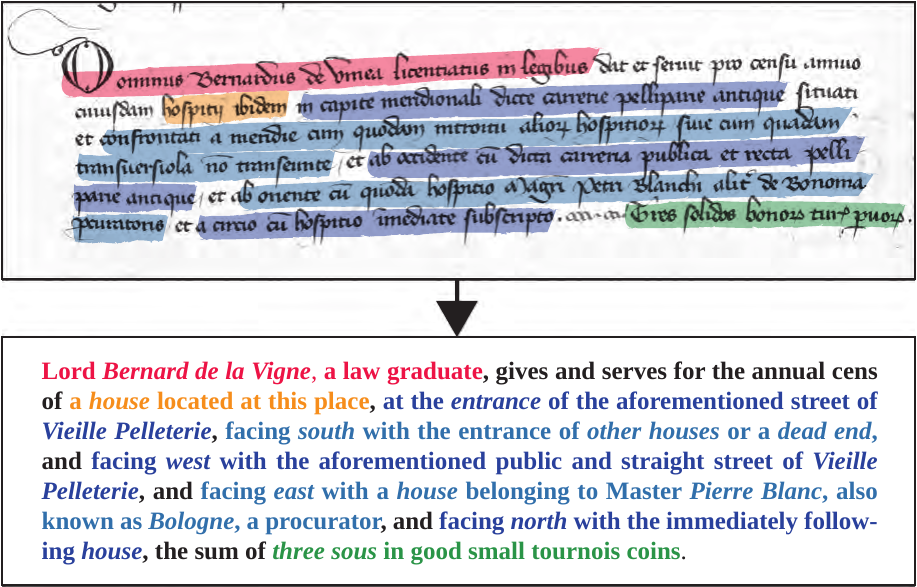}
    \\[3mm]
    \includegraphics[width=1\textwidth]{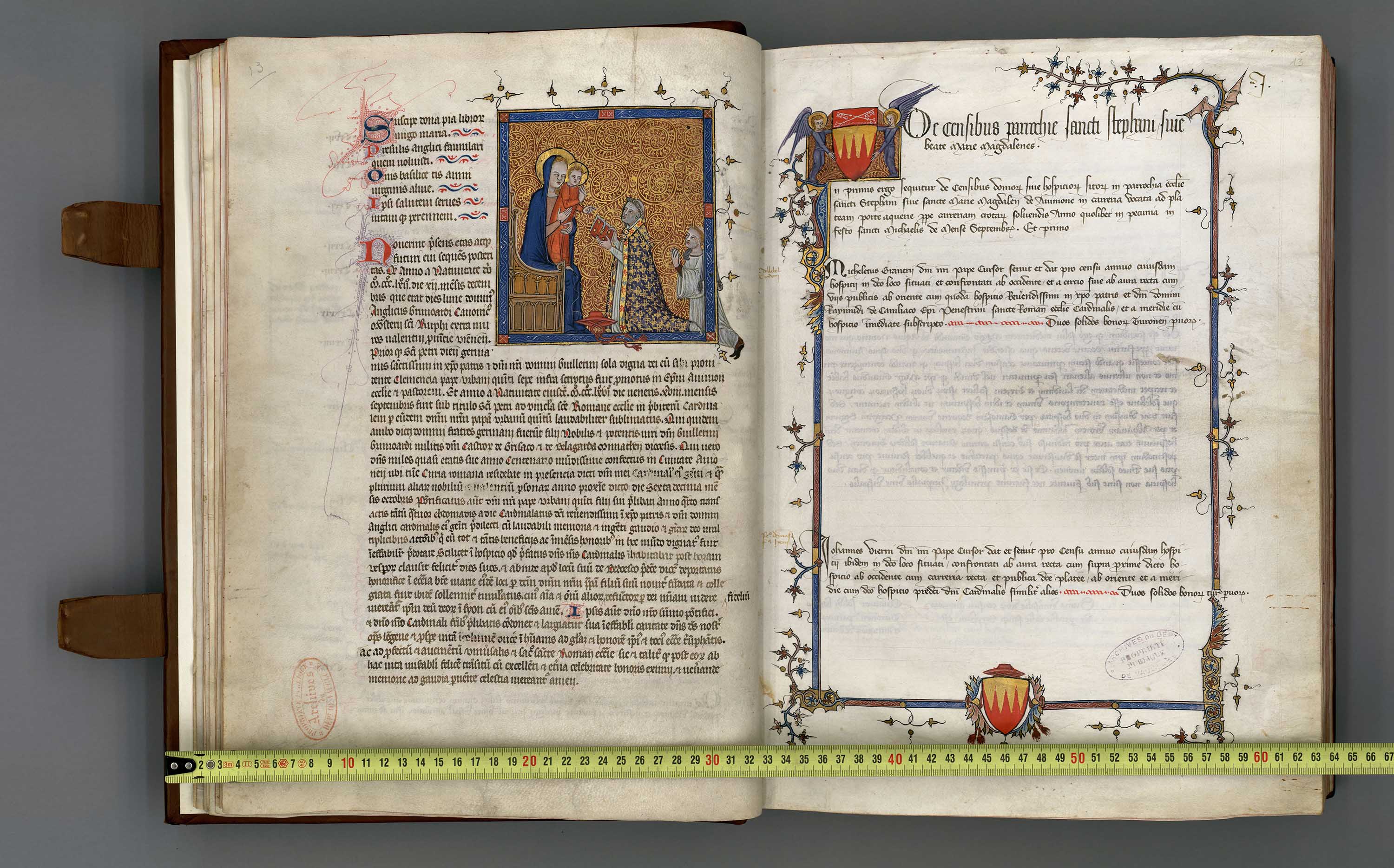}
    \caption{\textbf{Top:} Example of declaration retrieved from a terrier, Vaucluse Departmental Archives, 1G10 f.9v. It includes the original text (first frame) and its English translation (second frame). Each color represents a different piece of information: tenant (red), property (orange), location (blue, 5 different confronts here), and fees (green). Italics denote entities of interest. Diagram available at \href{http://doi.org/10.5281/zenodo.14175830}{10.5281/zenodo.14175830} under CC-BY license. \textbf{Bottom:} Terrier of Bishop Anglic Grimoard, Vaucluse Departmental Archives, 1G10 f.1.}
    \label{fig:ExDeclaration}
\end{figure}

Starting from the 14\textsuperscript{th} century, most landed lords commissioned the creation of land books. These were real tools intended to manage, authenticate, and protect the rights of landed lords on their properties. 
They list, in a serial and often stereotyped manner, the properties of those who had to pay a fee to a lord, giving a certain amount of information on the tenant, the land and the amount of the fees to be paid. Each land book consists of a series of declarations, presented as in Figure~\ref{fig:ExDeclaration} (left). Each declaration is constructed in the following way. The identity of the tenant is given first (shown in red in Figure~\ref{fig:ExDeclaration}), and can be defined in different ways. In addition to the person's first and last names, it may contain, for example, their occupation, social status and may be complemented by their family ties, familiarity (i.e. entourage) or professional ties. 

Next, the property held by the taxpayer is mentioned (in orange in the figure). In most cases, only the type of property is indicated (house, yard, garden, vineyard, undeveloped land). A more precise description of the plot is sometimes given (state of conservation of the property, building materials, size). The location of the property is then regularly indicated (in blue in the figure). From the 14\textsuperscript{th} century onwards, this location is increasingly precise, while always remaining \textit{relative}. The properties are first of all located in relation to the parish territory; then the street and the \textit{confronts}, i.e. the neighborhood, are mentioned. 
The due tax is generally indicated at the end of the entry (in green in the figure).
A number of different currencies can be used, and the census can even be specified in kind. This makes it difficult to estimate its exact value and to compare taxes from one entry to the other. 

Whatever the internal organization of the declarations in the documents, it most often testifies to the increased desire of the institutions to construct genuine management tools intended to be regularly used as legal proof of their possession. These documents are also intended to be archived. In addition to the legal function, most of these documents have a memorial function, which is illustrated above all by their appearance. Some of them are indeed large and beautiful registers, with an elaborate appearance, as shown in Figure~\ref{fig:ExDeclaration} (right).

In a city that had become the seat of Christendom, where the Pope had been the true political lord of the city since 1348, and in which numerous land lords held rights over the land, it was now essential that everyone be able to legally assert their possessions. 
These registers are therefore as much legal evidence as they are demonstrations of the power of the landed lords in relation to the other powers present in the city. 

It is impossible to provide an exhaustive list of landed lords who hold rights over urban land in Avignon. While some lords are well known and their rights can be partly identified through produced and preserved documentation, others remain completely unknown to us. Among the most well-known and important landed lords are ecclesiastical institutions, which have multiplied in the urban space of Avignon even before the arrival of the papacy in the city. Moreover, they have shown an early willingness to archive their documentary production. Thus, their documentation --especially that which is the focus of our study, namely the documentation produced by the bishopric, the cathedral chapter, the hospital of Saint-Jean de Jérusalem, the chapter of the Saint Pierre collegiate church, the Sainte-Catherine monastery, and the House of Repentants-- is still the best preserved and most accessible in the archive holdings today. Alongside these institutions are illustrious Avignonnais who have acquired significant land lordships through political and economic opportunities. This is the case, for example, of Jean Cabassole, knight and advisor to the Count of Provence Robert d'Anjou. The Count granted him the lordship of a strip of land parallel to the city walls in 1319. Like ecclesiastical land documentation, that produced by the urban community is of considerable richness, and testifies to the manifest desire to preserve the traces of its history and properties. Finally, when the Pope bought the city from the Countess of Provence Jeanne d'Anjou, he became the only political lord of Avignon. On another scale, that of land ownership, he was nonetheless one landed lord among others. The Pope did not have a land registry produced, but we know all of his properties and rights thanks to the accounts of the city's clavigers (a type of municipal officer). These accounts list, in the same way as land registries, all the leased properties belonging to the papal lordship and for which individuals paid a fee.

\subsection{Limitations and Difficulties}
\label{sec:HistoIssues}
Terriers are true multiple lists, whose content offers numerous possibilities of analysis. Historians and geographers, particularly those from the \textit{Annales School}, quickly became interested in studying them~\cite{Bloch1929}. However, they faced several obstacles in doing so: 1) quantity and incompleteness of the data; 2) lack of data standardization; and 3) geographic uncertainty.

\medskip\noindent\textbf{Data Incompleteness} 
The medieval city is a fragmented territory governed by various intertwined and often opposing powers. The documentation reflects this reality: it is often incomplete and fragmented, rarely covering the entirety of a territory. Studying a city based on these sources therefore requires cross-referencing and comparing the entirety of this particularly extensive documentation. However, the volume of data to process makes this work particularly tedious, and even challenging to approach using traditional historical methods. This extensive amount of data, previously studied in isolation, can now be approached and examined with a fresh perspective thanks to computer power and computational sciences. Moreover, researchers today have access to a substantial number of edited and sometimes digitized written sources, which they can approach in a serial manner.

In Avignon, the first cadastre covering the entire urban area was only created in the late 16\textsuperscript{th} century. Before this date, it was necessary to gather land books initiated by different lords, each covering different parts of the territory, to obtain the most comprehensive view possible of the urban space. Some areas of the city are better documented than others. Certain zones are not covered by the documentation, either because it has been lost or because it never existed. Some plots of land are indeed free of rights and are fully owned by their owners. In such cases, the holders do not have to pay any fees to a land lord, and these properties are therefore not listed in the land books mentioned above. In this sense, the systematic study of land documentation can also highlight free zones within the city. However, studying these sources systematically requires significant data preparation. 

\medskip\noindent\textbf{Data Heterogeneity} 
Although terriers share many similarities, the data they contain are not standardized. First, the documents are not all written in the same language. In the case of Avignon, for instance, some are written in Latin while others are in Provençal (the local language at the time). Therefore, the terminology used to refer to the same entity can be very different from one document to another. In the Sainte-Catherine registry, written in Provençal, we encounter Juan Teyseyre; but the same individual is named Johannes Textoris in the Anglic Grimoard registry written in Latin. Moreover, in the medieval period, spelling was not standardized. Significant variations exist even between documents written in the same language. While it is written as Juan Teyseyre in the Sainte-Catherine registry, in the city registry, also written in Provençal, we find the spelling Johan Teysseyre. Thus, considerable work is needed to extract the information from the sources and formalize the data for serial analysis.

\medskip\noindent\textbf{Geographic Uncertainty} 
One crucial point that complicates the analysis of this documentation lies in the geographic dimension of the information. Historians have long sought to reconstruct the land division, considering it an essential step in analyzing terriers. This is a very long and tedious task that can only be carried out for limited and particularly well-documented areas~\cite{Claveirole2001, Brunel2002}. If we want to consider the entire urban space, as we have seen, it is necessary to take into account a vast amount of data and cross-reference them: this work cannot be done manually. Moreover, due to the available information, reliably accomplishing this reconstruction is practically impossible. 
Indeed, property addressing, street name signage on panels, and street numbering only began to be standardized in the 18\textsuperscript{th} century, particularly following the Napoleonic conquests. Towards the end of the Middle Ages, informal street names became common; however, they were not standardized and could vary from one person to another or from one document to another. 
The emergence of street names primarily pertains to the roads that contribute to the political representation of the city and its economic prosperity; the chosen toponyms are a direct testament to this. Most of the names already in use in the 13\textsuperscript{th} century continued to be used at the end of the medieval period. Nevertheless, there were instances where they were no longer applied to the same street, or where they encompassed multiple streets. 

In the land documentation used for this study, the most precise information we have for locating a property, after the street, is that of its boundaries or \textit{confronts}. However, this information does not always allow us to determine the exact position of the property. It provides a \textit{relative} position. It was probably very clear to its contemporaries and left little doubt about the location of the declared parcels. However, today, out of its original context, it can sometimes be completely elusive to us. In fact, when the position of a property is given by its \textit{confronts} with properties owned by various individuals and not with a building, a street, or any other fixed point whose location is known to us, it can be extremely challenging, or even impossible, to precisely locate the properties mentioned in the sources.


\section{Geographic Database}
\label{sec:Data}
The information contained in terriers can hardly be exploited directly, due to their structural and textual heterogeneity, and to their complexity. In the context of this project, we design and apply natural language processing (NLP) tools to extract the relevant information and store it in a proper geographic database. The detail of this process, as well as the exact architecture of the resulting database, are out of the scope of this article: they are provided in~\cite{Ferrand2022}.

In this section, we first briefly summarize the process proposed to handle this task (Section~\ref{sec:DataNlp}), then we describe the obtained spatial objects (Section~\ref{sec:DataObjects}) and the spatial relations between them (Section~\ref{sec:DataRelations}), which populate our database. Finally, we explain how we complement this database with additional secondary sources (Section~\ref{sec:DataAdd}).

\subsection{Information Extraction}
\label{sec:DataNlp}
The comprehensive list of the historical sources used to constitute our database is available in Appendix~\ref{sec:ApdxHistSources}. A thorough initial reading of these land registers reveals a certain level of regularity in their structure and in the form of the information they contain. This observation, together with the amount of text to process, the substantial number of edited or already transcribed texts, and the unavailability of any appropriate tool able to handle Latin and Provençal, justifies the elaboration of a custom semi-automatic process aiming at identifying information of interest in the text, extracting it, and storing it in a relational database. Fully automating this process appears unnecessarily difficult, which is why we adopt a method consisting in automating as much of it as conveniently possible, while still relying on human verification and correction in order to control the quality of the data produced in the end.

The first step consists in scanning the edited or transcribed terriers, as no digital editions exist, before performing Optical Character Recognition (OCR) in order to get electronic versions of our historical sources. A manual intervention is necessary to solve the errors introduced by the OCR tool. 
Note that handwriting recognition tools were not yet sufficiently advanced at the beginning of our study, so we focused on edited or, at a minimum, transcribed documentation that was already substantial. However, handwriting recognition processes are becoming increasingly effective and should soon enable the consideration of even larger documentation.

At this stage, we can apply our NLP tool, called \textit{Auto-Annot}~\cite{Ferrand2022}. It relies on a fine-grained categorical model defined based on the studied sources. It is built using a dual symbolic approach, as it both takes advantage of a set of predefined rules describing recurrent patterns and external/internal clues, and of a manually constituted lexicon. The tool is designed to recognize multilingual documents, distinguishing between Latin and Provençal. Entity-wise, it detects the information previously described in Section~\ref{sec:HistoTerrier} (cf. Figure~\ref{fig:ExDeclaration}): individuals, fees, properties and other spatial objects. Some entities are described according to several traits, for instance individuals are likely to be described using a first name, family name, nickname, honorifics, hometown,  health or legal status. Relation-wise, our tool detects inter-individual links (e.g. family or professional relations) as well as spatial relationships that allow the relative positioning of spatial objects (e.g. being North of a church). The different types of entities and relations targeted in the text were identified during our first reading of the historical sources, and were later complemented iteratively until reaching a satisfying quality level. 

The lexical variability characteristic of our historical sources is a major difficulty to detect entities and their relations. The form of Latin used in the late Middle Ages tends to use fewer declensions, but they are still present. Furthermore, the language is marked by processes of vernacularization, including a greater flexibility in sentence construction. Finally, there is no orthographic standardization; quite the opposite. The liberties taken by the writers in choosing vocabulary and spelling are particularly significant. Not to mention the spelling of proper nouns, which sometimes offers surprising variations. It is not uncommon to encounter several different spellings for the same word, or different words used to refer to the same entity within a document. To solve these issues, our tool includes a custom lemmatizer, and a post-processing step leverages Levenshtein's distance to detect similar strings, and more specifically homonyms. 

The architecture of our database directly depends on the types of entities and relationships extracted from the text. In addition, it allows storing some information that is not present in the terriers, but rather comes from some secondary sources, such as the absolute location of certain objects (under the form of GPS coordinates). This database allows us to represent geographic information when we have sufficient indications about the location of a historical object and to query the extracted data from the sources more precisely. We next list and discuss the different types of entities and spatial relations between them, and how we exploit the secondary sources.


\subsection{Spatial Objects}
\label{sec:DataObjects}
At the end of the extraction process, our database contains nine types of spatial objects, as listed in Table~\ref{tab:ObjectTypes} and detailed in the following.

\begin{table}[!h]
    \caption{Types of spatial objects present in our database, with their main characteristics. Column \textit{Nbr}. shows their number of occurrences in the database, before graph extraction. Column \textit{Inv}. indicates whether the objects are invariants (cf. text).}
    \label{tab:ObjectTypes}
    \centering
    \resizebox{\linewidth}{!}{%
        \begin{tabular}{l r l l l}
            \hline
            \textbf{Type} & \textbf{Nbr.} & \textbf{Description} & \textbf{Inv.} & \textbf{Dimension}\\
            \hline
            Properties & 3,021 & Privately owned pieces of real-estate (declared and undeclared) & No & Punctual \\
            Parishes / sectors & 19 & Parochial territories and geographical sectors & No & Surface \\
            Boroughs & 59 & Extensions of urbanization outside the old city walls & Yes & Surface \\
            Defensive system & 7 & Moats, ramparts, walls and towers & Yes & Linear \\
            Gates & 32 & All forms of openings piercing the city walls & Yes & Punctual \\
            Liveries & 25 & Residences granted to the cardinals & Yes & Surface \\
            Geological landmarks & 6 & Points of reference such as rocks, rivers, and canals & Yes & Linear/Surface \\
            Streets & 326 & All types of roads and ways & Yes & Punctual/Linear \\
            Edifices & 152 & Institutional properties & Yes & Punctual \\
            \hline
        \end{tabular}
    }
\end{table}

\begin{figure}[!h]
    \centering
    \includegraphics[width=0.49\linewidth]{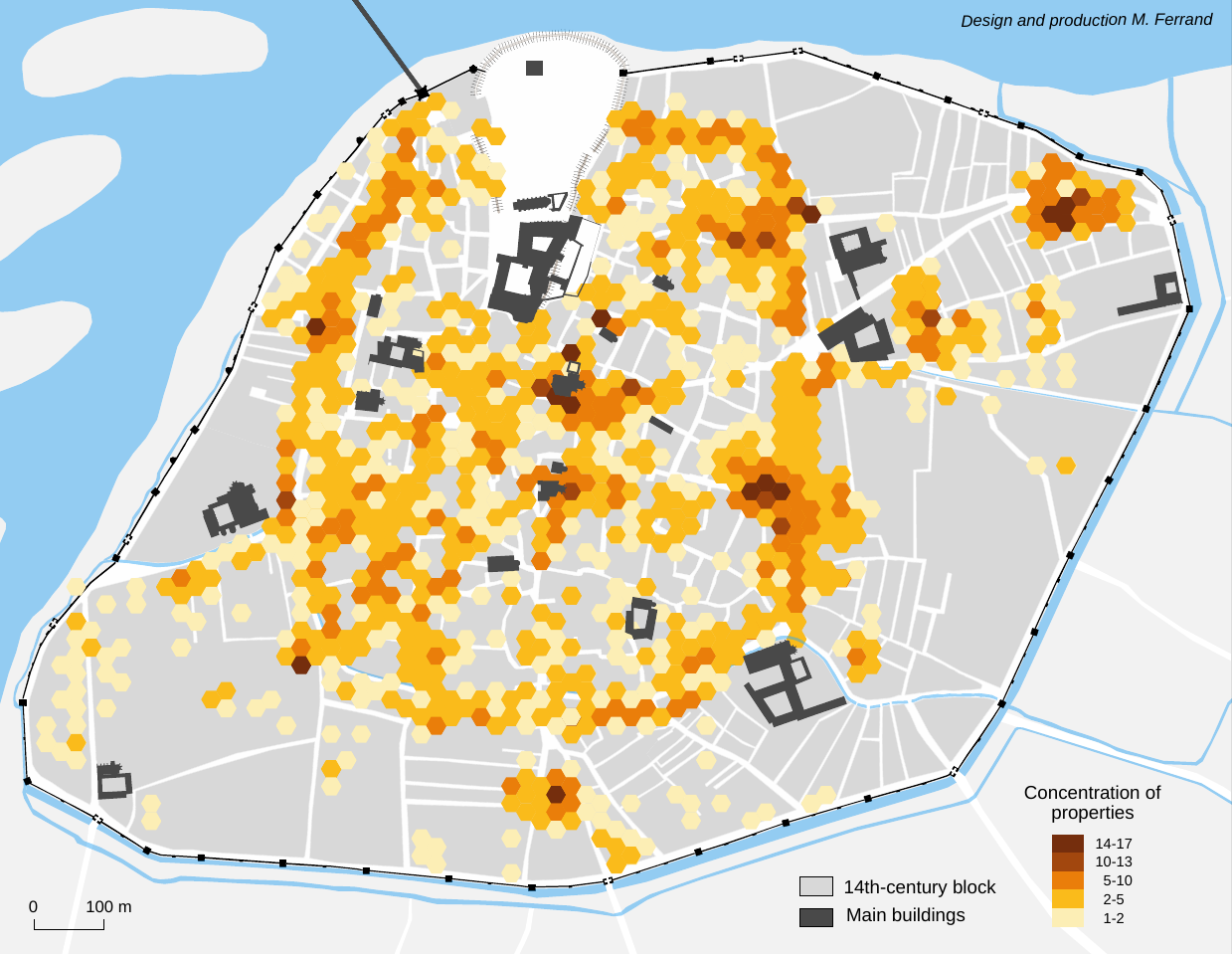}
    \hfill
    \includegraphics[width=0.49\linewidth]{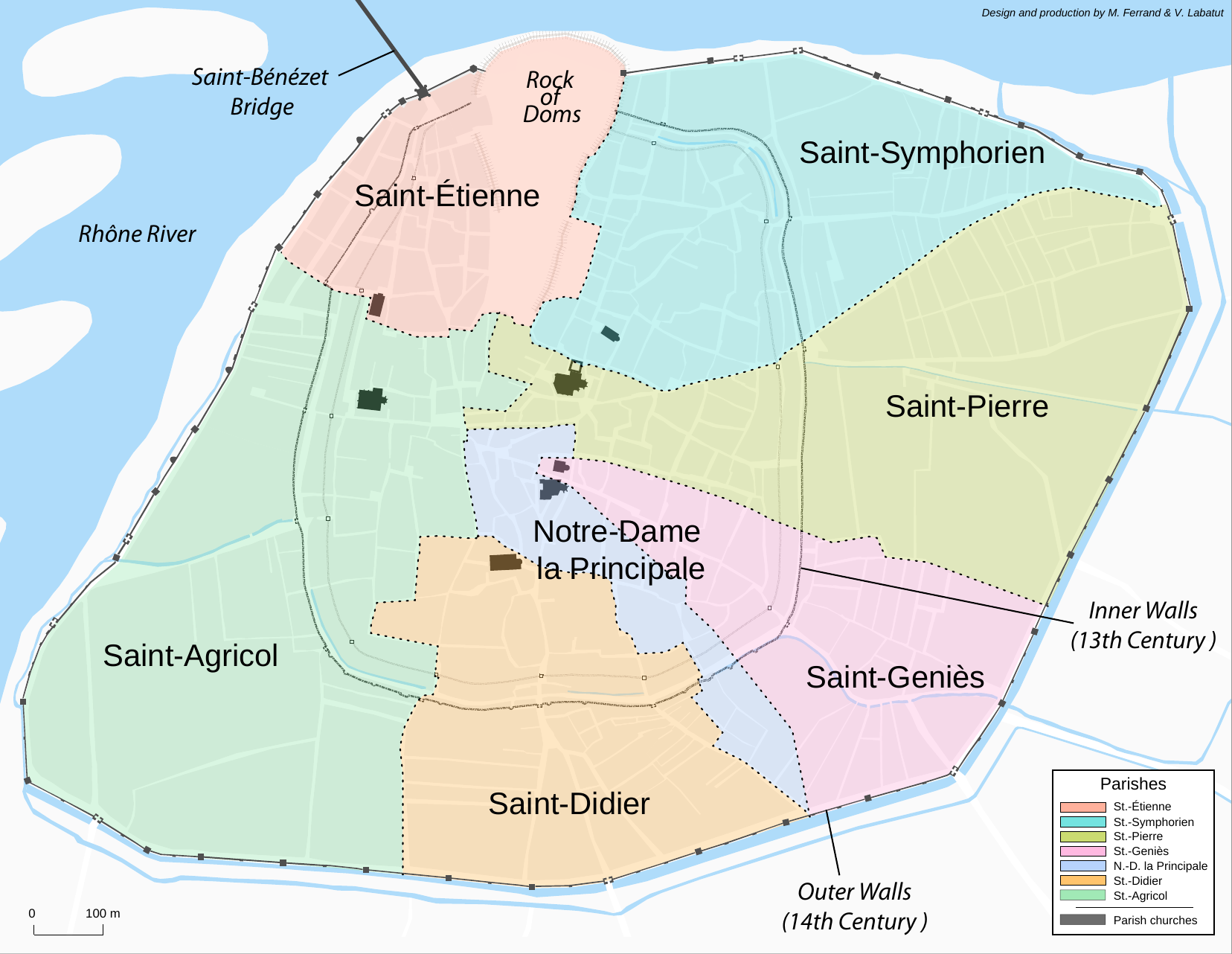}
    \caption{\textbf{Left:} Density map of properties (declared and undeclared) in our dataset; location by interpolation using the grid method. \textbf{Right:} seven parishes of Avignon, and main geological landmarks. Plots available at \href{http://doi.org/10.5281/zenodo.14175830}{10.5281/zenodo.14175830} under CC-BY license.}
    \label{fig:AvignonMap}
\end{figure}

The first type corresponds to private \textit{properties}, as their description is the main focus of the terriers. It includes houses, backyards, shops, gardens, vineyards, fields, barns, cellars, and others. Figure~\ref{fig:AvignonMap} (left) shows their distribution over the city, based on the historical sources. Some of these properties appear as proper entries in a terrier, as they are explicitly \textit{declared} by their tenants. But some properties are \textit{undeclared}, in the sense that they appear in the terriers only as confronts, i.e. as spatial points of reference used to locate declared properties in a relative way. They might be properly described in other, unavailable documents, though.

The second type of spatial object is \textit{Parishes and sectors}. In the late Middle Ages, Avignon was divided into seven parochial territories of varying sizes. The exact boundaries of these parochial territories are not known to us, as no document references them. However, research conducted by Pierre Pansier~\cite{Pansier1930} and Anne-Marie Hayez~\cite{Hayez1993} has allowed for a proposed reconstruction of these boundaries, as shown in Figure~\ref{fig:AvignonMap} (right). Their cartography reveals significant differences in territorial extent from one parish to another. With urban expansion and the construction of new ramparts in the second half of the 14\textsuperscript{th} century, some parochial territories experienced considerable enlargement (such as the parishes of Saint-Pierre and Saint-Agricol), while for others, expansion was notably limited by the natural landscape (such as the parish of Saint-Étienne). To locate the properties in most of the terriers, the scribes use parish affiliation. However, in some sources, other sectors are used, such as the Jewish quarter, or the space between two gates of the communal ramparts.

The seven remaining object types collectively constitute what we call \textit{invariants}, in the sense that they can be considered as the constants of the urban landscape~\cite{Rodier2012a}. This expression refers to all objects that can be referenced in confronts, excluding properties (declared or undeclared). This includes streets, buildings, geological landmarks such as rocks or canals, city gates, and others similar objects.

\textit{Boroughs} constitute our third object type. By definition, boroughs are ``extensions of urbanization outside the city walls.'' In her research, Anne-Marie Hayez identifies over seventy so-called ``bourgs'' in Avignon at the end of the 14\textsuperscript{th} century and locates them approximately. These are subdivisions, meaning that the initially often agricultural lands were later divided to create multiple plots (in terms of housing units). Their size can vary from a few plots to several hundred. Initially, these subdivisions primarily developed outside the walls of the 13\textsuperscript{th} century. However, in the second half of the 14\textsuperscript{th} century, they became incorporated into the urban space with the construction of the new city walls.

The fourth object type is the city \textit{defensive system}. The communal ramparts consist of double walls, an inner wall and an outer wall, likely constructed in the 13\textsuperscript{th} century. Between the two walls are the \textit{intermural} spaces. A moat is also present in front of the outer wall. In the mid-14\textsuperscript{th} century, new ramparts were built at the initiative of the papacy and the urban community. Each rampart is divided into multiple sections, and has towers. 
The city \textit{gates} constitute the fifth type of spatial objects. The outer wall of the double city ramparts is pierced by twelve gates. Eight of them have corresponding gates on the inner walls, and in those cases, the facing gates bear the same name. The 14\textsuperscript{th}-century ramparts are also pierced by twelve gates.

The sixth type corresponds to Cardinal's \textit{liveries}. Since 1316 and the regulation of housing, every cardinal who arrives in Avignon is granted a \textit{livery}. These are rented residences that are ``delivered,'' hence the name given to them. The space of a Cardinal's livery's is not limited to a single dwelling that accommodates a cardinal and a few of his entourage. It can be considerable and always contains multiple houses. The number of dwellings that a Cardinal's livery's represents depends on the size and needs of a cardinal's court. To delineate the space assigned to them, the cardinals place barriers, called \textit{cancels}, at the end of the streets that make up their livery. These wooden barriers explicitly enclose the territory of the cardinal's residence.

The seventh type gathers \textit{Geological landmarks}, which correspond to geological points of reference, including rivers, channels, and rocks. 
The eighth type is \textit{streets}, and it covers not only proper streets, but also alleyways, squares, dead-ends, pathways, etc.
Finally, the ninth type is \textit{edifices}. It gathers institutional properties, including administrative buildings, pyres, chapels, graveyards, convents, churches, hospitals, granaries, hospices, monasteries, bridges, wells, etc.

We also distinguish our spatial objects in terms of their \textit{dimensions}. First, some are \textit{punctual} (or 0-dimensional) in the sense that they occupy a relatively compact surface. Most of these are single buildings, such as houses, shops, churches, but also short streets. Second, some objects are \textit{linear} (or 1-dimensional), meaning they cannot be reduced to a single position in space, but rather to a sequence of contiguous positions. These are mainly longer streets, as well as walls, rivers and channels. Third, some objects are better modeled as shapes possessing a surface (2-dimensional), because they are very large and/or are constituted of several adjacent buildings: boroughs, parishes, cardinal's liveries, and certain landmarks. Table~\ref{tab:ObjectTypes} shows the dimensionality of each type of object.

\subsection{Spatial Relations}
\label{sec:DataRelations}
In the  historical sources, we identify no fewer than 42 distinct types of spatial relationships between these objects. Based on their semantics, one can gather them in six main categories:
\begin{itemize}
    \item \textit{Cardinal} relationships: being to the north, south, east or west of another object.
    \item \textit{Vertical} relationships: being above or below another object (e.g. an apartment topping a shop, in the same building).
    \item \textit{Horizontal} relationships: being behind an object, facing an object. 
    \item \textit{Street} relationships: being at the beginning or end of a street, at the intersection of two streets.
    \item \textit{Proximity} relationships: being around or near an object, adjacent to an object.
    \item \textit{Hierarchical} relationships: being inside an object (e.g. a building in a borough), surrounded by an object, just outside an object.
\end{itemize}

The \textit{Hierarchical} category corresponds approximately to one third of the relations present in the database. It typically indicates that some punctual object is part (or not) of some 2-dimensional object. By comparison, the remaining two thirds are \textit{flat} relationships, in the sense that they locate an object relative to another object without any notion of inclusion (e.g. \textit{West of}). 
A comprehensive list of all 42 spatial relation types is provided in Appendix~\ref{sec:ApdxConvRel} (Table~\ref{tab:ConvRel}).

\subsection{Additional Information}
\label{sec:DataAdd}
If all the objects described in the land registers, along with their relationships, are inventoried in the database, it is not possible to determine their absolute spatial location, due to insufficient information. Drawing on numerous studies focusing on the urban space of Avignon and using a regressive approach to the landscape based on available planimetric sources and archaeological data, the majority of the invariants, however, have been located in the literature. A proposal for the restitution of the parish territory~\cite{Hayez1993} and the medieval road network has been made~\cite{Pansier1930}, areas of boroughs~\cite{Hayez1977} and cardinalatial liveries~\cite{Hayez1993b} have been identified, and the extent and geographical position of the main buildings have been highlighted. We leveraged these results, and then refined and supplemented them.

We have georeferenced a large part of our spatial objects in a Geographic Information System: 50 out of 59 boroughs, all parishes and geographical sectors, the entire defensive system and city gates from different eras, all geological landmarks, all Cardinal's liveries, 162 our of 326 streets, and 144 out of 152 edifices. It should be noted that while we have georeferenced these objects, only a small portion of them are accurately located. The level of accuracy depends on whether they are still in place, which is the case of certain churches and buildings, and on whether textual, iconographic, and especially archaeological sources allow us to place them precisely. For the others, we rely primarily on a hypothetical location made possible by the compilation of various sources and historical studies.

We matched the declared properties identified in our corpus with the finest spatial reference for which we had information (parish, borough, street), and then arranged them manually based on additional information, particularly confronts. Out of 3,021 properties, 2,049 have been georeferenced. However, this property localization is highly uncertain because it relies on the localization of all the invariants, which is often quite hypothetical. Therefore, it is nearly impossible to locate the properties without errors. It should be noted that we do not have addresses but only relative positions. Attempting to precisely locate the properties inevitably leads to data degradation. 

As a result, we decided to move away from georeferencing the properties and focus on the main historical source of information, which is the topological relationship between the properties. Building on existing historical research and various sources, including planimetric sources used to spatialize certain objects, we were able to enrich our database and add certain relationships between objects that are not present in the land documentation. Data related to the road network, in particular, were enhanced with information about the connections between each street or square. Finally, we also added information about the positioning of the buildings, specifying the street in which each listed building is located.
To refine the information, the streets were then divided into segments, and each of these segments was linked in the database. Similarly, the Rock of the Doms, a very important and frequently used landmark, was broken down into several pieces that were interconnected in the database. This allowed for a more detailed topological refinement of the original information. 

In the rest of the document, we refer to the information extracted from the land registries as \textit{primary} data, whereas that produced from secondary sources through the process described above is called \textit{additional} data.




\section{Graph-Based Modeling}
\label{sec:GraphJustif}
To address the issues of data incompleteness and geographic uncertainty that affect terriers, we leverage graph theory to model the urban space. In this section, we justify this methodological approach (Section~\ref{sec:GraphJustifWhy}) before discussing the challenges it implies (Section~\ref{sec:GraphJustifHow}).

\subsection{Advantages of a Graph Model}
\label{sec:GraphJustifWhy}
Graphs are designed to represent relational information, therefore they are particularly well-suited in our case, where the retrieved data are essentially a collection of spatial objects with spatial relations between them. These objects can be modeled as vertices, and these relationships as edges between them. Moreover, graphs enable us to overcome the need for precise property locations. If the absolute position of a spatial object could be estimated, then this information can be included as coordinates attached to the corresponding vertex, making the graph \textit{spatial}. Otherwise, this vertex simply does not have any known position. In addition, graphs even allow us to handle certain contradictions present in the historical sources, e.g. some building being located to the North of another in some terrier entry, whereas the latter is also declared to the North of the former in another entry. When one does not know which piece of information is correct, keeping both seems to be a reasonable approach. Finally, another important advantage of graphs, compared to the more traditional modeling approach, is that they exempt us from the tedious (or even impossible) task of reconstructing parcel divisions, mentioned in Section~\ref{sec:HistoIssues}.

Not only does graph-based modeling not require reconstructing parcel divisions, but it even allows partitioning the urban space at a higher-level. In the Middle Ages, one of Avignon's characteristics is the absence of any administrative subdivisions other than the parish. The parish is indeed the religious subdivision of the city, but it is also the only civil division used simultaneously from an administrative, fiscal, and military perspective. However, it is a fairly large territory, directly affected by urban expansion, and ultimately not very representative of socio-spatial relationships. Two individuals can belong to the same parish without necessarily living in proximity. Two people can be attached to the same parish without any actual connections. To study the socio-spatial relationships of proximity and the resulting land uses, we must therefore use another scale that is closer to the individual himself. 
A graph extracted from our database defines a certain representation of the urban space, very close to the declarers' perception, as it is based on spatial relationships declared by the tenants themselves. In this regard, we will see that partitioning such a graph by the means of a community detection method~\cite{Fortunato2010} allows for a more refined scale of analysis. In this case, communities correspond to \textit{ad hoc} neighborhoods, built upon the confront relations. Here, we use the word \textit{neighborhood} in the sense of a lived space, that is, as it is perceived and practiced by individuals~\cite{Levy2003}; its definition is based on the topological relation defined in the sources. The neighborhood is therefore primarily understood as the everyday living space in which relationships between individuals are the densest.

\subsection{Graph Extraction Challenges}
\label{sec:GraphJustifHow}
As we will see in Section~\ref{sec:Extraction}, due to the richness of the data collected in our database, there are a number of possible ways to extract graphs representing medieval Avignon. They differ mainly in the data that they leverage: which part of the primary historical data to keep or remove, and which additional secondary information to use. In order to make a choice, it is important to define a method to compare them and identify the most appropriate for our use. 

The first goal is to have a graph containing as many vertices as possible, so that it covers the largest possible area. Put technically, we want to maximize the graph order. More precisely, we want to maximize the number of properties it contains, as these spatial objects constitute our main interest in this study. At the same time, we want this graph to be as dense as possible, so that its structure reliably reflects the spatial organization of the city. Yet, as was observed empirically, complex networks tend to be sparser as they get larger~\cite{Blagus2012}, which means that these two goals of coverage and reliability may be opposed. 

Finally, it is worth stressing that our database describes a number of very different spatial relationships, with very different levels of reliability. Therefore, the extracted edges may not all represent the spatial structure of the city equally, in terms of semantics, granularity, precision, etc. Thus, another challenge is to determine how to build a graph structure that constitutes a good approximation of the spatial structure, in order to produce a relevant partition of the city as a final result.

\section{Graph Extraction Methods}
\label{sec:Extraction}
\begin{figure}[!h]
    \centering
    \includegraphics[scale=0.30]{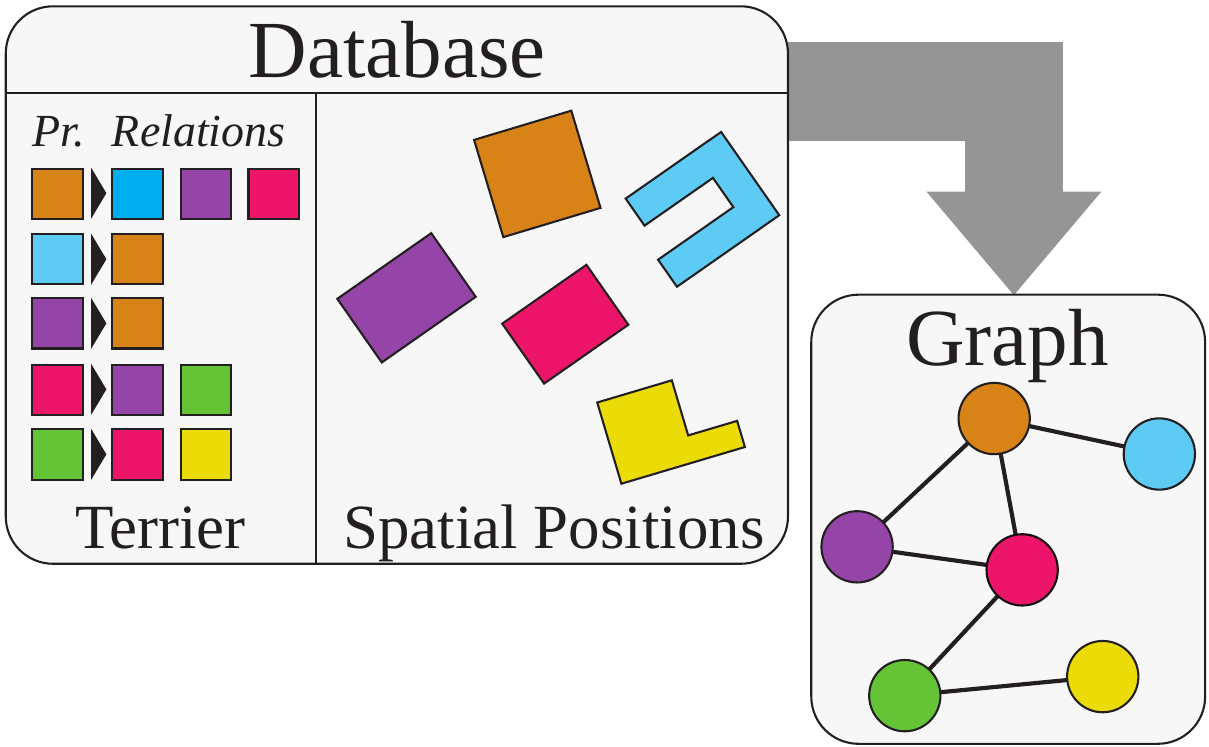}
    \caption{Straightforward extraction of a graph, from our database. Each colored shape represents a spatial object (e.g. a building). The edges depend on the spatial relations described in the historical sources. Figure available at \href{http://doi.org/10.5281/zenodo.14175830}{10.5281/zenodo.14175830} under CC-BY license.}
    \label{fig:ExtrBasic}
\end{figure}

The most straightforward method to extract graphs from our database is to leverage the whole information, and model each object by a vertex, and each relationship by an edge. This is illustrated by a simple example in Figure~\ref{fig:ExtrBasic}. The left part of the \textit{Database} block represents the spatial relations described in the terrier, whereas its right part shows the absolute locations of the listed objects, when available. Each color represents a unique spatial object, some of which are properties. In the \textit{Terrier} part, the left-hand squares are properties, associated (on their right) to their list of declared confronts. Some objects appear only because they are cited as confronts, e.g. the yellow one. The absolute spatial location of certain objects is unknown, e.g. the green property does not appear in the right part of the \textit{Database} block. Graph edges do not necessarily depend on spatial proximity, since they reflect the confronts, i.e. the relationships that are \textit{explicitly} described in the database (left part). For instance, the magenta and yellow objects are spatially close, but not connected by any confront, so there is no edge between the corresponding vertices.  

Our database describes various types of objects (e.g. buildings, cemeteries, walls, gardens, parishes), spatially related in various ways (e.g. being near another object, containing another object). Leveraging the entirety of this information results in a very heterogeneous graph, as it is simultaneously spatial~\cite{Barthelemy2011}, attributed~\cite{Wasserman1994}, multimode~\cite{Borgatti1997}, and multilayer~\cite{Kivelae2013}. It is shown in Appendix~\ref{sec:ApdxGraphs} (Figure~\ref{fig:GraphFull}), for reference. 
However, not all of this information may be useful to produce a graph that would provide a good approximation of the spatial distance. In fact, the preliminary experiments that we conducted revealed that some of this information can even be detrimental. Consequently, one of our methodological challenges consists in identifying which part of the data should be discarded. On the contrary, it is possible to leverage some additional information besides that strictly coming from the terriers, by using our secondary historical sources, and possibly improve the quality of the extracted graphs. We implement these different approaches under the form of several distinct extraction steps. These are not mutually exclusive, they can be chained in various ways, resulting in a number of possible combinations. We consider some of these steps to be \textit{compulsory}, because the extracted graph is hardly usable at all without them: we describe them in Section~\ref{sec:ExtractionSyst}. Some are \textit{optional}, and we later study how they affect the extracted graph: these are the object of Section~\ref{sec:ExtractionOpt}. Finally, we provide the reader with a convenient summary of these methods in Section~\ref{sec:ExtractionOverview}.

\subsection{Systematic Steps}
\label{sec:ExtractionSyst}
Both steps described below are systematically applied to all extracted graphs. They concern the way we normalize relationships, and how we handle disconnected graphs.

\subsubsection{Relationship Types}
The number of distinct relationship types (42) is an issue, as they complicate the interpretation and analysis of the network. Moreover, some of them are very infrequent, occurring only once or twice out of the 6,129 relationships listed in the database. For this reason, we normalize these types to make them coarser, retaining only 7 main types: being to the \textit{North}/\textit{South}/\textit{East}/\textit{West of} an object (i.e. so-called cardinal relationships), being \textit{Inside}/\textit{Outside} an object, and being \textit{Around} an object (a general relationship gathering all the remaining types). The detail of this normalization is provided in Appendix~\ref{sec:ApdxConvRel} (Table~\ref{tab:ConvRel}).

\subsubsection{Minor Components}
Generally speaking, real-world complex networks often contain a so-called \textit{giant component}, i.e. a very large maximally connected subgraph gathering almost all the vertices and edges, whereas the rest are spread over several much smaller separate subgraphs~\cite{Newman2003b}. It is also the case for all the networks extracted from our database, whatever the extraction method. In this situation, the standard approach adopted in the literature is to focus the analysis only on the giant component, or on the few largest components, as the rest of the data are implicitly considered as less complete or reliable.
In our case, more particularly, approximating the spatial distance on these remote and sparser parts of the network is likely to result in a poor estimation. For this reason, for each extracted network, we filter out all vertices and edges located in minor components, including isolates. For our data, we empirically determine that a lower threshold of 25 vertices is appropriate to remove noisy data while preserving relevant information.

\subsection{Optional Steps}
\label{sec:ExtractionOpt}
We perform our extraction procedure by alternatively including and discarding the following steps, in order to assess how they affect the extracted graphs. While describing these steps, we also introduce the notation used later to refer to them and their different variants.

\subsubsection{Non-Punctual Objects}
\label{sec:ExtractionOptNonpunct}
\begin{figure}[!h]
    \centering
    \includegraphics[scale=0.30]{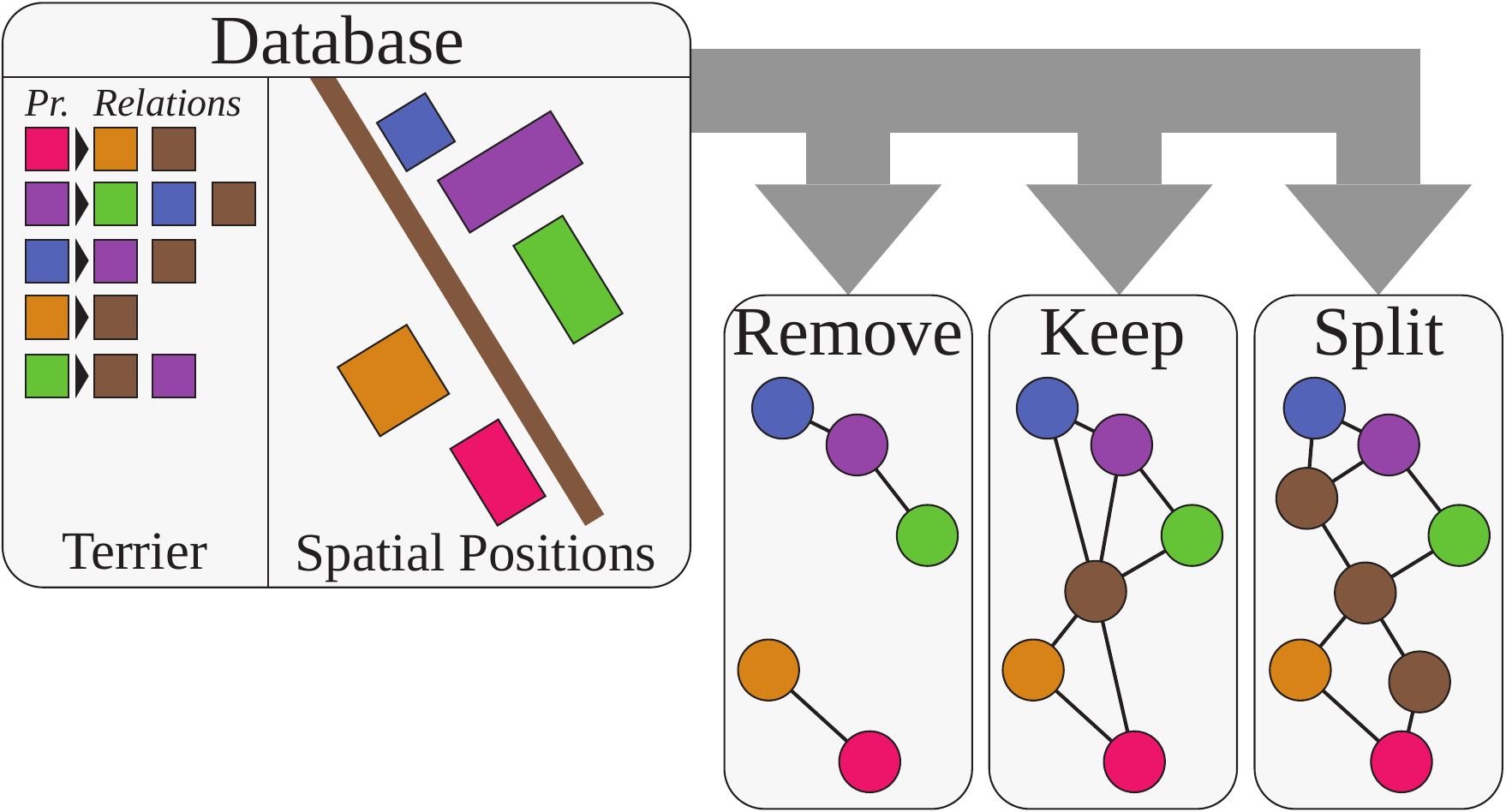}
    \caption{The three strategies proposed to handle 1- and 2-dimensional objects such as the street in this figure (shown in brown): removing it by not representing it at all in the graph, keeping it as it is by modeling it through a single vertex, or splitting it and representing its constituting pieces with several linearly connected vertices. Figure available at \href{http://doi.org/10.5281/zenodo.14175830}{10.5281/zenodo.14175830} under CC-BY license.}
    \label{fig:ExtrSplit}
\end{figure}

The vertices representing 1- and 2-dimensional objects are likely to act as shortcuts in the network. For instance, two buildings located at opposite ends of a street are just two hops away in terms of graph distance, independently of the street length. The blue and pink buildings in Figure~\ref{fig:ExtrSplit} illustrate this situation (\textit{Keep} graph). This could strongly affect the way our confront graph models the urban space. We propose two modifications of the extraction process to deal with this issue: outright removing the concerned vertices (\textit{Remove} graph in Figure~\ref{fig:ExtrSplit}), or splitting them (\textit{Split} graph).

\medskip\noindent\textbf{Removal} 
Removing \textit{all} non-punctual objects is the most straightforward approach. However, our experiments reveal that this causes the networks to be very segmented, i.e. it breaks them down into many small components. Consequently, they may not be suitable to produce acceptable approximations of the urban space.

A less radical approach consists in removing only the vertices representing the longest/largest objects. Of course, this requires possessing the required spatial information, which is not always the case. Our database contains the length of certain streets, and we are confident that this information is missing only for the most minor streets, which are also the shortest. In addition, a few relationships also specifically concern parts of streets (angle, beginning, end, etc.). 

We experiment with two removal strategies\footnote{We actually considered a third strategy consisting in removing only long streets and keeping short streets and parts of streets. It is not presented here because the results are quite similar to the \texttt{k} method, but these are included in our data repository.}: 1) keep all streets (a strategy we denote \texttt{streets}); 2) remove the $k$ longest streets, and keep the rest of the streets (noted \texttt{k}). 
In both strategies, we outright remove the other types of 1- and 2-dimensional objects.

\medskip\noindent\textbf{Split} 
Alternatively, we consider splitting non-punctual objects instead of removing them, i.e. breaking them down into smaller parts, each one represented by a specific vertex. This requires injecting more knowledge in the database than the previous method, though. First, if the spatial location of the original object is known, it is necessary to estimate those of its constituting pieces. Second, to preserve continuity, we must introduce artificial relationships between the adjacent pieces of the split object\footnote{We iteratively remove vertices possessing exactly one such artificial relationship, though, as these leaves make the graph unnecessarily larger without bringing any useful information.}. Third and finally, relationships that involve split objects must be adjusted by identifying which piece of the original object they concern.

In our case, we are able to manually split all linear objects (streets, rivers, channels, walls) in segments sufficiently small to be considered as punctual. We also break down the largest 2-dimensional object (the so-called \textit{Rock of the Doms}). We experiment with two splitting strategies: 1) split all streets (noted \texttt{streets}); and 2) split the $k$ longest streets, and keep the rest of the streets as single vertices (noted \texttt{k}). Like before, the rest of the non-punctual objects are removed. 

We use the letters \texttt{W} (\textit{whole}) and \texttt{S} (\textit{split}) to denote whether or not a graph underwent such a splitting step during extraction.

\subsubsection{Hierarchical Relationships}
\label{sec:ExtractionOptHierar}
\begin{figure}[!h]
    \centering
    \includegraphics[scale=0.30]{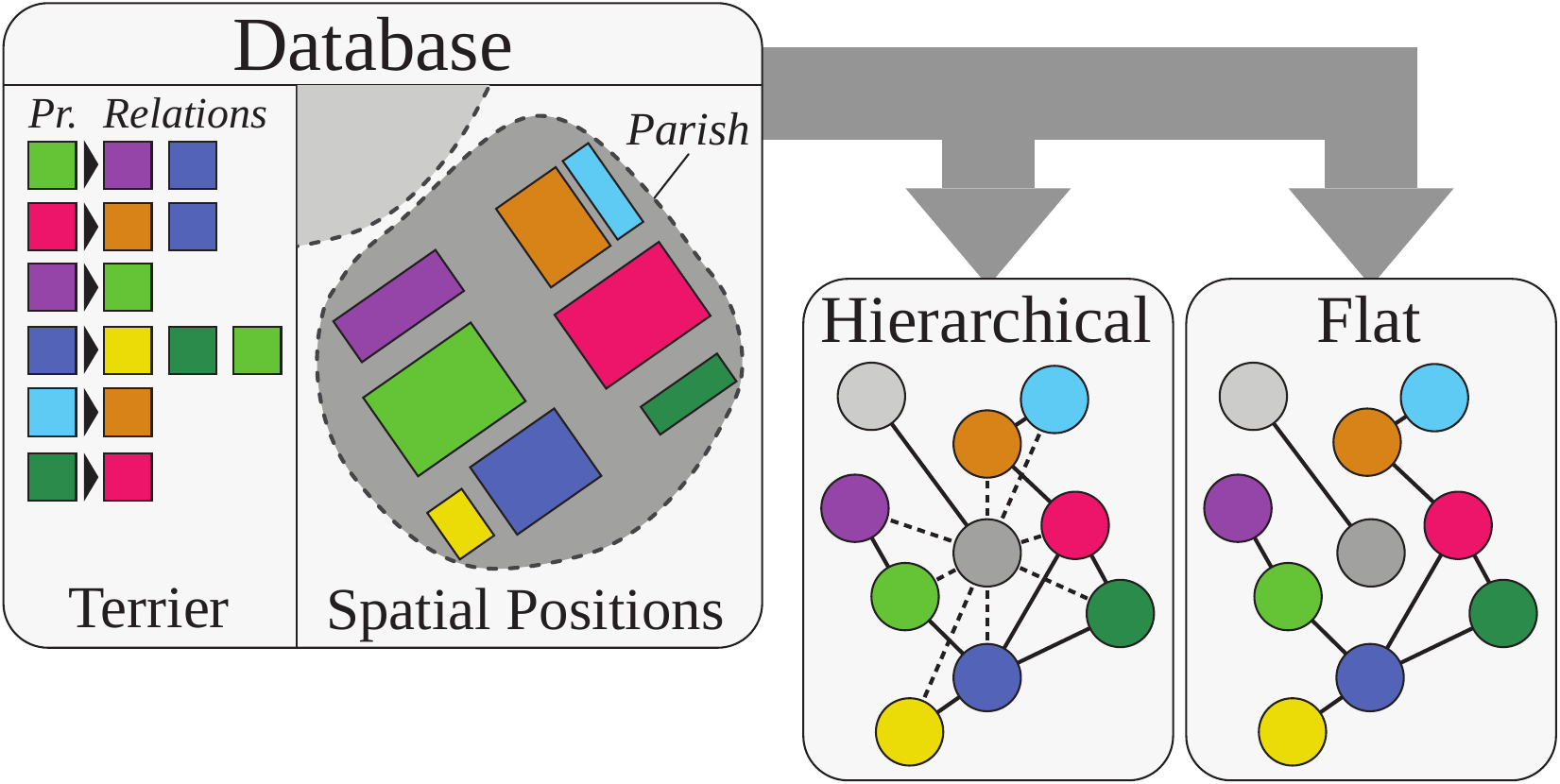}
    \caption{The two strategies proposed to handle hierarchical relationships, such as a parish containing eight buildings in this figure: keep them and make the graph hierarchical, or remove them and make it flat. The dotted edges show the hierarchical relationships. Figure available at \href{http://doi.org/10.5281/zenodo.14175830}{10.5281/zenodo.14175830} under CC-BY license.}
    \label{fig:ExtrHier}
\end{figure}

As explained in Section~\ref{sec:DataRelations}, our database includes hierarchical and flat relationships. The issue here is that 2-dimensional objects, which are involved in the former type of relations, are all very large in our case. Including hierarchical relationships in the extracted graph results in the presence of hubs corresponding to these objects, connected to many punctual objects. Some of these are very far from each other in terms of spatial distance, but are very close in the graph, as one is only two hops away from the other. Consequently, the graph is likely to constitute a poor approximation of the urban space, as already observed when discussing the split of linear objects. This is illustrated in Figure~\ref{fig:ExtrHier}, which shows a parish (in gray) containing eight buildings (in colors). Independently of the spatial distance between them, considering the hierarchical relationships leads to a graph where they are  all two hops away from each other, as they are all connected to the parish vertex. 

We propose to experiment by outright deleting all hierarchical relationships when extracting graphs, a strategy which we note \texttt{F} for \textit{flat}, whereas keeping all relationships is noted \texttt{H} for \textit{hierarchical}.
It is worth stressing that even when focusing on flat relationships only, the membership information can still be retained in a non-structural way, by integrating it in the graph under the form of vertex attributes. For instance, a given vertex can be labeled with the district, parish or borough to which it belongs.

\subsubsection{Additional Relationships}
\label{sec:ExtractionOptAdd}
As explained in Section~\ref{sec:HistoTerrier}, the terriers list and describe only certain declared properties, and only indirectly mention other objects for the purpose of locating them. They therefore ignore objects that were not recognized during this period –-these could be properties belonging to another lord for which no such document is preserved, or properties free from feudal jurisdiction-– or objects that were never mentioned in these documents (such as certain religious and civil buildings, for example). Furthermore, a relationship extracted from a terrier necessarily concerns a property declared in the terrier and some object mentioned in the same terrier. This description is likely incomplete: the authors may not list all possible relationships but only the few necessary to accurately locate the property. Additionally, there is no relationship between two objects that are not declared in these documents, for example, two streets. In summary, our primary historical sources provide an incomplete view of the city.

\begin{figure}[!h]
    \centering
    \includegraphics[scale=0.30]{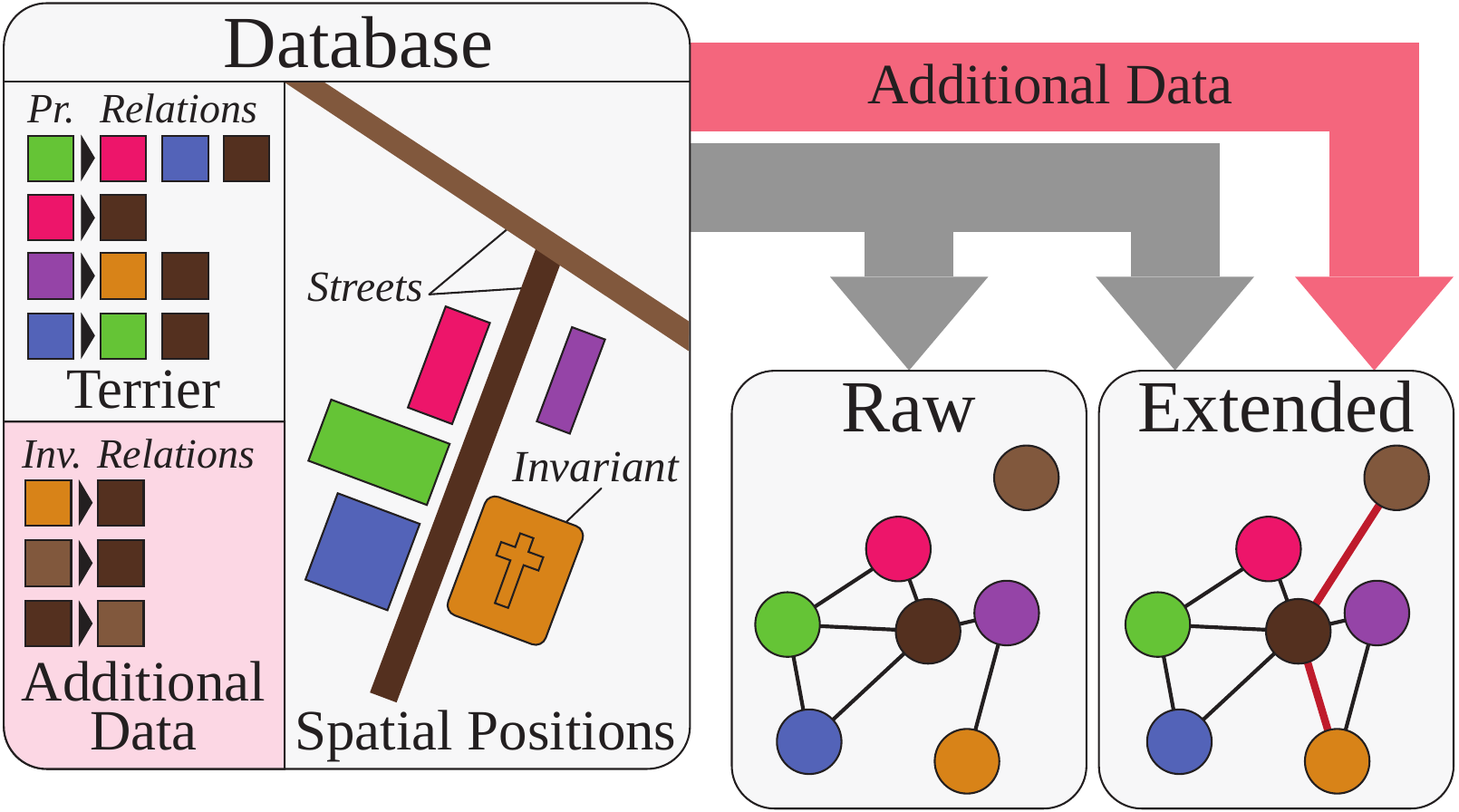}
    \caption{Extending the perimeter of the considered historical sources allows including additional relationships, represented here in red, between streets and other invariants. Figure available at \href{http://doi.org/10.5281/zenodo.14175830}{10.5281/zenodo.14175830} under CC-BY license.}
    \label{fig:ExtrExt}
\end{figure}

The question is to know whether this is enough to get a reliable estimation of the spatial distance through the extracted network. In order to answer this question, we leverage the additional information described in Section~\ref{sec:DataAdd}, in order to complement our graphs with two types of missing relationships: between adjacent streets, and between edifices and adjacent streets. Figure~\ref{fig:ExtrExt} shows how this affects the graphs. The pink part of the \textit{Database} block corresponds to the additional data used to extract additional edges, shown in red in the \textit{Extended} version of the graph. Our goal here is to study the effect of this extra information on the estimation of the distance. We note \texttt{R} (for \textit{Raw} data) the version of our networks without this information, and \texttt{E} (for \textit{Extended} data) the version that includes it.

\begin{table}[!h]
    \caption{List of the different methods proposed to extract spatial graphs from our database, with the processing steps they rely upon. The letters in the code stand for: use only raw data (\texttt{R}) vs. extend them with additional relationships (\texttt{E}); use only flat relationships (\texttt{F}) vs. include also hierarchical ones (\texttt{H}); split large objects (\texttt{S}) vs. keep them whole (\texttt{W}). 
    }
    \label{tab:ExtrMethods}
    \centering
    \resizebox{\textwidth}{!}{%
        \begin{tabular}{l l l l}
            \hline
            \textbf{Method} & \textbf{Additional} & \textbf{Hierarchical} & \textbf{Non-Punctual} \\
            \textbf{Code} & \textbf{Relationships} & \textbf{Relationships} & \textbf{Objects} \\
            \hline
            \texttt{RHW\_all} & No & Keep all & Keep all \\
            \texttt{RFW\_all} & No & Remove all & Keep all remaining non-punctal objects \\
            \texttt{RFW\_streets} & No & Remove all & Keep only streets, remove other non-punctual objects \\
            \texttt{RFW\_k} & No & Remove all & Remove $k$ longest streets and other non-punctual objects \\
            \hline
            \texttt{EHW\_all} & Yes & Keep all & Keep all \\
            \texttt{EFW\_all} & Yes & Remove all & Keep all remaining non-punctal objects \\
            \texttt{EFW\_streets} & Yes & Remove all & Keep only streets, remove other non-punctual objects \\
            \texttt{EFW\_k} & Yes & Remove all & Remove $k$ longest streets and other non-punctual objects \\
            \hline
            \texttt{RHS\_all} & No & Keep all & Split all \\
            \texttt{RFS\_all} & No & Remove all & Split all remaining non-punctal objects \\
            \texttt{RFS\_streets} & No & Remove all & Split only streets, remove other non-punctual objects \\
            \texttt{RFS\_k} & No & Remove all & Split $k$ longest streets, keep other streets, remove the rest \\
            \hline
            \texttt{EHS\_all} & Yes & Keep all & Split all \\
            \texttt{EFS\_all} & Yes & Remove all & Split all remaining non-punctal objects \\
            \texttt{EFS\_streets} & Yes & Remove all & Split only streets, remove other non-punctual objects \\
            \texttt{EFS\_k} & Yes & Remove all & Split $k$ longest streets, keep other streets, remove the rest \\
            \hline
        \end{tabular}
    }
\end{table}

\subsection{Overview}
\label{sec:ExtractionOverview}
The optional processing steps described in this section are not mutually exclusive: on the contrary, they are meant to be combined. This results in a number of possible methods to extract a network from our databases. The goal of our experimental work is to identify which one is the most in line with our objectives. Table~\ref{tab:ExtrMethods} summarizes the combinations of these steps that we consider in our experiments, with the code that we use in the rest of the article to refer to them. We do not consider all possible combinations, because some result in relatively similar effects, even if not exactly identical. For instance, removing all 2-dimensional objects is very similar to removing all hierarchical relationships, as all of the latter involve the former. In the rest of the article, we use a wildcard $\dotp$ to denote any variant over some column(s) of Table~\ref{tab:ExtrMethods}. For instance, \texttt{R$\dotp$W\_all} denotes collectively \texttt{RHW\_all} and \texttt{RFW\_all}.

\section{Method Comparison and Selection}
\label{sec:MethComp}
We apply all 16 variants of the extraction process described in Table~\ref{tab:ExtrMethods}, and get as many versions of the network representing medieval Avignon. All of them are shown in Appendix~\ref{sec:ApdxGraphs} (Figures~\ref{fig:GraphsWholeLambert} and~\ref{fig:GraphsSplitLambert}), as well as the full graph (Figure~\ref{fig:GraphFull}), for reference. The latter is the graph extracted when using the raw dataset, without any vertex or edge removal, or any transformation. Table~\ref{tab:GraphStats} shows the main topological properties of these graphs. In the following, we first discuss methods to measure the coverage of the extracted graphs, and their ability to reliably model the spatial structure of the data (Section~\ref{sec:MethCompGeneral}). We then assess separately the effect of each optional processing step described in Section~\ref{sec:ExtractionOpt}: hierarchical relationship filtering (Section~\ref{sec:MethCompHier}), non-punctual object filtering (Section~\ref{sec:MethCompStreets}), vertex splitting (Section~\ref{sec:MethCompSplit}), and data extension (Section~\ref{sec:MethCompExtend}). Finally, based on these results, we select the most appropriate graph to perform our analysis (Section~\ref{sec:MethCompChoice}).

\begin{table}[!h]
    \caption{Main topological characteristics of the networks extracted from our database: numbers of vertices ($n$) and edges ($m$), density ($\delta$), number of property vertices and proportion of such vertices relative to the database (\textit{Properties}), number of components ($C(G)$), diameter ($d_{\max}$), harmonic mean of the graph distance ($\langle d \rangle$), Spearman's correlation between the graph and spatial distances ($\rho_d$).}
    \label{tab:GraphStats}
    \centering
    \resizebox{\textwidth}{!}{%
        \begin{tabular}{l r r r r@{\hspace{0.5cm}}r r r r r}
            \hline
            \textbf{Method} & $n$ & $m$ & $\delta$ & \multicolumn{2}{r}{\# Properties} & $C(G)$ & $d_{\max}$ & $\langle d \rangle$ & $\rho_d$ \\
            \hline
            Full graph            & 3,173 & 6,619 & 0.0007 & 2,693 & 100,00\% & 110 & 16 &  6.75 & 0.22 \\
            \hline
            \texttt{RHW\_all}     & 2,867 & 6,415 & 0.0008 & 2,397 &  89.01\% &   1 & 16 &  5.51 & 0.29 \\
            \texttt{RFW\_all}     & 2,174 & 4,290 & 0.0009 & 1,807 &  67.10\% &   5 & 45 & 13.12 & 0.03 \\
            \texttt{RFW\_streets} & 2,003 & 3,980 & 0.0010 & 1,673 &  62.12\% &  12 & 33 & 38.22 & 0.48 \\
            \texttt{RFW\_k}       & 1,903 & 3,750 & 0.0010 & 1,597 &  59.30\% &  14 & 40 & 45.10 & 0.49 \\
            \hline
            \texttt{EHW\_all}     & 2,919 & 6,840 & 0.0008 & 2,427 &  90.12\% &   1 & 15 &  5.46 & 0.34 \\
            \texttt{EFW\_all}     & 2,390 & 4,895 & 0.0009 & 1,959 &  72.74\% &   1 & 26 &  8.28 & 0.48 \\
            \texttt{EFW\_streets} & 2,268 & 4,647 & 0.0009 & 1,862 &  69.14\% &   1 & 30 &  9.37 & 0.74 \\
            \texttt{EFW\_k}       & 2,167 & 4,382 & 0.0009 & 1,782 &  66.17\% &   2 & 31 & 11.28 & 0.80 \\
            \hline
            \texttt{RHS\_all}     & 3,074 & 6,630 & 0.0007 & 2,397 &  89.01\% &   1 & 21 &  6.19 & 0.47 \\
            \texttt{RFS\_all}     & 2,381 & 4,505 & 0.0008 & 1,807 &  67.10\% &   5 & 75 & 21.57 & 0.27 \\
            \texttt{RFS\_streets} & 2,032 & 4,017 & 0.0010 & 1,673 &  62.12\% &  12 & 34 & 40.79 & 0.49 \\
            \texttt{RFS\_k}       & 2,020 & 3,578 & 0.0009 & 1,673 &  62.12\% &  12 & 33 & 40.32 & 0.49 \\
            \hline
            \texttt{EHS\_all}     & 3,146 & 7,072 & 0.0007 & 2,427 &  90.12\% &   1 & 22 &  6.13 & 0.48 \\
            \texttt{EFS\_all}     & 2,617 & 5,127 & 0.0007 & 1,959 &  72.74\% &   1 & 51 & 11.25 & 0.74 \\
            \texttt{EFS\_streets} & 2,317 & 4,701 & 0.0009 & 1,862 &  69.14\% &   2 & 34 & 11.86 & 0.69 \\
            \texttt{EFS\_k}       & 2,294 & 4,208 & 0.0008 & 1,862 &  69.14\% &   1 & 37 & 10.32 & 0.80 \\
            \hline
        \end{tabular}
    }
\end{table}

\subsection{Measuring Coverage and Reliability}
\label{sec:MethCompGeneral}
The numbers of vertices ($n$) and edges ($m$) vary significantly between the different versions of the graph. On the contrary, their densities ($\delta$) are comparable, showing a level of sparsity that is on par with what is often observed in general real-world networks of the same order~\cite{Blagus2012, Estrada2011a}. The \textit{Properties} column contains two values: the number of vertices representing properties in the considered graph, and the proportion of properties in the whole database that this number amounts to. As such, these statistics are related to our \textit{graph coverage} criterion described in Section~\ref{sec:GraphJustifHow}: we want these values to be as large as possible. The number of connected components ($C(G)$) is largely affected by the filtering step, as shown by the marked difference between the full graph (110 components) and the other graphs (at most 14 components). A large number of components reflects a very segmented representation of the city space, so this metric is related to how reliably the graph models this spatial organization. 

At first sight, the diameter ($d_{\max}$), which is the largest geodesic distance observed in the graph, could be considered as related to graph coverage: a large value means the graph is ``wide''. However, it is not the case for two reasons. First, this width is only topological, and not necessarily spatial: removing certain vertices can make a graph more linear, thereby increasing its diameter without covering more space. Second, the geodesic distance is traditionally computed by discarding pairs of vertices located in different components, as they are conventionally considered as infinitely distant. Therefore, removing certain vertices from a network can also result in splitting it in several components, thereby decreasing the diameter without necessarily changing its spatial coverage. Both situations appear in Table~\ref{tab:GraphStats}, which confirms that the diameter is not an appropriate metric in our case. The average geodesic distance ($\langle d \rangle$) is more interesting, provided one computes the \textit{harmonic} mean instead of the traditional arithmetic mean. As noted by Newman~\cite{Newman2003b}, it allows taking into account infinite distance values in the computation.

The simplest way to assess how reliably the graph models the spatial structure of the data is probably to measure the statistical association between the spatial distance, i.e. the Euclidean distance between the spatial objects themselves, and the graph distance, i.e. the distance on the graph between the vertices representing these objects. We experiment with Kendall's $\tau$ and Spearman's $\rho$ coefficients, both widespread in the literature, because such rank correlation measures allow taking infinite values into account (unlike the more widespread Pearson's coefficient). We get qualitatively similar results with both of them, so in the rest of the article we only present the results obtained with the latter, which is faster to compute. 
These values are shown in the last column of the table ($\rho_d$), and it appears that the graphs exhibit a wide range of correlation values. The full graph, which, by construction, has maximal coverage, reaches only a $0.22$ score, whereas some of the smaller graphs get $0.80$ scores. 

In the following, we discuss how each extraction step presented in Section~\ref{sec:ExtractionOpt} affect the coverage and reliability of the graphs.

\subsection{Flat vs. Hierarchical Relationships}
\label{sec:MethCompHier}
We first consider whether to keep or remove hierarchical relationships, i.e. edges that model the inclusion of some spatial object into some other object, e.g. a building belonging to a parish. In each of the four parts of Table~\ref{tab:GraphStats}, the first two rows compare these two options (\texttt{$\dotp$H$\dotp$\_all} vs. \texttt{$\dotp$F$\dotp$\_all}). Independently of the other extraction steps, retaining the hierarchical relationships results in a much higher number of properties, i.e. a much better coverage. For the reliability, on the contrary, it depends on whether we stick to the raw data or leverage the additional relationships (\texttt{R$\dotp$$\dotp$\_all} vs. \texttt{E$\dotp$$\dotp$\_all}): in the former case, the distance correlation, and thus the reliability of the graph, is higher when using hierarchical relationships, whereas in the latter case it is lower. But even when the hierarchical relationships bring a better distance correlation, we must stress that this value is not satisfying, as it is closer to zero than to one.

Explaining these observations is quite straightforward by looking at the other statistics. When using only the raw data, removing the hierarchical relationships segments the graphs in several separated components (see $C(G)$). Many of them are small, and thus discarded at the filtering step, which causes the lower coverage. Moreover, the removal of the hierarchical edges also breaks shortcuts, as shown by the largely increased diameter ($d_{\max}$) and average distance ($\langle d \rangle$). This is quite visible when considering the full graph (Figure~\ref{fig:GraphFull}), as it shows how the pink edges modeling these relationships connect local hubs (parishes and boroughs) to their neighborhoods. The graph representations from Figure~\ref{fig:GraphsWholeLambert} use a geographic instead of an algorithmic layout, and shows that these relationships have a very long range, spatially speaking. This means that they connect spatially remote vertices, making them closer in the graph than they should be, as hypothesized in Section~\ref{sec:ExtractionOptNonpunct}. This is the cause for the relatively low distance correlation. 

In summary, hierarchical relationships allow a very good coverage, but seriously hinder the reliability of the graph. We conclude that it is preferable to ignore this type of relationships during graph extraction. In the rest of this section, we focus our discussion on the flat graphs (\texttt{$\dotp$F$\dotp$\_$\dotp$}).

\subsection{Keeping vs. Removing Non-Punctual Objects}
\label{sec:MethCompStreets}
The next extraction step is the possible removal of non-punctual objects. As a consequence, we focus on the top half of Table~\ref{tab:GraphStats}, which corresponds to whole vertices (\texttt{$\dotp$FW\_$\dotp$}), by opposition to split vertices in the bottom half, because splitting induces a different processing of non-punctual objects. Such objects include a few 2D entities (e.g. parishes), but there are predominantly linear objects (i.e. 1D), and more particularly streets (see Table~\ref{tab:ObjectTypes}).

We first compare between retaining all such objects (\texttt{$\dotp$FW\_all}) vs. removing all non-punctual objects except streets (\texttt{$\dotp$FW\_streets}). This removal causes a significant decrease of the graph coverage: this is expected, as deleting parishes and other 2D objects indirectly results in the removal of the hierarchical edges, and therefore yields the same effect as observed in Section~\ref{sec:MethCompHier}. This effect is even stronger as one removes more vertices and edges. When using only the raw data (\texttt{RFW\_streets}), the number of components and the average distance increase notably, which is not the case when using the additional relationships (\texttt{EFW\_streets}). In both cases though, the distance correlation is much higher compared to the graphs obtained when retaining all non-punctual objects ($0.03$ vs. $0.48$, and $0.48$ vs. $0.74$, respectively). The explanation is the same as before: non-punctual vertices are likely to connect objects that are spatially very distant, thereby decreasing the graph reliability. The Rhône river constitutes an extreme case of this issue, see the top-left vertex in the graphs of Figure~\ref{fig:GraphsWholeLambert}.

Some of the streets are quite long, and likely to cause the same problem, which is why we experiment with their controlled removal through methods \texttt{$\dotp$FW\_streets}. As explained in Section~\ref{sec:ExtractionOptNonpunct}, they consist in removing the $k$ longest streets (in addition to the other non-punctual objects) while keeping the shorter streets. In order to determine the best value of $k$, we rank all streets by decreasing length and iteratively remove an increasing number of streets while computing the same metrics as in Table~\ref{tab:GraphStats}. Simultaneously maximizing coverage and distance correlation is a bi-objective optimization problem. The corresponding plots are provided in Appendix~\ref{sec:ApdxStatsStreetProc} (Figure~\ref{fig:ParetoK}). The most appropriate solutions appear to be the removal of the $k = 6$ and $7$ longest streets for \texttt{RFW\_k} and \texttt{EFW\_k}, respectively. When comparing the former to \texttt{RFW\_streets} (i.e. keeping all streets), the difference is hardly noticeable: a small decrease in coverage ($-5 \%$) and a small improvement in distance correlation ($+2 \%$). However, when comparing \texttt{EFW\_k} to \texttt{EFW\_streets}, we see a comparable decrease in coverage ($-4 \%$) and a much improved correlation ($+8 \%$).

In summary, removing all the non-punctual objects except streets leads to a slightly lower coverage, but a much improved reliability. Removing also the longest streets has not much effect when dealing with the raw data, but improves the results further when using the extended data.



\subsection{Whole vs. Split Vertices}
\label{sec:MethCompSplit}
Next, we consider the splitting process described in Section~\ref{sec:ExtractionOptNonpunct}: it consists in using several separate vertices to represent portions of 1D and 2D objects, instead of just removing them outright. We focus on the bottom half of Table~\ref{tab:GraphStats}, that shows the results involving this optional step (\texttt{$\dotp$FS$\_\dotp$}). A street is sometimes the only vertex connecting a property to the rest of the graph. Thus, removing a vertex representing a street may indirectly lead to the deletion of properties from the network. For this reason, our expectation is that splitting instead of removing will allow improving the coverage. Moreover, splitting a spatially large object into several vertices increases the length of the concerned paths on the graph, which is likely to improve the distance correlation, too.

We first consider both methods that split all non-punctual objects (\texttt{$\dotp$FS\_all}). As expected, when compared to their counterparts that keep all these objects whole (\texttt{$\dotp$FW\_all}), it appears that splitting has no noticeable effect on coverage, as the proportions of properties are identical: $67.10\%$ for both \texttt{RFW\_all} and \texttt{RFS\_all}, and $72.74\%$ for both \texttt{EFW\_all} and \texttt{EFS\_all}. On the contrary, there is a clear increase of the distance correlation: $0.03$ vs. $0.27$ (\texttt{RF$\dotp$\_all}), and $0.48$ vs. $0.74$ (\texttt{EF$\dotp$\_all}). This is because splitting object such as long streets increases the length of certain paths on the graph, as exemplified by the ramparts that appear clearly in Figure~\ref{fig:GraphsSplitLambert} (as yellow edges around the city). When we split only the streets and remove the other non-punctual objects (\texttt{$\dotp$FS\_streets}), the observations are roughly the same.

We now focus on the last methods (\texttt{$\dotp$FS\_k}), which consist in splitting the $k$ longest streets, keeping the shorter ones, and removing the rest of the non-punctual objects. Unlike with the methods based on vertex removal (\texttt{$\dotp$FW\_$\dotp$}), estimating the best value of $k$ is not a bi-objective optimization problem, because splitting an increasing number of streets does not affect the coverage (cf. Figure~\ref{fig:EvolutionK}). Consequently, we just select the values that maximize distance correlation: $k = 6$ (\texttt{RFS\_k}) and $k = 7$ (\texttt{EFS\_k}). Compared to their counterpart methods that remove the longest streets instead of splitting them, we observe a better coverage while retaining the same distance correlation.

To conclude, our experiments show that splitting non-punctual objects instead of removing them allows preserving the property coverage, and sometimes even improving it, while increasing the distance correlation when it is low, or preserving it when it is already high.

\subsection{Raw vs. Extended Data}
\label{sec:MethCompExtend}
It is possible to complement the data extracted from the original sources with some additional information (cf. Section~\ref{sec:DataAdd}) in order to insert extra relationships in the graphs (cf. Section~\ref{sec:ExtractionOptAdd}). However, this operation has a cost. Hence, the interest of assessing whether such effort allows improving the extracted graph. For this purpose, in this section we compare the 1\textsuperscript{st} and 3\textsuperscript{rd} parts of Table~\ref{tab:GraphStats} (\texttt{R$\dotp \dotp$\_k}) with their extended counterparts in the 2\textsuperscript{nd} and 4\textsuperscript{th} parts (E$\dotp \dotp$\_k) of the same table.

The effects are the same independently of the other optional steps, and they are quite strong. First, using the extended dataset considerably improves the coverage: the increase ranges from 5 to 7 percentage points. Second, this operation also greatly improves the reliability, as the increase in distance correlation ranges from $0.26$ to $0.45$. This improvement also shows in the number of components ($C(G)$), diameter ($d_{\max}$), and average graph distance ($\langle d \rangle$), which systematically decrease.

In conclusion, our experimental results show that making the effort of gathering the additional data describing spatial relationships between objects that are not properties, allows to significantly improve the extracted graph, whatever other optional steps are used during this process.

\subsection{Concluding Remarks}
\label{sec:MethCompChoice}
Based on the analysis conducted in this section, we can now identify the best graph extraction methods among the proposed variants. Our main observations are as follows:
\begin{enumerate}
    \item Hierarchical relationships are largely detrimental to network reliability, and should be removed.
    \item Additional relationships always improve coverage and reliability, and should be used when available.
    \item Removing non-punctual objects strongly improves reliability while only slightly decreasing coverage, provided streets are kept whole.
    \item Splitting non-punctual objects instead of removing them leads to much better reliability, for the same coverage.
    \item Splitting only the longest streets, keeping the remaining streets whole, and removing the rest of the non-punctual objects, seems to be the best trade-off \textit{when using additional relationships}.
\end{enumerate}

\begin{figure}[!h]
    \centering
    \includegraphics[width=0.49\linewidth]{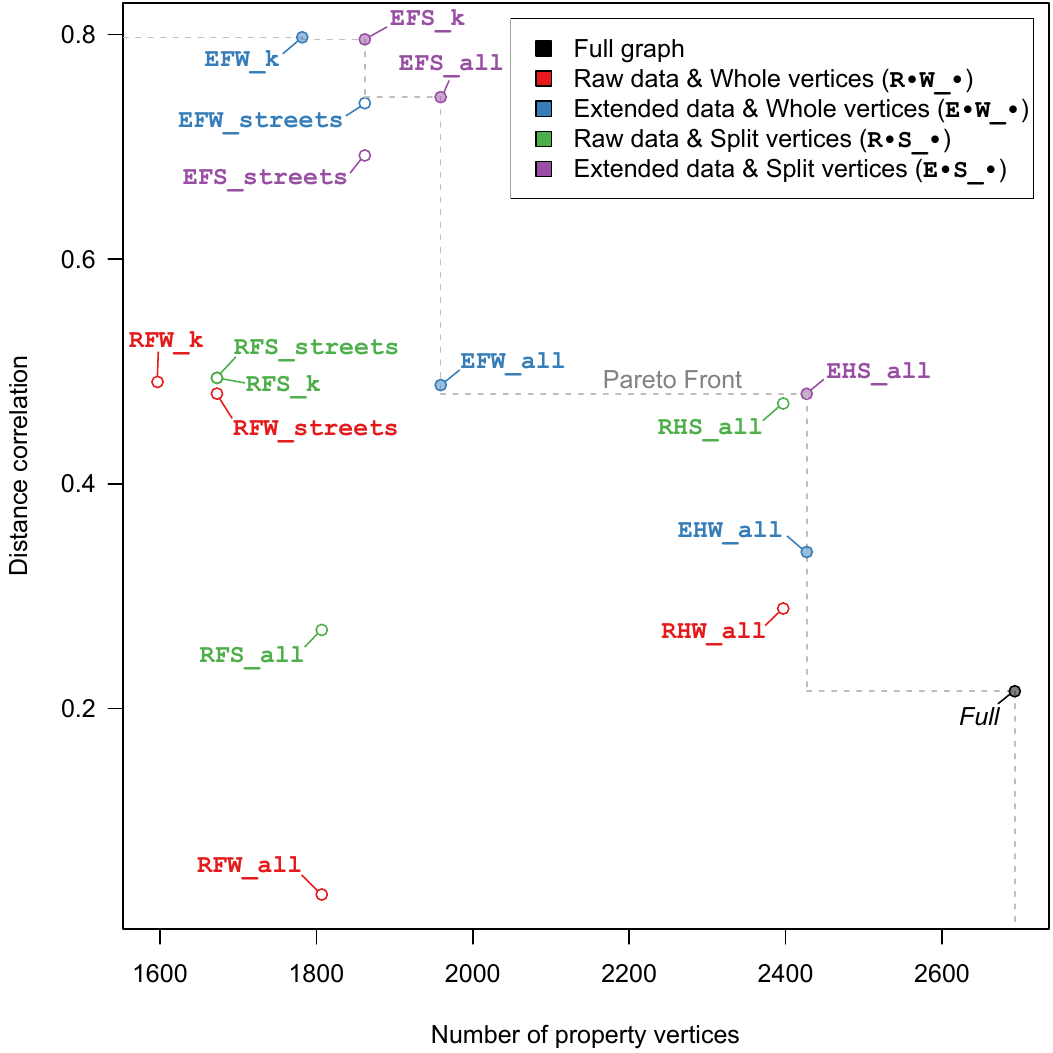}
    \hfill
    \includegraphics[width=0.49\linewidth]{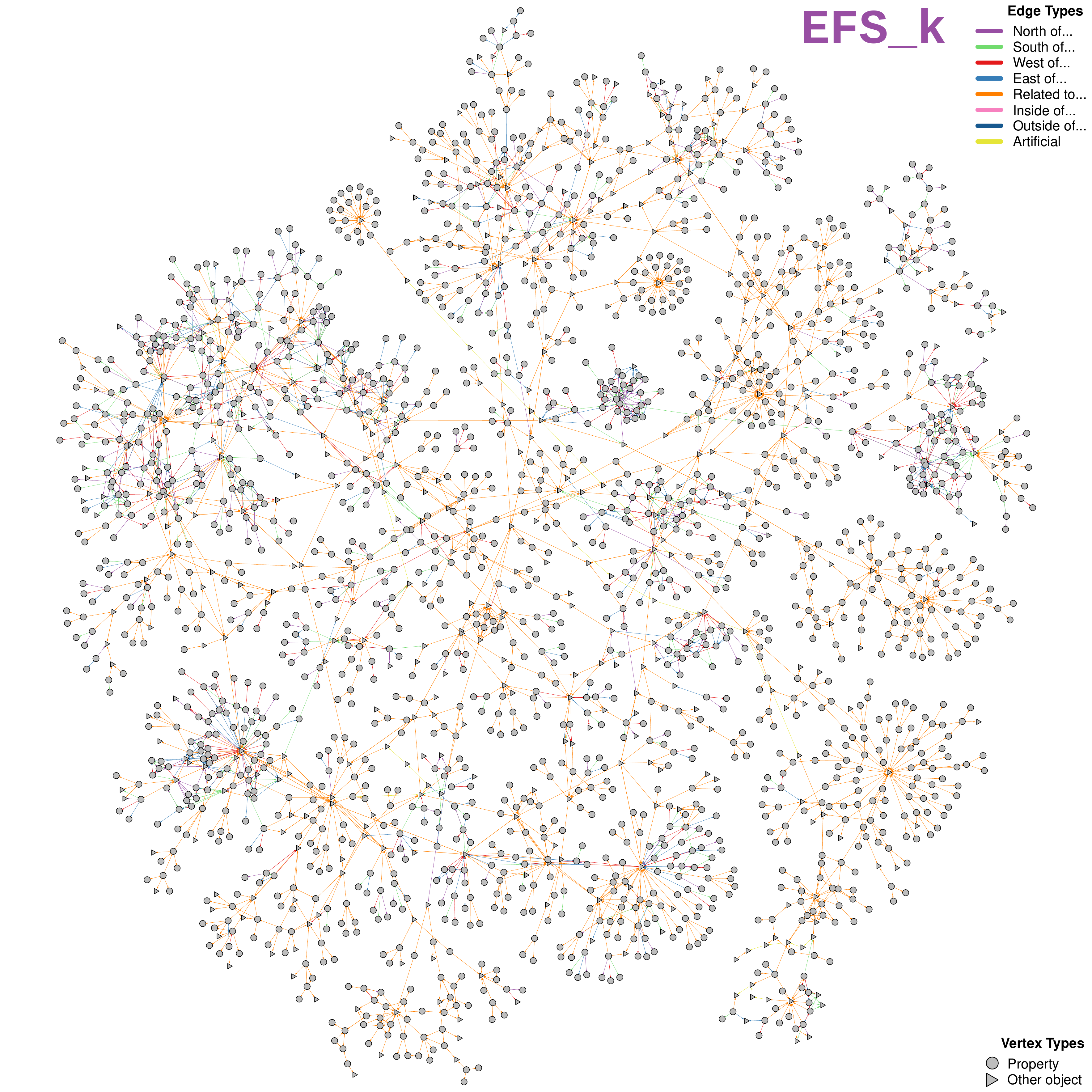}
    \caption{\textbf{Left:} Comparison of the methods from Table~\ref{tab:GraphStats} based on coverage (number of vertices representing properties) and reliability (distance correlation). \textbf{Right:} best confront graph according to our criteria, shown using the Yifan-Hu layout~\cite{Hu2006}. Figure available at \href{http://doi.org/10.5281/zenodo.14175830}{10.5281/zenodo.14175830} under CC-BY license.}
    \label{fig:CompPareto}
\end{figure}

We use Figure~\ref{fig:CompPareto}, which shows how each method performs in terms of coverage (number of properties in the graph) and reliability (distance correlation) as a visual aid to make our final decision. To go beyond distance correlation, Figure~\ref{fig:DistComp} (Appendix~\ref{sec:ApdxStatsDistComp}) shows the relation between graph and spatial distances. We consider that the method exhibiting the best trade-off overall is \texttt{EFS\_k} (extended relationships, flat relationships only, split the longest streets, keep the shortest, and remove the rest of the non-punctual object), and we will discuss the corresponding graph in the next section. However, this method requires some extra data, both to split the streets and to include additional relationships. In case one is not able to perform the split step (methods shown in blue in the figure), we consider that the next best choice is \texttt{EFW\_k}. If no additional relationship is available (in green), then graph quality seriously drops, and the most conservative approach seems to be the best choice: \texttt{RHS\_all}. Finally, if it is not possible to split nor use the extended relationships, then \texttt{RFW\_streets} seems to be the best method, even if it requires removing many vertices before reaching a passable reliability.

\section{Segmentation of the Spatial Graph}
\label{sec:ComStruct}
At this stage, we have identified the most appropriate method (\texttt{EFS\_k}, see Section~\ref{sec:MethCompChoice}) to extract a spatial graph representing the city of Avignon in medieval times. We now proceed with the analysis of this graph at a meso scale, i.e. from the perspective of its community structure. Our goal here is to partition the confront network into spatial zones representative of the tenants' perception and of the use of urban space. For this purpose, we rely on the traditional modularity function~\cite{Newman2004e} to assess the quality of the partition, but we also consider more subjective aspects, in order to take into account our objective of interpreting the detected communities. First, we want to get a reasonable number of communities, as it would not be practically feasible to interpret hundreds of spatial segments. Second, we want the community size to be relatively uniform: getting a single giant community and many very small ones would be of any help.

\begin{table}[!h]
    \caption{Main topological characteristics of the communities identified in the selected spatial graph: numbers of vertices ($n$) and edges ($m$), density ($\delta$), number of property vertices and proportion of such vertices relative to the community (\textit{Properties}), diameter ($d_{\max}$), harmonic mean of the graph distance ($\langle d \rangle$), Spearman's correlation between the graph and spatial distances ($\rho_d$).}
    \label{tab:ComStats}
    \centering
        \begin{tabularx}{\textwidth}{X r r r r@{\hspace{0.5cm}}r r r r}
            \hline
            \textbf{Com.} & $n$ & $m$ & $\delta$ & \multicolumn{2}{r}{\# Properties} & $d_{\max}$ & $\langle d \rangle$ & $\rho_d$ \\
            \hline
             $1$ & $156$ & $429$ & $0.0177$ & $132$ & $84.62 \%$ & $11$ & $3.78$ & $0.75$ \\
             $2$ &  $89$ & $239$ & $0.0305$ &  $73$ & $82.02 \%$ &  $8$ & $2.92$ & $0.52$ \\
             $3$ &  $90$ & $160$ & $0.0200$ &  $62$ & $68.89 \%$ & $10$ & $3.60$ & $0.50$ \\
             $4$ &  $20$ &  $24$ & $0.0632$ &  $16$ & $80.00 \%$ &  $6$ & $2.27$ & $0.32$ \\
             $5$ & $163$ & $449$ & $0.0170$ & $132$ & $80.98 \%$ &  $9$ & $3.55$ & $0.57$ \\
             $6$ &  $31$ &  $70$ & $0.0753$ &  $24$ & $77.42 \%$ &  $6$ & $2.28$ & $0.56$ \\
             $7$ &  $52$ &  $82$ & $0.0309$ &  $42$ & $80.77 \%$ & $10$ & $3.22$ & $0.36$ \\
             $8$ & $126$ & $283$ & $0.0180$ & $105$ & $83.33 \%$ &  $8$ & $2.97$ & $0.34$ \\
             $9$ & $168$ & $278$ & $0.0099$ & $135$ & $80.36 \%$ & $10$ & $3.79$ & $0.48$ \\
            $10$ &  $41$ &  $54$ & $0.0329$ &  $32$ & $78.05 \%$ &  $7$ & $3.04$ & $0.50$ \\
            $11$ &  $33$ &  $49$ & $0.0464$ &  $29$ & $87.88 \%$ &  $6$ & $2.63$ & $0.21$ \\
            $12$ & $132$ & $210$ & $0.0121$ & $105$ & $79.55 \%$ &  $9$ & $3.72$ & $0.53$ \\
            $13$ &  $31$ &  $34$ & $0.0366$ &  $27$ & $87.10 \%$ &  $3$ & $1.93$ & $0.27$ \\
            $14$ &  $61$ & $100$ & $0.0273$ &  $55$ & $90.16 \%$ &  $7$ & $2.91$ & $0.22$ \\
            $15$ & $115$ & $144$ & $0.0110$ & $105$ & $91.30 \%$ &  $9$ & $3.56$ & $0.03$ \\
            $16$ &  $55$ &  $56$ & $0.0189$ &  $39$ & $70.91 \%$ & $13$ & $3.89$ & $0.78$ \\
            $17$ & $120$ & $154$ & $0.0108$ &  $90$ & $75.00 \%$ & $12$ & $4.35$ & $0.61$ \\
            $18$ &  $91$ & $120$ & $0.0147$ &  $78$ & $85.71 \%$ & $10$ & $3.88$ & $0.75$ \\
            $19$ &  $99$ & $148$ & $0.0153$ &  $89$ & $89.90 \%$ & $10$ & $3.57$ & $0.38$ \\
            $20$ &  $28$ &  $35$ & $0.0463$ &  $23$ & $82.14 \%$ &  $8$ & $2.69$ & $0.11$ \\
            $21$ &  $91$ & $111$ & $0.0136$ &  $69$ & $75.82 \%$ & $12$ & $4.21$ & $0.52$ \\
            $22$ & $106$ & $153$ & $0.0137$ &  $95$ & $89.62 \%$ &  $6$ & $3.09$ & $0.29$ \\
            $23$ &  $96$ & $179$ & $0.0196$ &  $86$ & $89.58 \%$ &  $7$ & $3.26$ & $0.48$ \\
            $24$ &  $13$ &  $21$ & $0.1346$ &  $11$ & $84.62 \%$ &  $5$ & $1.97$ & $0.31$ \\
            $25$ &  $49$ &  $68$ & $0.0289$ &  $33$ & $67.35 \%$ & $13$ & $3.62$ & $0.24$ \\
            $26$ &  $26$ &  $27$ & $0.0415$ &  $20$ & $76.92 \%$ &  $7$ & $2.93$ & $0.60$ \\
            $27$ &  $28$ &  $30$ & $0.0397$ &  $20$ & $71.43 \%$ &  $9$ & $3.14$ & $0.36$ \\
            $28$ &  $40$ & $136$ & $0.0872$ &  $36$ & $90.00 \%$ &  $5$ & $1.97$ & $0.54$ \\
            $29$ &  $21$ &  $21$ & $0.0500$ &  $16$ & $76.19 \%$ &  $8$ & $2.77$ & $0.63$ \\
            $30$ &  $68$ & $126$ & $0.0277$ &  $46$ & $67.65 \%$ & $10$ & $3.24$ & $0.63$ \\
            $31$ &  $55$ &  $91$ & $0.0306$ &  $37$ & $67.27 \%$ & $10$ & $3.41$ & $0.32$ \\
            \hline
        \end{tabularx}
\end{table}

We assess seven standard community detection algorithms implemented in the \texttt{igraph}~\cite{Csardi2006} R library: Edge-betweenness~\cite{Newman2004e}, FastGreedy~\cite{Clauset2004}, Leading Eigenvectors~\cite{Newman2006b}, LabelPropagation~\cite{Raghavan2007}, WalkTrap~\cite{Pons2006}, InfoMap~\cite{Rosvall2008}, and Louvain~\cite{Blondel2008}. According to our criteria, the Louvain method provides the most relevant partition of the graph. It reaches a modularity of 0.93, and contains 31 communities, whose main characteristics are described in Table~\ref{tab:ComStats}. Each statistic is computed by considering only the subgraph induced by a community. Community size ranges from 13 to 168 vertices, including 67\% to 91\% of properties. The average distance is relatively homogeneous, ranging from 1.93 to 4.35, but the diameter is more heterogeneous, ranging from 3 to 13. The distance correlation is very heterogeneous, ranging from 0.03 to 0.78, showing that space is not modeled in all communities with the same reliability.

In the following, in order to illustrate the historical relevance of the communities detected in our confront network, we discuss a few particularly interesting examples, both in terms of how they are distributed over the city space (Section~\ref{sec:ComStructPos}), and of the types of spatial objects that constitute them (Section~\ref{sec:ComStructComp}).

\subsection{Organization of the Communities}
\label{sec:ComStructPos}
\begin{figure}[!h]
    \centering
    \includegraphics[width=1\linewidth]{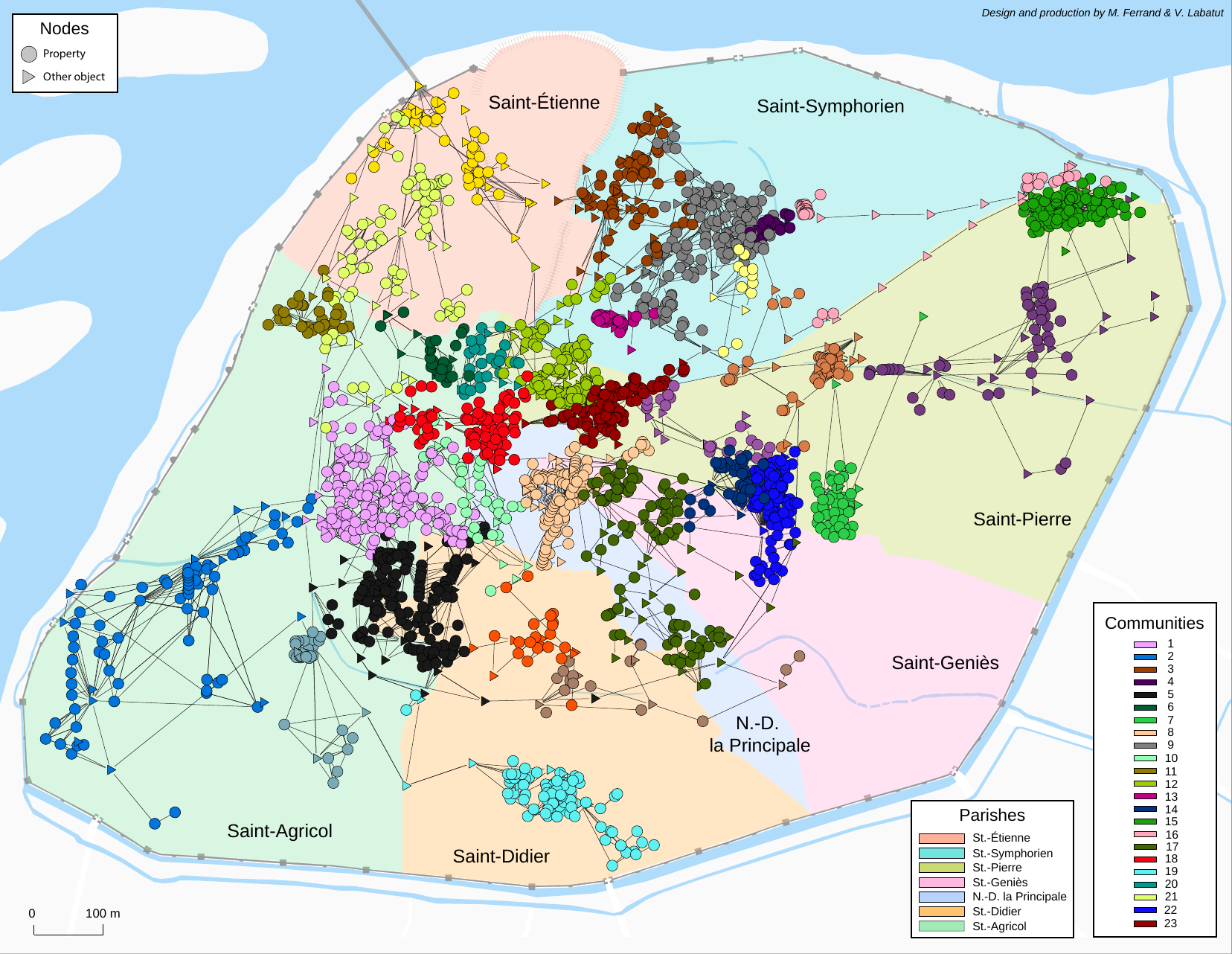}
    \caption{Distribution of the communities over the 7 historical parishes of Avignon. Each vertex color represents a specific community. The vertices are placed depending on their spatial position (known or estimated). Figure available at \href{http://doi.org/10.5281/zenodo.14175830}{10.5281/zenodo.14175830} under CC-BY license.}
    \label{fig:ComParish}
\end{figure}

The spatial distribution of communities generally aligns with the logic of parochial territory, as shown in Figure~\ref{fig:ComParish}. 
However, while some communities consist of properties that depend on a single parish, others are quite revealing of uncertain boundaries or the interweaving of parishes. One community, which spreads through the heart of the city, comprises properties associated with five different parishes (C18), while three communities include properties from three distinct parishes (C8, C10, and C17), and nine from two parishes. See Figure~\ref{fig:ComParishPies} in Appendix~\ref{sec:ApdxComs} for further details.

Here, it is not the parish that defines the neighborhood, but primarily the spatial proximity and what it implies. Parish boundaries do not segment the perceived space of the tenants. They are not physically visible in the city, which partly explains why they do not reflect in the spatial distribution of some communities. These boundaries are inevitably less significant than a very concrete and physical division of urban space, such as the old city walls, even if they have disappeared for several generations. In this sense, the mental division of urban space, inherited from the old city delineation, is quite perceptible in the structure of the detected communities.

\begin{figure}[!h]
    \centering
    \includegraphics[width=0.49\linewidth]{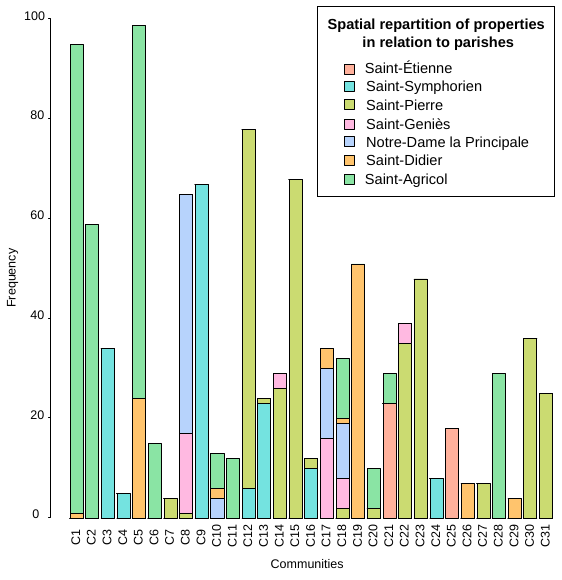}
    \hfill
    \includegraphics[width=0.49\linewidth]{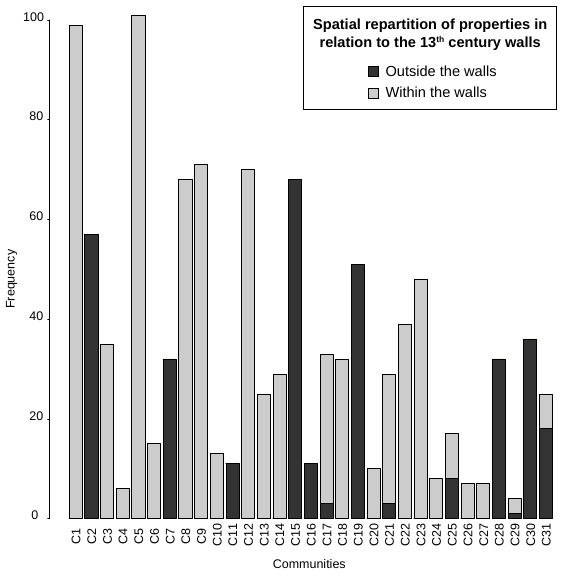}
    \caption{Distribution of property location over communities, in terms of parochial membership (left) and relative to the old walls (right). Figure available at \href{http://doi.org/10.5281/zenodo.14175830}{10.5281/zenodo.14175830} under CC-BY license.}
    \label{fig:ComHisto}
\end{figure}

Out of the 31 communities, only five include properties located both within the old intra-muros and beyond the old city walls (C17, C21, C25, C29, C31), as shown in Figure~\ref{fig:ComHisto}. These are located at the border of the two spaces. The perception of individuals does not depend on a pre-established order based on the official division of the city. It is primarily founded on the spatial proximity of properties and, by extension, of individuals. More than its relation to the administrative subdivision of the city, the division highlighted by the communities aligns with that drawn by the oldest and most structuring streets of the city, constrained by the very topography of the places and inherited spaces. Thus, the distribution of communities is constrained by certain elements that concretely segment the space: the topography with the Rock of the Doms, the canals, and the old walls. These elements clearly separate properties located near each other. When they are not connected by bridges, they hinder communication and, consequently, sociability between individuals.

Beyond spatial proximity, it is also the interactions of individuals around common places that truly create the community, in this case, the neighborhood.

\subsection{Composition of the Communities}
\label{sec:ComStructComp}
\begin{figure}[!h]
    \centering
    \includegraphics[width=1\linewidth]{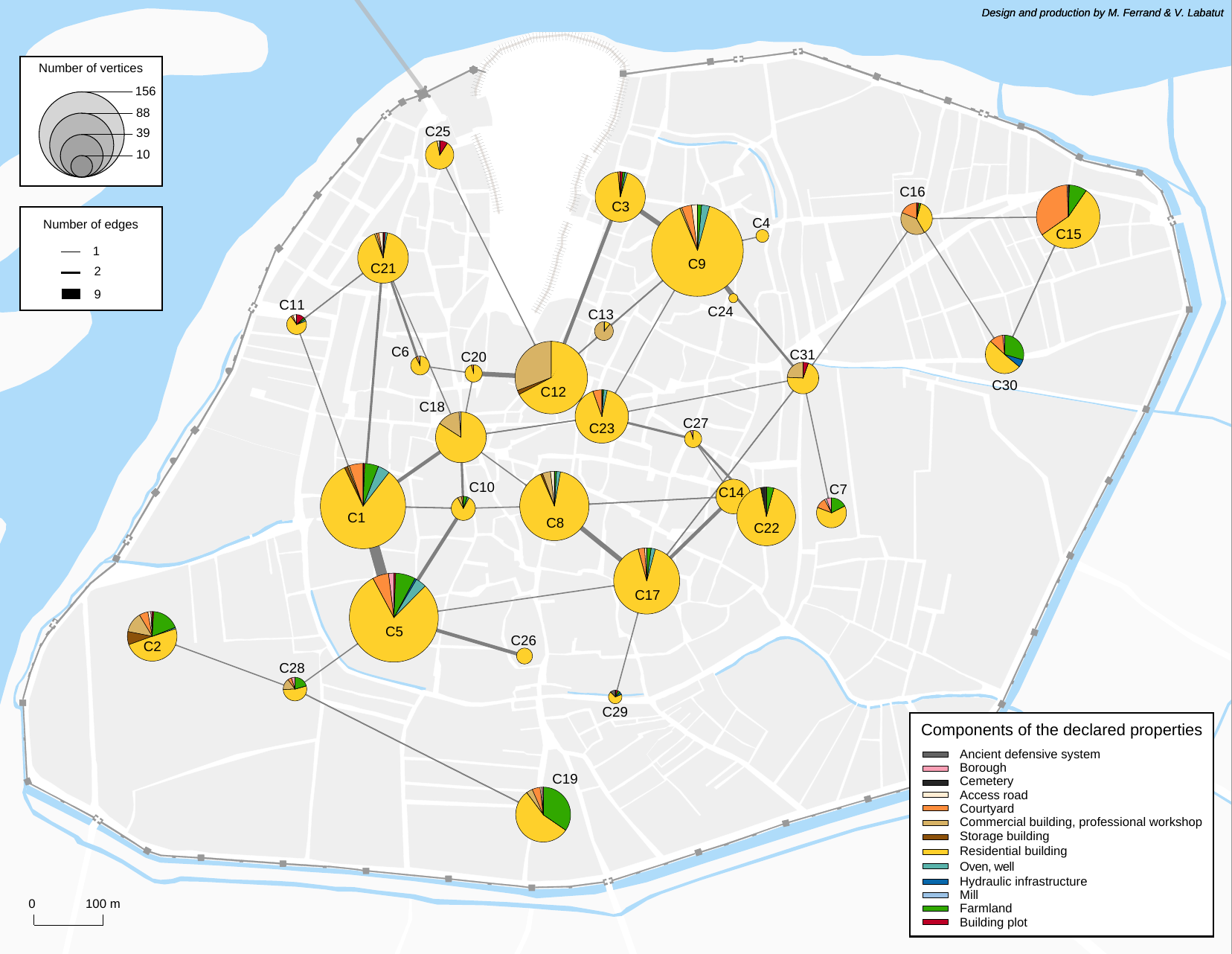}
    \caption{Simplified representation of the communities: each node represents a community from the original network. Node size and link width are proportional to the numbers of vertices constituting the original communities, and to the number of edges between them in the original graph, respectively. The pie charts show the composition of the communities in terms of types of spatial objects. Figure available at \href{http://doi.org/10.5281/zenodo.14175830}{10.5281/zenodo.14175830} under CC-BY license.}
    \label{fig:ComComposition}
\end{figure}

The communities exhibit a wide diversity of composition. As shown in Table~\ref{tab:ComStats} and illustrated in Figure~\ref{fig:ComComposition}, they contain mainly properties, whose characteristics may vary from one neighborhood to another, and invariants –- more or less numerous and attractive depending on the case. Different rationales of construction can be identified, that are primarily related to the uses of space, their evolution, or, on the contrary, their persistence and stability.

\subsubsection{Within the Old Walls}
\label{sec:ComStructCompIn}
Depending on the neighborhoods, certain elements are predominantly used by the owners to locate and identify their properties, occupying a central place in the construction of individuals' lived spaces. In the old intra-muros of the city, that of the 13\textsuperscript{th} century, it is often the main roads of the city, those that have structured the city since antiquity. These are main arteries that traverse the neighborhood, punctuating exchanges and social interactions. Many communities are built around these streets, though not all of these communities have the same characteristics. While some communities are more residential, others are resolutely oriented towards craftsmanship and commerce, or even towards accommodating travelers. The places that unite them are often revealing of their typology.

Given its location near the Pont Saint-Bénézet and the Rhône river, C21, for example, has many inns, which are significant social places and important topographical landmarks, often with very conspicuous signs. Like the inns, the bathhouses are places of social interaction and notable landmarks in a neighborhood. Alongside the most structuring streets of the community and the inns, it is thus a bathhouse that holds one of the highest degrees in this community.

Neighborhoods with a much more residential aspect than the one we just mentioned also aggregate around many roads. Community C5 is a good example. Rue des Ortolans plays an important role in the life of the neighborhood and in the perception the owners have of it. It has long been dominated by the presence of a prominent Avignonese family, who gave its name to the street and who settled in the city well before the arrival of the popes. The residential character of this neighborhood is notably based on the fact that it is a recently constructed area, at least more recent than the heart of the city further north.

A bit further north of C5, community detection reveals a particularly imposing neighborhood both in the number of nodes it contains and its extent (C1). It has an obvious residential character and includes notably beautiful and large residences. Until 1370, one house is particularly used by contemporaries to situate themselves in this area: it is that of Tymburge Vayrane, a property owner with a rich real estate portfolio throughout the city. This house is mentioned as a landmark more often than some streets. This illustrates how much the neighborhood and social interactions directly influence the perception of the users.

Like the ones we just mentioned, C8 is a very good example of the centrality of streets in the composition and definition of neighborhoods. The community has a relatively small diameter (8) given the number of its nodes (126). Spatially speaking, it is particularly dense, which is quite revealing of the very genesis of the neighborhood and its development from the 11\textsuperscript{th} century. The extent of this community corresponds very likely to that of the bishop's boroughs called Scofaria and Pelliparia. The regularity of the plots around the churches, especially to the south of the main street, suggests that the bishop, holder of the borough, had certainly intervened very precisely in the subdivision of certain parts of the area.

Community C18 has a very different characteristic, much less residential. It is located in the heart of the city, spanning several parishes. It is built around the most important and oldest commercial areas and marketplaces of the city. This neighborhood is resolutely commercial, and this is particularly evident in the profile of the individuals one encounters and the types of properties they own. Thus, in addition to their houses, the owners declare numerous shops or stalls, two types of properties not found in more residential neighborhoods like C5. Conversely, certain features of residences found in other neighborhoods are completely absent here: there are no gardens or courtyards; the space is primarily used by and for commerce.

If the streets are at the heart of neighborhood construction, it is primarily because they are the sites of social interactions. These streets are organized around a significant and unifying household, host commercial and artisanal activities, and are central to the city's political representation. Each neighborhood also has essential elements for the daily lives of individuals, such as wells, ovens, and cemeteries, which contribute to the construction of individuals' perception of places. These are undeniable landmarks of the urban landscape, places of sociability, exchanges, and meetings. As such, they are often the vertices with the highest degree in communities. In the city, burial sites are thus privileged and central spaces for exchanges and sociability. They must also be associated with public places such as streets or even more so, squares. 
Community 13 is a very good example. It is very small and particularly concentrated (31 vertices for a diameter of 3). All these properties are connected to the Saint-Symphorien cemetery. The degree of centralization of the community is, in this respect, quite high. Here, the cemetery is the spatial reference par excellence, and the neighborhood is entirely built around the burial site.

\subsubsection{Beyond the Old Walls}
\label{sec:ComStructCompOut}
Within the old walls, communities are thus constructed by the proximity relationship between properties and individuals and by the presence of topographical landmarks that shape the perception and usage of individuals in the neighborhood. 
The neighborhoods located outside the walls of the 13\textsuperscript{th} century follow very diverse logics, which differ in several ways from the communities studied so far. 

The largest neighborhood detected in this area corresponds to C2. At first glance, it might seem to follow the same construction logics as those located within the old walls. However, it has certain characteristics specific to its location and its distance from the historical center of the city. 
To locate their properties, the owners refer not only to the streets but also to recent urbanization: the numerous boroughs developed in this area on former agricultural land. Alongside these boroughs, many gardens are still present. People frequently use these gardens as reference points. This neighborhood has another unique characteristic: in addition to its semi-rural aspect, created by the number of gardens adjoining the new houses of the boroughs, it houses the dependencies of many services from the old intra-muros. It is likely that this community also hosts the pleasure homes of certain prelates, as suggested by the number of residences described as large and beautiful, accompanied by gardens, courtyards, or wells. This neighborhood also has a significant presence of artisans and merchants. 
The presence of large gardens still cultivated at the end of the 14\textsuperscript{th} century must have given a rural appearance to the place, sharply contrasting with the old intra-muros, which was very densely built and populated. As such, it is overall not densely built. 
Two other communities (C19 and C30), located outside the 13extsuperscript{th} century walls, share characteristics with C2 and could thus be described as semi-rural neighborhoods. They retain the marks of significant rurality while being partly occupied by recent urbanization in certain areas. 

Finally, a specific typology of community outside the 13extsuperscript{th} century walls must be highlighted. These are what could be called the borough communities, such as C15. Here, the borough forms the neighborhood. This is quite noticeable when looking at the elements to which individuals refer to situate themselves in this space. Firstly, it is the straight street of the borough that is used: it unites the community built around it, suggesting the planning of the road in the borough's subdivision project. Although less used, another street is mentioned; it is called the ``vicinal street'' of the borough; there is also a cross street, a dead-end, and a square. In all cases, the attachment of these elements to the borough is specified. These details remind us that while these streets are not structuring at the city level, they are strongly so at the neighborhood and borough level. Whereas the boundaries of the previously studied neighborhoods were more difficult to perceive, especially for those within the 13\textsuperscript{th} century walls, here the neighborhood space is clearly delineated. It is no longer just a perceived, subjective space with fuzzy boundaries, constructed by the proximity and sociability of individuals around various places and inherited spaces. In the case of the "borough neighborhood," the boundaries are tangible and inherent to the very genesis of the place. Proximity and sociability are indeed present, as evidenced by the presence of a large borough well to which the owners sometimes refer. However, the neighborhood is built here around a predefined space, not an unplanned process inherited from an "already existing space." The typology of the declared properties once again highlights this, forming a particularly homogeneous neighborhood.

\bigskip
The graph extracted using our method is a sufficiently good approximation of spatial distance for the communities to correspond to spatial segments representative of real neighborhoods in the sense of lived space. 
In this regard, the detection of communities in the graph shows a partition of the city consistent with social practices. The elements that define the communities are multiple and complementary (livability, buildings, neighborhoods, commerce) and they are essentially places of sociability or symbolic markers of the urban landscape. They pace and define spatial perceptions. These perceptions are primarily based on the layering of heritages, with evolutions being integrated only slowly.

\section{Conclusion}
\label{sec:Conclusion}
In this article, we tackled the problem of segmenting a historical urban space based on incomplete data. We focused on the specific case of medieval Avignon, during the papacy. On the one hand, historical sources rarely provide the exact locations of the buildings constituting the city, but on the other hand, they describe how they are located relative to each other. We adopted a graph-based approach to take advantage of this relational information while modeling the city, and proposed several methods to extract such confront networks. We designed two conflicting objective criteria to select the optimal extraction method: data coverage and distance correlation. It turns out the best method requires discarding some of the information provided by the original historical sources in order to preserve the spatial proximity encoded in the graph. Moreover, our results show that leveraging certain information from secondary sources can considerably improve the quality of the graph. Finally, we partitioned the best confront network through community detection, in order to perform a qualitative analysis of the resulting segmentation of the urban space. Our discussion showed that the communities are completely consistent with historical knowledge, and help understand how people perceived urban space at the time.

We identify mainly two perspectives to extend our work. First, the land registries that we used in this article are very common in historical studies, especially during the Medieval Ages. However, they do not always take exactly the same form, and can also vary in content from one city to the other, depending on the local practices. We want to assess the robustness of our extraction approach by applying it to other cities, which will require some adaptation. For example, in the case of the city of Orleans~\cite{Fianu2011}, the spatial relationships partly differ from those described in Section~\ref{sec:DataRelations} and Table~\ref{tab:ConvRel} for Avignon: there are only very few cardinal relationships. But this could be compensated by leveraging other types of information, for instance regarding the notaries and families involved, and their social relationships. 
The second perspective concerns the estimation of the missing absolute positions. In the case of Avignon, the position of certain spatial objects is known with reasonable certainty, and can be leveraged together with the spatial relationships encoded in the confront networks, to interpolate the position of the other objects. A simple baseline consists in averaging the position of the neighboring objects, but this does not account for the semantics of the spatial relationships. We want to take advantage of Graph Neural Networks~\cite{Hamilton2020} to perform a better prediction, possibly also using the many vertex attributes at our disposal.

\section*{Resource Availability}
The source code written to implement our extraction methods and experiments is available online at \url{https://github.com/CompNet/MedievalAvignon}. The input data are included in this repository, whereas all the files produced during their processing are available at \url{https://doi.org/10.5281/zenodo.14175830}.

\section*{Acknowledgment}
This research benefited from the support of the \textit{Agorantic} interdisciplinary research federation (FR 3621), through the funding of Margot Ferrand's PhD. thesis and of the \textit{HistoGraph} research project.


\bibliographystyle{plainnat}
\bibliography{Ferrand2023_biblio.bib}

\begin{thebibliography}{34}
\providecommand{\natexlab}[1]{#1}
\providecommand{\url}[1]{\texttt{#1}}
\expandafter\ifx\csname urlstyle\endcsname\relax
  \providecommand{\doi}[1]{doi: #1}\else
  \providecommand{\doi}{doi: \begingroup \urlstyle{rm}\Url}\fi

\bibitem[Arnaud(2008)]{Arnaud2008}
J.-L. Arnaud.
\newblock \emph{Analyse spatiale, cartographie et histoire urbaine}.
\newblock Parcours m\'{e}diterran\'{e}en. Parenth\`{e}ses/MMSH,
  Marseille/Aix-en-Provence, FR, 2008.
\newblock URL
  \url{https://www.editionsparentheses.com/analyse-spatiale-cartographie-et}.

\bibitem[Barth\'{e}lemy(2011)]{Barthelemy2011}
M.~Barth\'{e}lemy.
\newblock Spatial networks.
\newblock \emph{Physics Reports}, 499\penalty0 (1-3):\penalty0 1--101, 2011.
\newblock \doi{10.1016/j.physrep.2010.11.002}.

\bibitem[Bertrand(2015)]{Bertrand2015}
P.~Bertrand.
\newblock \emph{Les \'{e}critures ordinaires : sociologie d’un temps de
  r\'{e}volution documentaire (entre royaume de {F}rance et {E}mpire,
  1250--1350)}.
\newblock Histoire ancienne et m\'{e}di\'{e}vale. Publications de la Sorbonne,
  Paris, FR, 2015.
\newblock URL
  \url{http://www.editionsdelasorbonne.fr/fr/livre/?GCOI=28405100607730}.

\bibitem[Blagus et~al.(2012)Blagus, \v{S}ubelj, and Bajec]{Blagus2012}
N.~Blagus, L.~\v{S}ubelj, and M.~Bajec.
\newblock Self-similar scaling of density in complex real-worldnetworks.
\newblock \emph{Physica A}, 391\penalty0 (8):\penalty0 2794--2802, 2012.
\newblock \doi{10.1016/j.physa.2011.12.055}.

\bibitem[Bloch(1929)]{Bloch1929}
M.~Bloch.
\newblock Les plans parcellaires en {F}rance.
\newblock \emph{Annales d'histoire {\'{e}}conomique et sociale}, 1\penalty0
  (1):\penalty0 60--70, 1929.
\newblock \doi{10.3406/ahess.1929.1039}.

\bibitem[Blondel et~al.(2008)Blondel, Guillaume, Lambiotte, and
  Lefebvre]{Blondel2008}
V.~D. Blondel, J.-L. Guillaume, R.~Lambiotte, and E.~Lefebvre.
\newblock Fast unfolding of communities in large networks.
\newblock \emph{Journal of Statistical Mechanics}, 2008\penalty0 (10):\penalty0
  P10008, 2008.
\newblock \doi{10.1088/1742-5468/2008/10/P10008}.

\bibitem[Borgatti and Everett(1997)]{Borgatti1997}
S.~P. Borgatti and M.~G. Everett.
\newblock Network analysis of 2-mode data.
\newblock \emph{Social Networks}, 19\penalty0 (3):\penalty0 243--269, 1997.
\newblock \doi{10.1016/S0378-8733(96)00301-2}.

\bibitem[Brunel et~al.(1998)Brunel, Guyotjeannin, and Moriceau]{Brunel2002}
G.~Brunel, O.~Guyotjeannin, and J.-M. Moriceau, editors.
\newblock \emph{Terriers et plans-terriers du {XIII}e au {XVIII}e si\`ecle :
  Actes du colloque de {P}aris}, Paris, FR, 1998. Association d’Histoire des
  Soci\'et\'es Rurales / \'Ecole Nationale des Chartes.
\newblock URL \url{https://books.google.fr/?id=mjRpmXRfNkYC}.

\bibitem[Clauset et~al.(2004)Clauset, Newman, and Moore]{Clauset2004}
A.~Clauset, M.~E.~J. Newman, and C.~Moore.
\newblock Finding community structure in very large networks.
\newblock \emph{Physical Review E}, 70\penalty0 (6):\penalty0 066111, 2004.
\newblock \doi{10.1103/PhysRevE.70.066111}.

\bibitem[Claveirole and P\'elaquier(1999)]{Claveirole2001}
A.~Claveirole and \'E. P\'elaquier, editors.
\newblock \emph{Le compoix et ses usages -- Actes du colloque de {N}\^imes},
  Montpellier, FR, 1999. Publications de l’Universit\'e Paul
  Val\'ery-Montpellier 3.
\newblock URL \url{https://books.google.fr/?id=JNO8bwAACAAJ}.

\bibitem[Csardi and Nepusz(2006)]{Csardi2006}
G.~Csardi and T.~Nepusz.
\newblock The igraph software package for complex network research.
\newblock \emph{InterJournal}, 1695\penalty0 (Complex Systems), 2006.
\newblock URL \url{http://igraph.sf.net}.

\bibitem[Estrada(2011)]{Estrada2011a}
E.~Estrada.
\newblock \emph{The Structure of complex networks: Theory and applications}.
\newblock Oxford University Press, 2011.
\newblock \doi{10.1093/acprof:oso/9780199591756.001.0001}.

\bibitem[Ferrand(2022)]{Ferrand2022}
M.~Ferrand.
\newblock \emph{Usages et repr\'esentations de l'espace urbain m\'edi\'eval :
  Approche interdisciplinaire et exploration de donn\'ees g\'eo-historiques
  d’Avignon \`a la fin du Moyen \^Age}.
\newblock Phd thesis, Avignon Universit\'e, 2022.
\newblock URL \url{https://www.theses.fr/2022AVIG1002}.

\bibitem[Fianu(2011)]{Fianu2011}
K.~Fianu.
\newblock L'utilisation des actes d'apr\`{e}s les registres notari\'{e}s
  orl\'{e}anais du xve si\`{e}cle.
\newblock \emph{Cahiers de recherches m\'{e}di\'{e}vales et humanistes},
  22:\penalty0 457--479, 2011.
\newblock \doi{10.4000/crm.12625}.

\bibitem[Fortunato(2010)]{Fortunato2010}
S.~Fortunato.
\newblock Community detection in graphs.
\newblock \emph{Physics Reports}, 486\penalty0 (3-5):\penalty0 75--174, 2010.
\newblock \doi{10.1016/j.physrep.2009.11.002}.

\bibitem[Fossier(1978)]{Fossier1978}
R.~Fossier.
\newblock \emph{Polyptiques et censiers}.
\newblock Brepols, Turnhout, BE, 1978.
\newblock URL
  \url{https://bibliothequearchives.seinemaritime.fr/Default/doc/SYRACUSE/243138/polyptyques-et-censiers-par-robert-fossier}.

\bibitem[Hamilton(2020)]{Hamilton2020}
William~L. Hamilton.
\newblock \emph{Graph Representation Learning}, volume~46 of \emph{Synthesis
  Lectures on Artificial Intelligence and Machine Learning}.
\newblock Morgan \& Claypool, 2020.
\newblock \doi{10.2200/s01045ed1v01y202009aim046}.

\bibitem[Hautefeuille(2016)]{Hautefeuille2016}
F.~Hautefeuille.
\newblock G\'eolocalisation des sources fiscales pr\'e-r\'{e}volutionnaires :
  la quadrature du cercle.
\newblock \emph{Bulletin du Centre d'\'Etudes M\'edi\'{e}vales d'Auxerre},
  HS-9, 2016.
\newblock \doi{10.4000/cem.13800}.

\bibitem[Hayez(1977)]{Hayez1977}
A.-M. Hayez.
\newblock Les bourgs avignonnais au {XIVe} si\`ecle.
\newblock \emph{Bulletin philologique et historique du Comit\'e des Travaux
  Historiques et Scientifiques}, pages 77--102, 1977.
\newblock URL
  \url{https://bibliotheques.avignon.fr/in/faces/details.xhtml?id=p::usmarcdef\_0000359794}.

\bibitem[Hayez(1993{\natexlab{a}})]{Hayez1993}
A.-M. Hayez.
\newblock \emph{Le terrier avignonnais de l'\'ev\^eque Anglic Grimoard
  (1366-1368) r\'edig\'e par Sicard de Fraisse}.
\newblock Comit\'e des Travaux Historiques et Scientifiques, Paris, FR,
  1993{\natexlab{a}}.
\newblock URL \url{https://gallica.bnf.fr/ark:/12148/bpt6k6431727c.texteImage}.

\bibitem[Hayez(1993{\natexlab{b}})]{Hayez1993b}
A.-M. Hayez.
\newblock Livr\'ees avignonnaises de la p\'eriode pontificale.
\newblock \emph{M\'emoires de l'Acad\'{e}mie de Vaucluse}, 8\penalty0
  (3):\penalty0 33--89, 1993{\natexlab{b}}.
\newblock URL
  \url{https://bibliotheques.avignon.fr/in/faces/details.xhtml?id=p::usmarcdef\_0000119299}.

\bibitem[Hu(2006)]{Hu2006}
Y.~Hu.
\newblock Efficient, high-quality force-directed graph drawing.
\newblock \emph{Mathematica Journal}, 10:\penalty0 37--71, 2006.
\newblock URL \url{http://yifanhu.net/PUB/graph_draw.pdf}.

\bibitem[Kivel\"{a} et~al.(2014)Kivel\"{a}, Arenas, Barth\'{e}lemy, Gleeson,
  Moreno, and Porter]{Kivelae2013}
M.~Kivel\"{a}, A.~Arenas, M.~Barth\'{e}lemy, J.~P. Gleeson, Y.~Moreno, and
  M.~A. Porter.
\newblock Multilayer networks.
\newblock \emph{Journal of Complex Networks}, 2\penalty0 (3):\penalty0
  203--271, 2014.
\newblock \doi{10.1093/comnet/cnu016}.

\bibitem[Levy and Lussault(2003)]{Levy2003}
J.~Levy and M.~Lussault.
\newblock \emph{Dictionnaire de la g\'{e}ographie et de l’espace des
  soci\'{e}t\'{e}s}.
\newblock Belin, Paris, FR, 2003.
\newblock URL \url{https://shs.hal.science/halshs-01252959}.

\bibitem[Mazel(2011)]{Mazel2011}
F.~Mazel.
\newblock Pouvoir comtal et territoire : r\'eflexion sur les partages de
  l'ancien comt\'e de provence au {XIIe} si\`ecle.
\newblock \emph{M\'elanges de l'\'Ecole Fran\c{c}aise de Rome}, 123\penalty0
  (2):\penalty0 467--48, 2011.
\newblock \doi{10.4000/mefrm.634}.

\bibitem[Newman(2003)]{Newman2003b}
M.~E.~J. Newman.
\newblock The structure and function of complex networks.
\newblock \emph{SIAM Review}, 45:\penalty0 167--256, 2003.
\newblock \doi{10.1137/S003614450342480}.

\bibitem[Newman(2006)]{Newman2006b}
M.~E.~J. Newman.
\newblock Finding community structure in networks using the eigenvectors of
  matrices.
\newblock \emph{Physical Review E}, 74\penalty0 (3):\penalty0 036104, 2006.
\newblock \doi{10.1103/PhysRevE.74.036104}.

\bibitem[Newman and Girvan(2004)]{Newman2004e}
M.~E.~J. Newman and M.~Girvan.
\newblock Finding and evaluating community structure in networks.
\newblock \emph{Physical Review E}, 69\penalty0 (2):\penalty0 026113, 2004.
\newblock \doi{10.1103/PhysRevE.69.026113}.

\bibitem[Pansier(1930)]{Pansier1930}
P.~Pansier.
\newblock \emph{Dictionnaire des anciennes rues d'Avignon}.
\newblock Roumanille, Avignon, FR, 1930.
\newblock URL
  \url{https://bibliotheques.avignon.fr/in/faces/details.xhtml?id=p::usmarcdef\_0000031742}.

\bibitem[Pons and Latapy(2006)]{Pons2006}
P.~Pons and M.~Latapy.
\newblock Computing communities in large networks using random walks.
\newblock \emph{Journal of Graph Algorithms and Applications}, 10\penalty0
  (2):\penalty0 191--218, 2006.
\newblock \doi{10.7155/jgaa.00124}.

\bibitem[Raghavan et~al.(2007)Raghavan, Albert, and Kumara]{Raghavan2007}
U.~N. Raghavan, R.~Albert, and S.~Kumara.
\newblock Near linear time algorithm to detect community structures in
  large-scale networks.
\newblock \emph{Physical Review E}, 76\penalty0 (3):\penalty0 036106, 2007.
\newblock \doi{10.1103/PhysRevE.76.036106}.

\bibitem[Rodier et~al.(2012)Rodier, Le~Cou\'edic, Jouve, Hautefeuille, Leturcq,
  and Fieux]{Rodier2012a}
X.~Rodier, M.~Le~Cou\'edic, B.~Jouve, F.~Hautefeuille, S.~Leturcq, and
  E.~Fieux.
\newblock De l'espace aux graphes. mesurer les dynamiques spatiales des
  terroirs villageois.
\newblock In \emph{XLIIIe Congr\`es de la Soci\'{e}t\'{e} des historiens
  m\'edi\'evistes de l’Enseignement sup\'erieur public}, pages 99--118.
  \'Editions de la Sorbonne, 2012.
\newblock URL \url{https://books.openedition.org/psorbonne/28550}.

\bibitem[Rosvall and Bergstrom(2008)]{Rosvall2008}
M.~Rosvall and C.~T. Bergstrom.
\newblock Maps of random walks on complex networks reveal community structure.
\newblock \emph{Proceedings of the National Academy of Sciences}, 105\penalty0
  (4):\penalty0 1118, 2008.
\newblock \doi{10.1073/pnas.0706851105}.

\bibitem[Wasserman and Faust(1994)]{Wasserman1994}
S.~Wasserman and K.~Faust.
\newblock \emph{Social Network Analysis: Methods and Applications}, volume~8 of
  \emph{Structural Analysis in the Social Sciences}.
\newblock Cambridge University Press, Cambridge, US, 1994.
\newblock \doi{10.1017/CBO9780511815478}.

\end{thebibliography}
%








\appendix
\newpage

\section{Historical Sources}
\label{sec:ApdxHistSources}
Table~\ref{tab:Sources} lists of all the documents used as primary historical sources in the present work. The conservation reference numbers include the following acronyms:
\begin{itemize}
    \item AAV: Apostolic Archives of the Vatican;
    \item AMA: Avignon Municipal Archives;
    \item DABdR: Departmental Archives of Bouches-du-Rhône;
    \item DAV: Departmental Archives of Vaucluse;
    \item PACB: Private Archives of the Château de Barbentane.
\end{itemize}
The Accounts of the City's Clavigers lists the properties belonging to the Pope since 1348.


\begin{table}[!htbp]
    \caption{List of documents used as the main historical sources in the present work. See the main text for the meaning of the acronyms used in column \textit{CRN} (Conservation Reference Numbers).}
    \label{tab:Sources}
    \centering
    \resizebox{\textwidth}{!}{%
        \begin{tabular}{l r l l}
        \hline
        \textbf{Document} & \textbf{Date} & \textbf{CRN} & \textbf{Language} \\ 
        \hline
        Terrier of Bishop Anglic Grimoard & 1366 & DAV, 1G10 & Latin \\
        Terrier of the Provost of the Cathedral & 1366 & DAV, G536 & Latin \\
        Terrier of Sainte-Catherine & 1366 & DAV, 71H5 & Provençal \\
        Terrier of the Urban Community & 1362 & AMA & Latin \\
        Terrier of the Repenties of Notre-Dame des Miracles & 1395 & DAV, 107H15 & Latin \\
        Terrier of the Collegiate Chapter of Saint-Pierre & 1380 & DAV, 9G2 & Latin \\
        Accounts of the City's Clavigers & 1384 & AAV & Latin \\
        Terrier of the Hospital of Saint-John of Jerusalem & 1350 & DABdR & Provençal \\
        Terrier of the Cabassoles & 1319 & PACB & Latin \\
        \hline
        \end{tabular}%
    }
\end{table}

\section{Relationship Normalization}
\label{sec:ApdxConvRel}
Table~\ref{tab:ConvRel} lists the relationships between objects that appear in the historical sources, with their English translation, and how we normalize them to ease graph extraction. 

For certain relationships, the normalization depends on whether the considered object is two-dim\-ensional, by opposition to punctual (building, shop...) and linear (streets, channels...) objects. Two-dimensional objects include boroughs, parishes, and cardinal's liveries. 

Relationship \textit{Egal} (i.e. \textit{Same as}) is particular, as it does not appear explicitly in the land registers, but is inferred manually. It connects two surface forms representing the same real-world entity: we merge them during the normalization.

\begin{table}[!htbp]
    \caption{Types of relationships appearing in the land registers to position properties relatively to other spatial objects, their translation in English, and how they are normalized before graph extraction.}
    \label{tab:ConvRel}
    \centering
    \resizebox{\textwidth}{!}{%
        \begin{tabular}{r l l l}
        \hline
        \textbf{Occurrences} & \textbf{Relationship} & \textbf{Translation} & \textbf{Normalized Form} \\
        \hline
        73 & \textit{Iuxta} & Near... & Related to... \\
        250 & \textit{Juxta} &  &  \\
        82 & \textit{Prope} &  &  \\
        2 & \textit{Proxime} &  &  \\
        \hline
        3 & \textit{In Angulo} & At the corner of... & Street: Related to... \\
        1 & \textit{In Cantono} &  & 2D objects: Inside of... \\
        13 & \textit{In Compito Sive Cantono} &  &  \\
        \hline
        8 & \textit{In Introytu} & At the entrance of... & Street: Related to... \\
         &  &  & Others: Inside of... \\
        \hline
        91 & \textit{Extra} & Outside of... & Outside of... \\
        \hline
        2,009 & \textit{In} & Inside of... & 2D objects: Inside of... \\
        4 & \textit{Intra} &  & Others: Related to... \\
        \hline
        2 & \textit{Ab Opposito} & Opposite to... & Related to... \\
        1 & \textit{Ex Opposit} &  &  \\
        \hline
        10 & \textit{In Capite} & At the beginning of... & 2D objects: Inside of... \\
         &  &  & Others: Related to... \\
        \hline
        18 & \textit{Super} & Above... & Related to... \\
        9 & \textit{Supra} &  &  \\
        \hline
        521 & \textit{A Orient} & West of... & West of... \\
        \hline
        535 & \textit{A Occident} & East of... & East of... \\
        \hline
        511 & \textit{A Circio} & South of... & South of... \\
        1 & \textit{Ab Aura Recta} &  &  \\
        \hline
        548 & \textit{A Meridie} & North of... & North of... \\
        \hline
        2 & \textit{A Una Part} & One side facing... & Related to... \\
        1,208 & Ab Una Part &  &  \\
        \hline
        4 & \textit{A Duabus Part} & Two sides facing... & Related to... \\
        \hline
        1 & \textit{A Tribus Part} & Three sides facing... & Related to... \\
        \hline
        21 & \textit{A Parte Retro} & Rear side facing... & Related to... \\
        \hline
        17 & \textit{A Part Ante} & Front side facing... & Related to... \\
        \hline
        1 & \textit{A Part Inferiori} & Lower side facing... & Related to... \\
        \hline
        2 & \textit{A Parte Lateris} & Lateral side facing... & Related to... \\
        \hline
        5 & \textit{A Part Posteriori} & Rear side facing... & Related to... \\
        \hline
        1 & \textit{Sive Ab Una Part} & One side may face... & Related to... \\
        \hline
        1 & \textit{Conjuncto} & Adjacent to... & Related to... \\
        15 & \textit{Contigu} &  &  \\
        2 & \textit{Contiguo} &  &  \\
        \hline
        35 & \textit{Retro} & Behind... & Related to... \\
        \hline
        59 & \textit{Ante} & Before... & Related to... \\
        \hline
        3 & \textit{Egal} & Same as... & (Merge vertices) \\
        \hline
        6 & \textit{Infra} & Below... & Related to... \\
        29 & \textit{Subtus} &  &  \\
        \hline
        8 & \textit{Ad} & Towards... & Related to... \\
        4 & \textit{Apud} &  &  \\
        13 & \textit{Versus} &  &  \\
        \hline
        \end{tabular}
    }
\end{table}

\section{Extracted Graphs}
\label{sec:ApdxGraphs}
\begin{figure}[!htbp]
    \centering
    \includegraphics[width=1\linewidth]{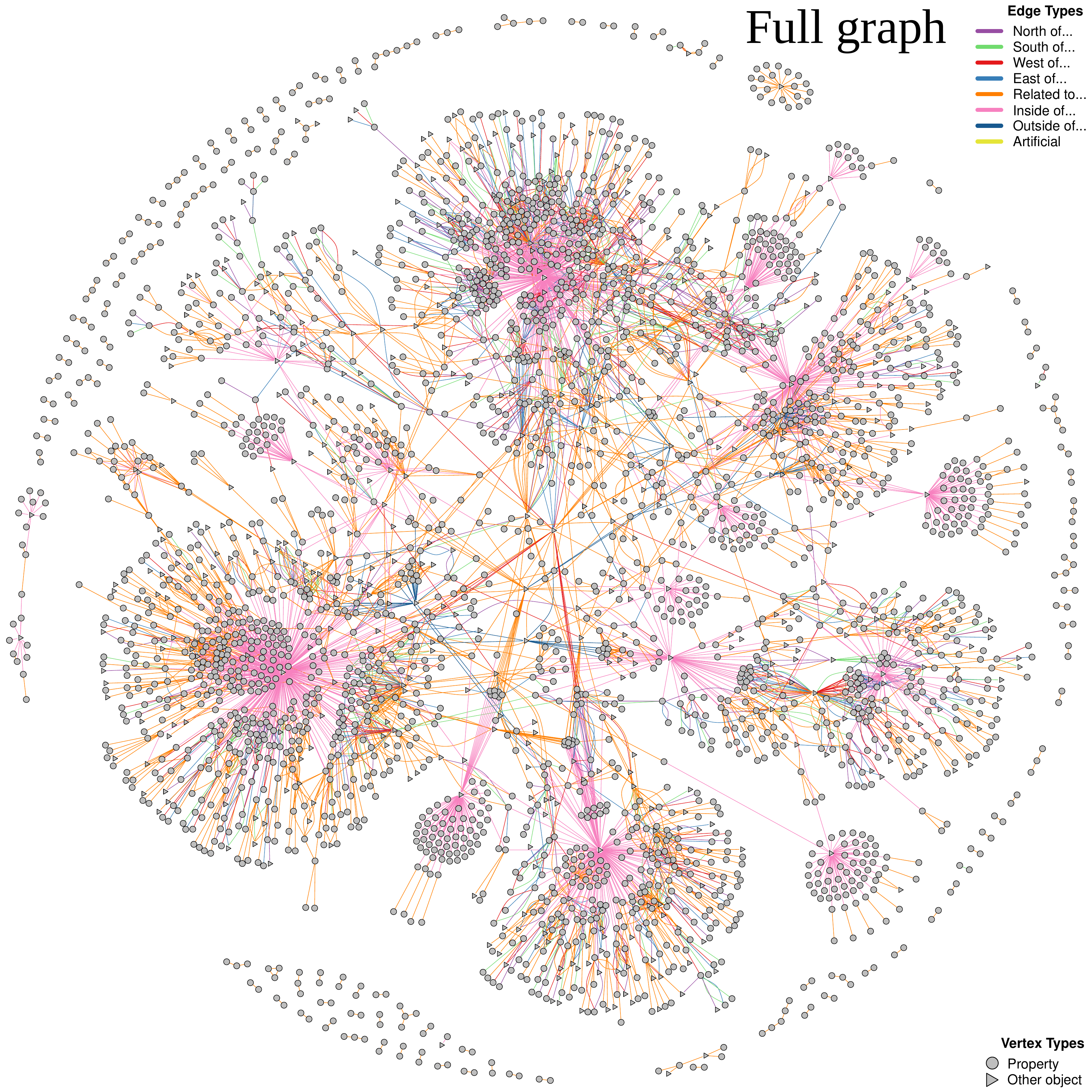}
    \caption{Full graph extracted from our database. It contains all the available raw data (but no additional relationships), without any vertex filtering. The layout is obtained using the Yifan-Hu method~\cite{Hu2006}. Figure available at \href{http://doi.org/10.5281/zenodo.14175830}{10.5281/zenodo.14175830} under CC-BY license.}
    \label{fig:GraphFull}
\end{figure}

Figure~\ref{fig:GraphFull} shows the full graph, i.e. the graph extracted without discarding any vertex or edge, and without inserting any additional relationship. Vertex location is determined using the Yifan-Hu layouting algorithm~\cite{Hu2006}.

\begin{figure}[!htbp]
    \centering
    \hfill
    \includegraphics[width=0.40\linewidth]{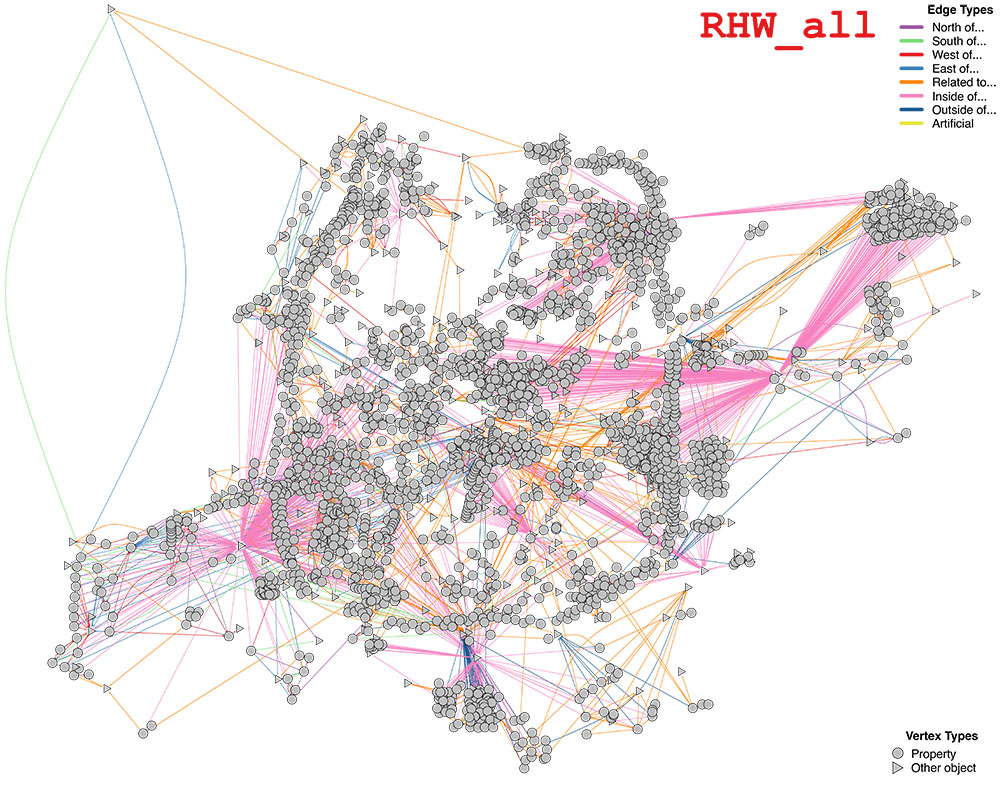}
    \hfill
    \includegraphics[width=0.40\linewidth]{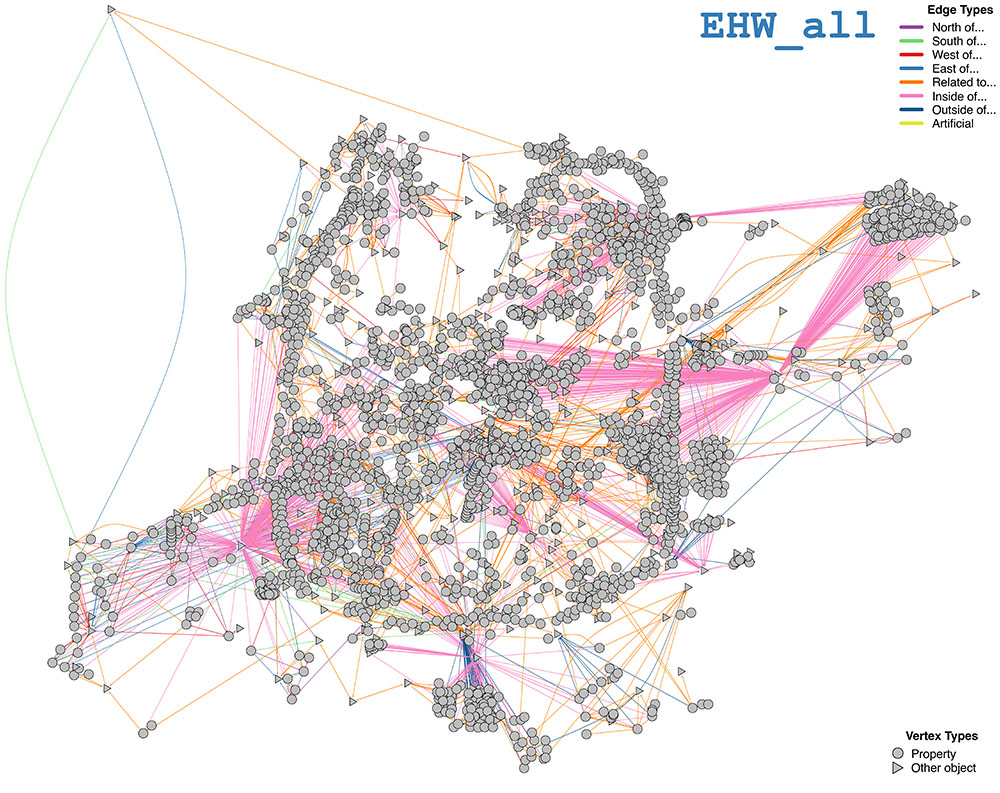}
    \hfill~\\[0.2cm]
    \hfill
    \includegraphics[width=0.40\linewidth]{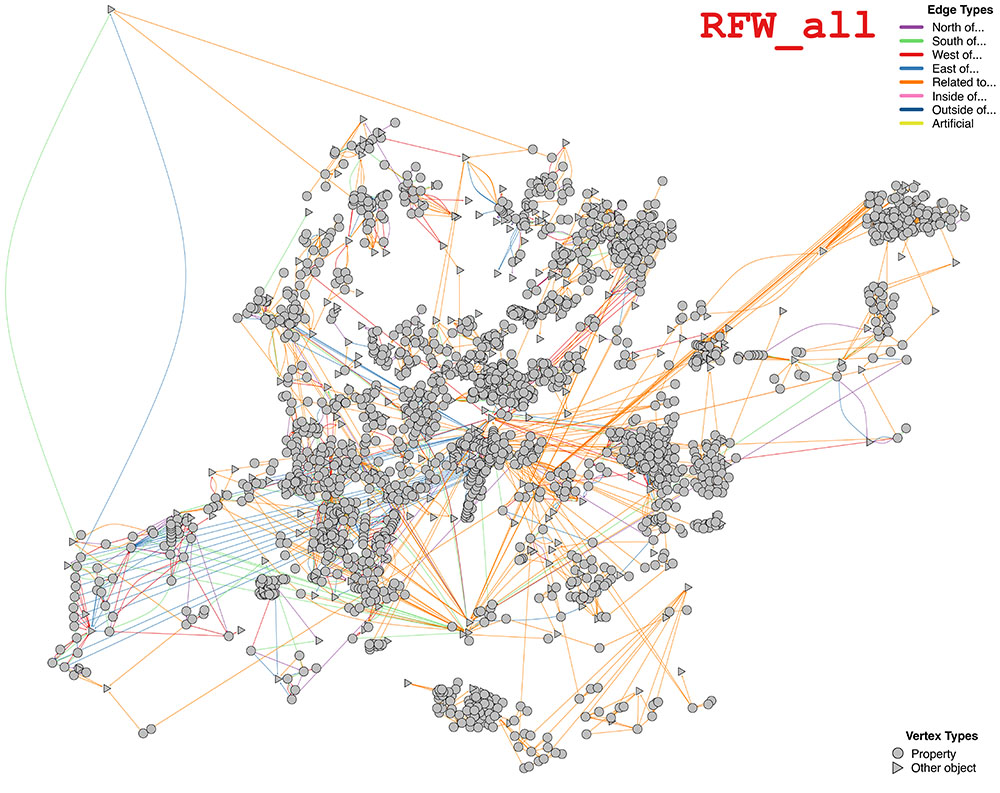}
    \hfill
    \includegraphics[width=0.40\linewidth]{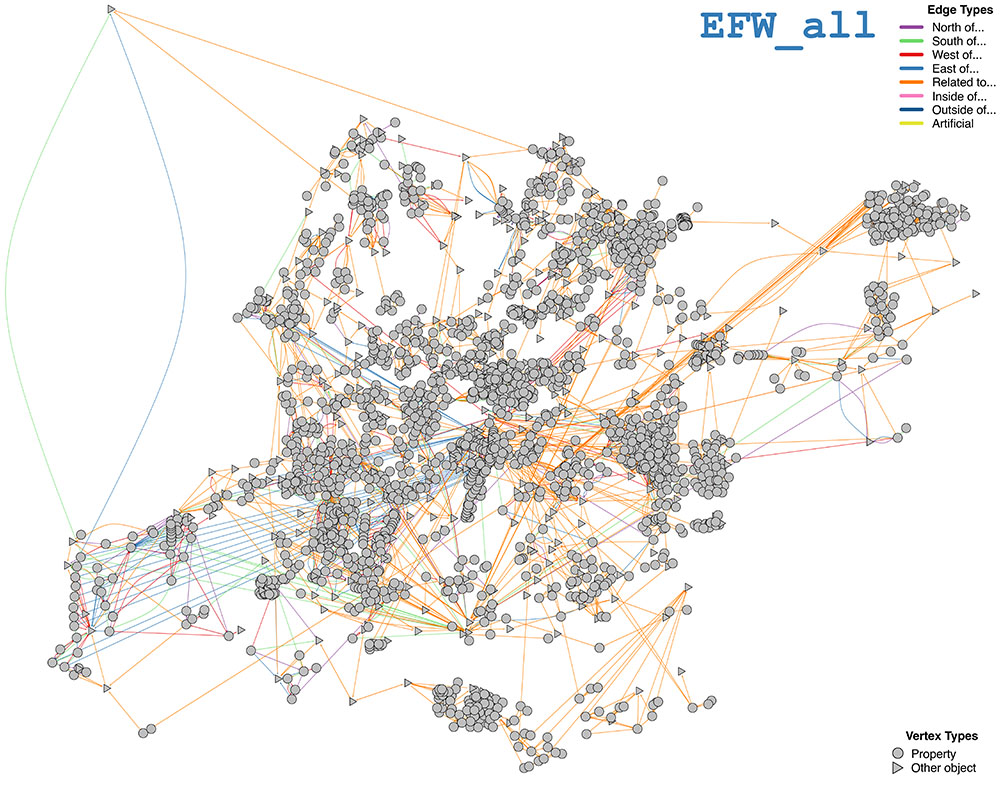}
    \hfill~\\[0.2cm]
    \hfill
    \includegraphics[width=0.40\linewidth]{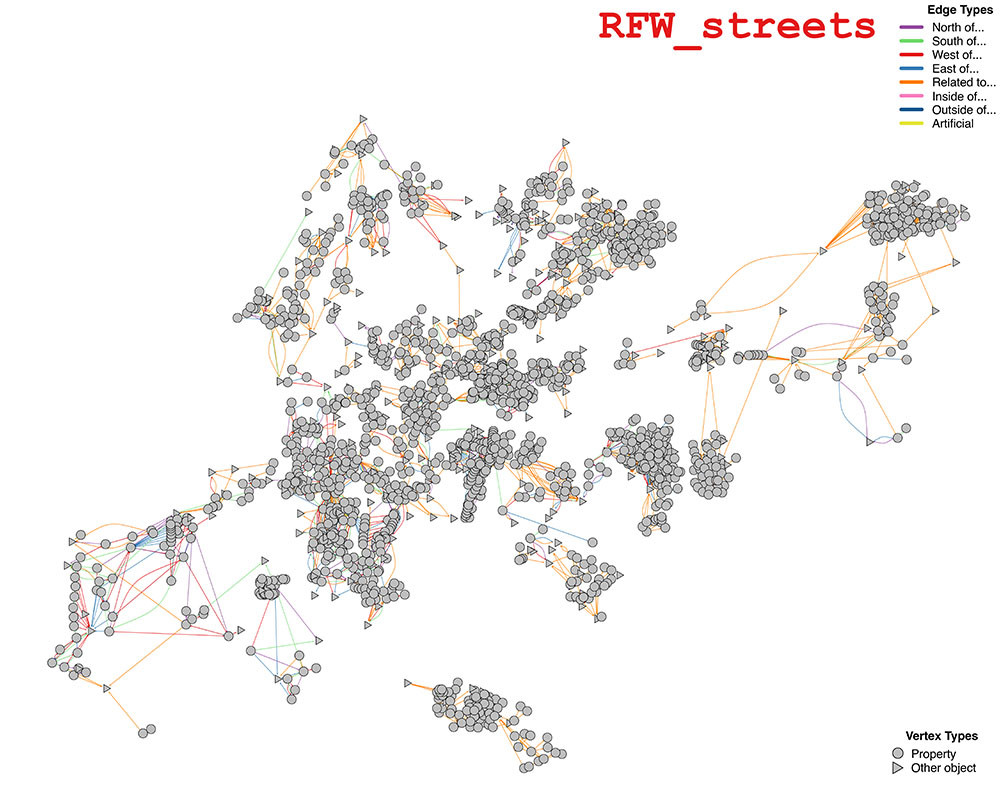}
    \hfill
    \includegraphics[width=0.40\linewidth]{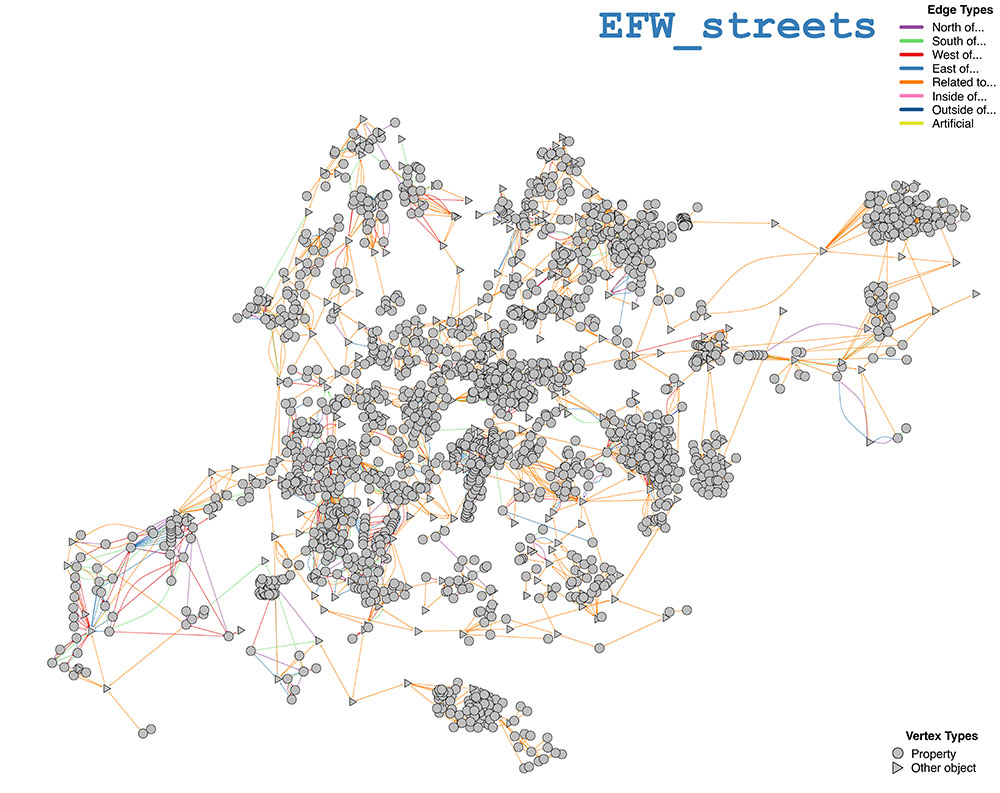}
    \hfill~\\[0.2cm]
    \hfill
    \includegraphics[width=0.40\linewidth]{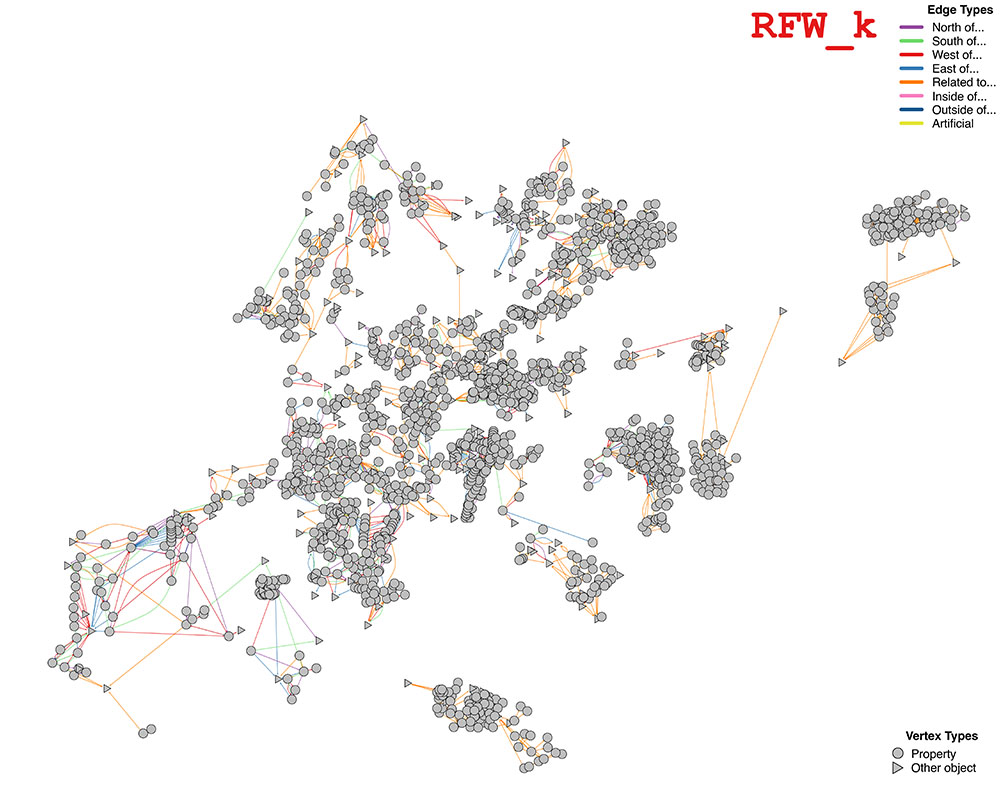}
    \hfill
    \includegraphics[width=0.40\linewidth]{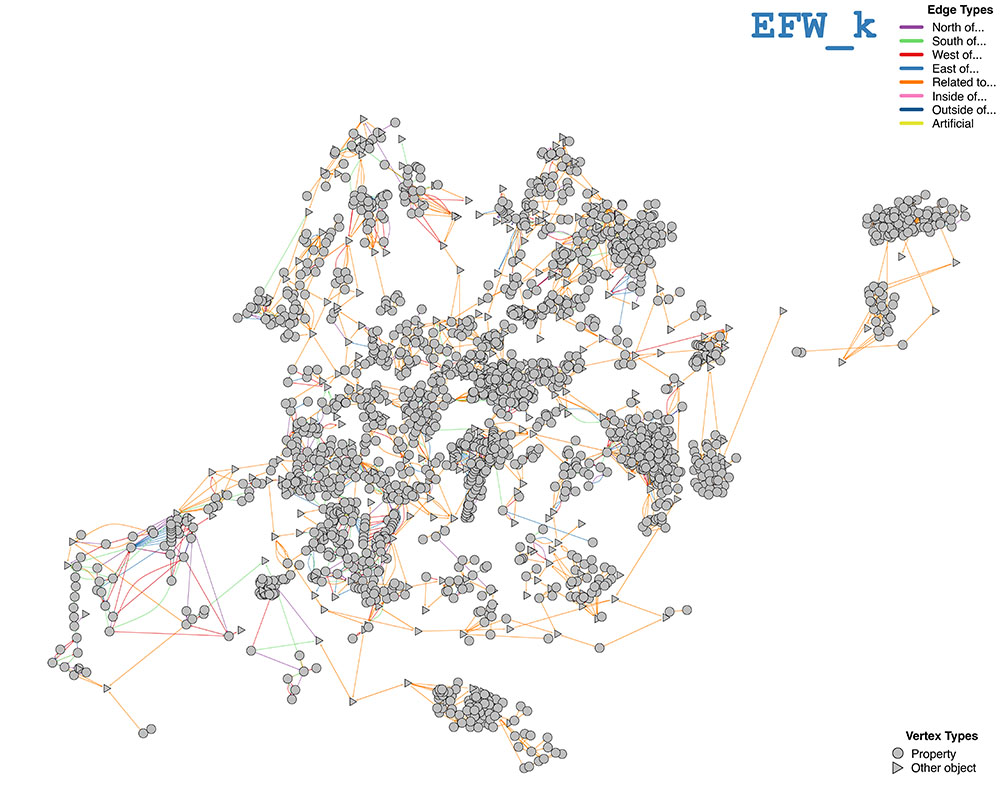}
    \hfill~\\
    \caption{Graphs extracted using the whole vertex approaches (\texttt{W}). Vertices are positioned depending on their geographic location. Those whose spatial position is unknown are not shown. The left-hand column shows the graphs based on the raw data (\texttt{R}), whereas the right-hand one contains the extended versions that leverage additional relationships (\texttt{E}). The two top rows show the graphs that focus on \texttt{all} types of vertices, with both flat and hierarchical relations (\texttt{H}) for the first and only flat edges for the second (\texttt{F}). The two bottom rows also show flat graphs, but without the non-punctual objects except the streets (\texttt{street}) and the shorter streets (\texttt{k}), respectively. Figure available at \href{http://doi.org/10.5281/zenodo.14175830}{10.5281/zenodo.14175830} under CC-BY license.}
    \label{fig:GraphsWholeLambert}
\end{figure}

\begin{figure}[!htbp]
    \centering
    \hfill
    \includegraphics[width=0.40\linewidth]{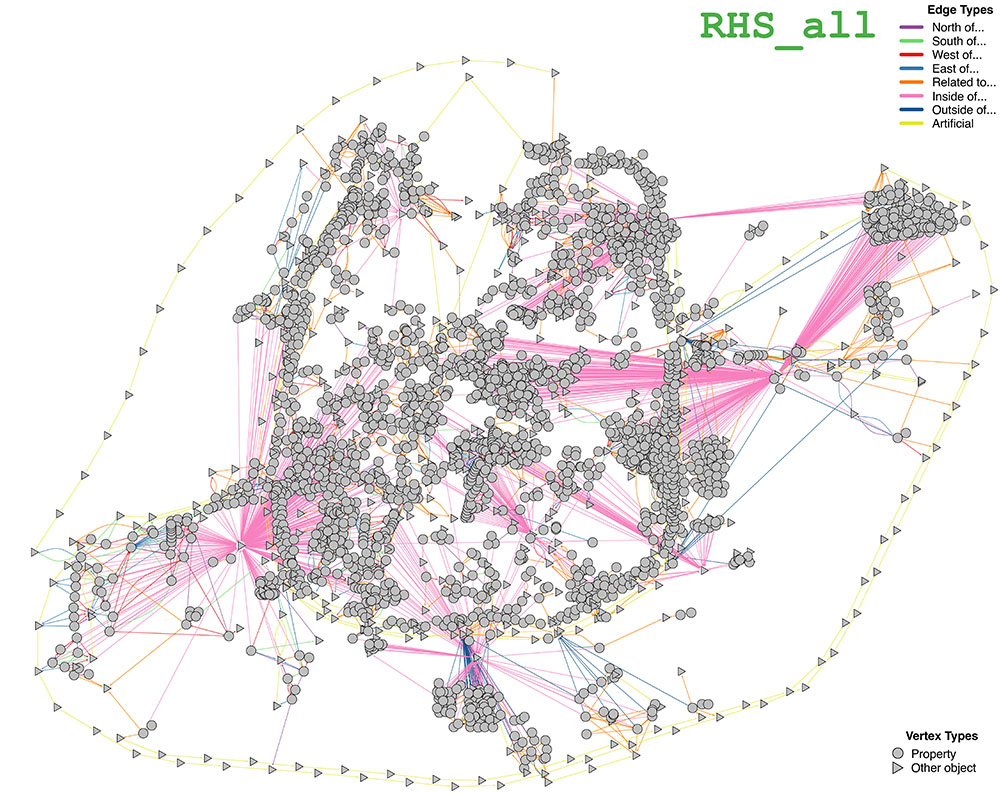}
    \hfill
    \includegraphics[width=0.40\linewidth]{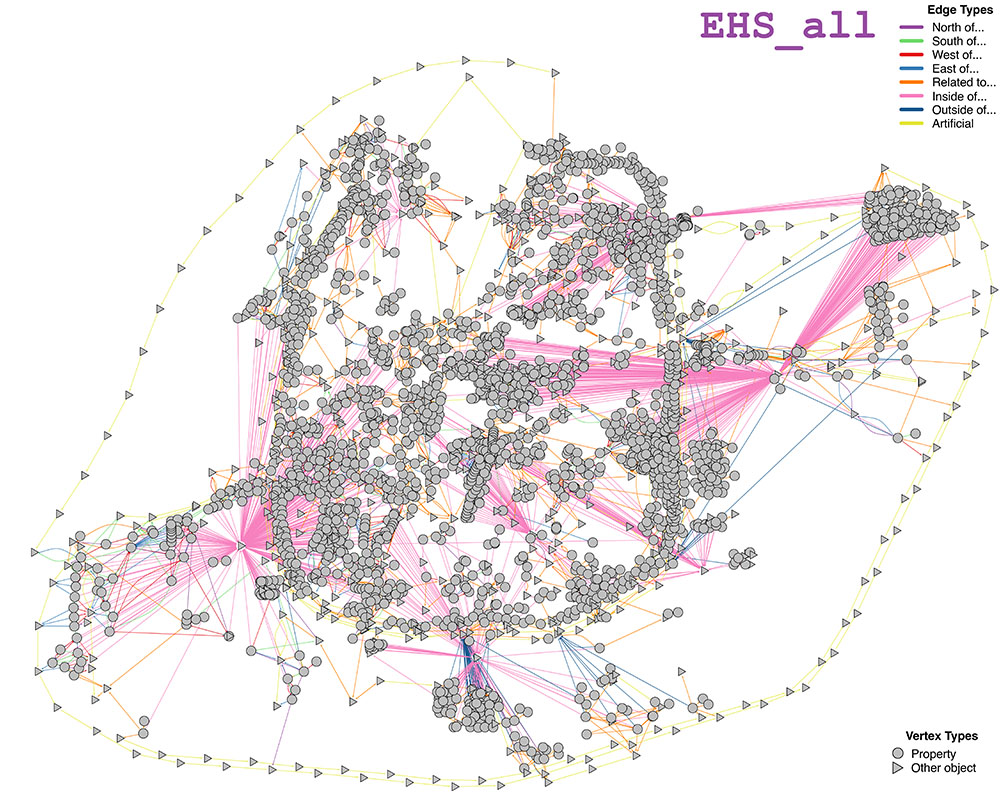}
    \hfill~\\[0.2cm]
    \hfill
    \includegraphics[width=0.40\linewidth]{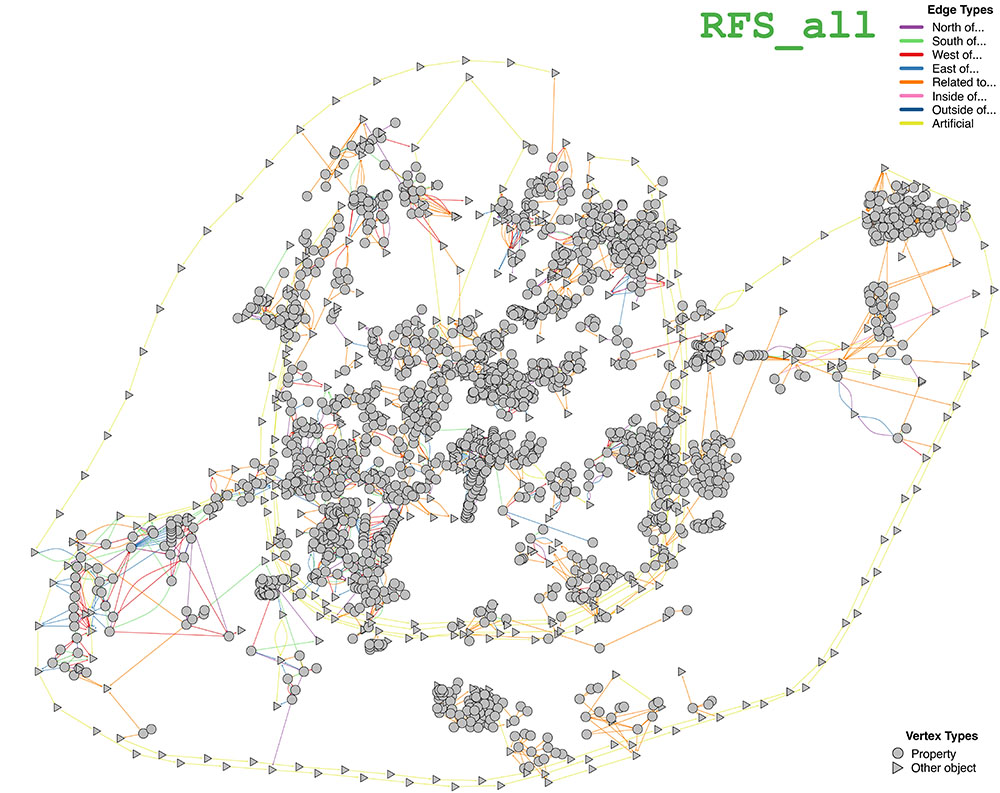}
    \hfill
    \includegraphics[width=0.40\linewidth]{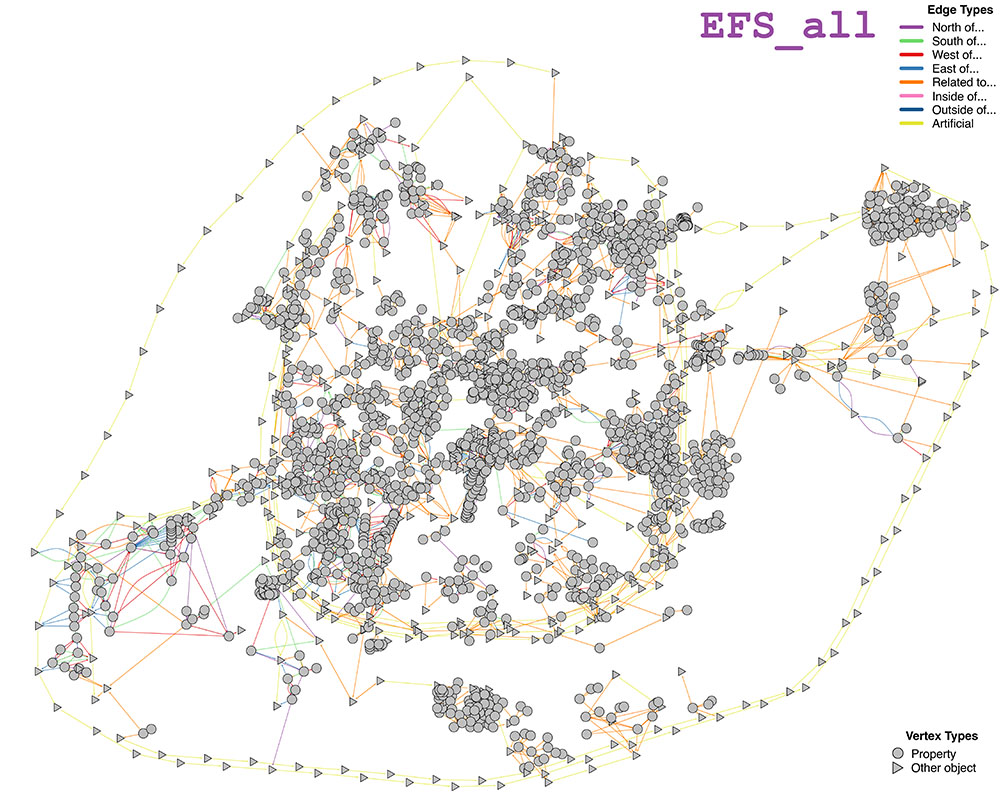}
    \hfill~\\[0.2cm]
    \hfill
    \includegraphics[width=0.40\linewidth]{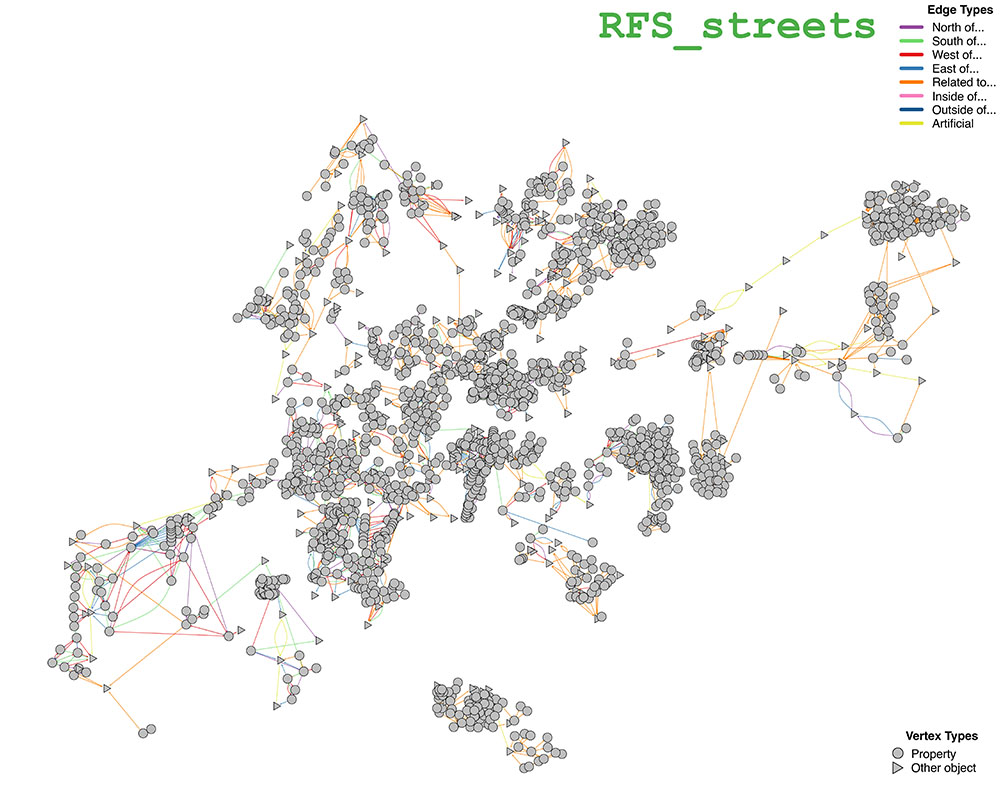}
    \hfill
    \includegraphics[width=0.40\linewidth]{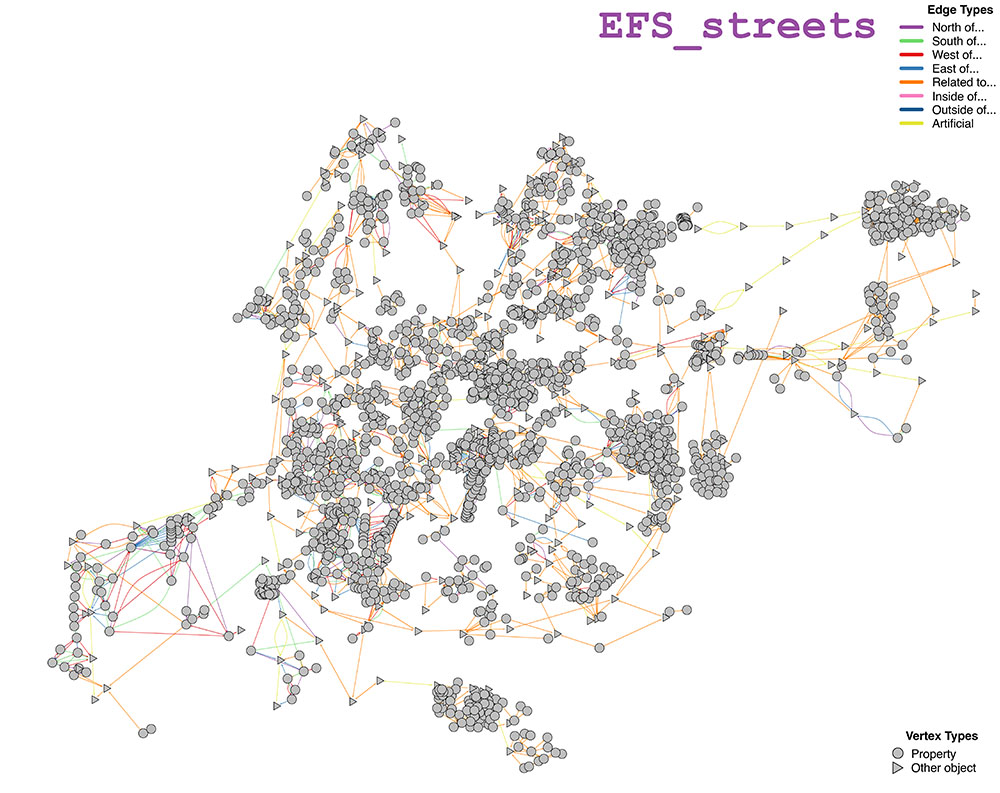}
    \hfill~\\[0.2cm]
    \hfill
    \includegraphics[width=0.40\linewidth]{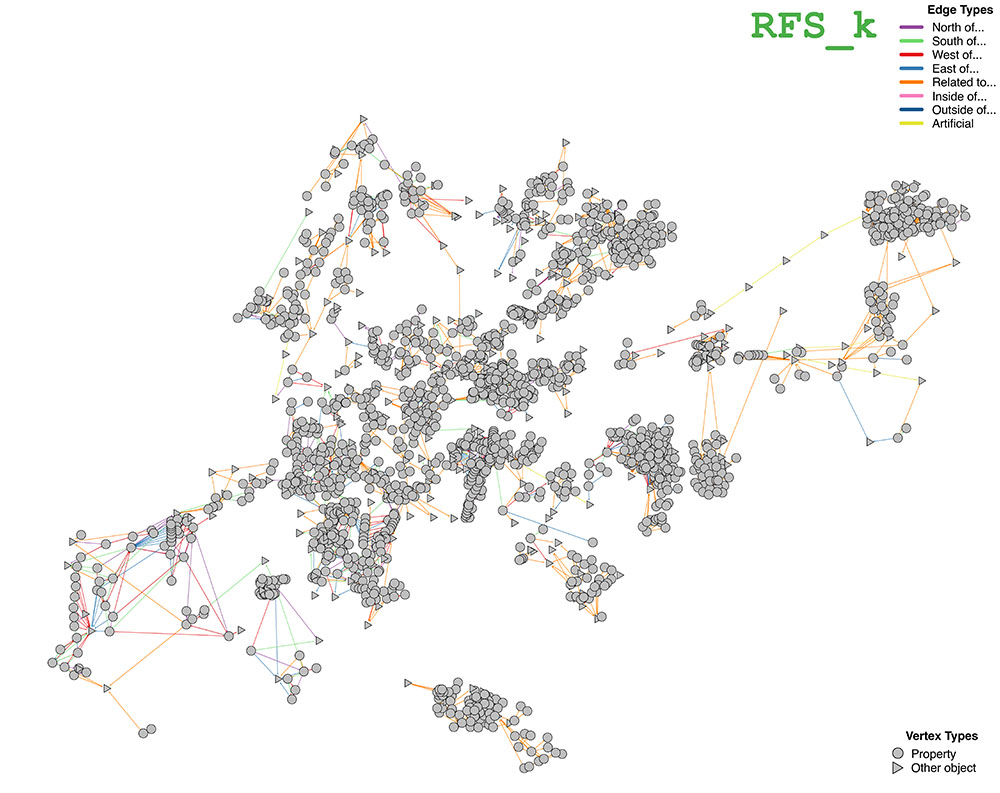}
    \hfill
    \includegraphics[width=0.40\linewidth]{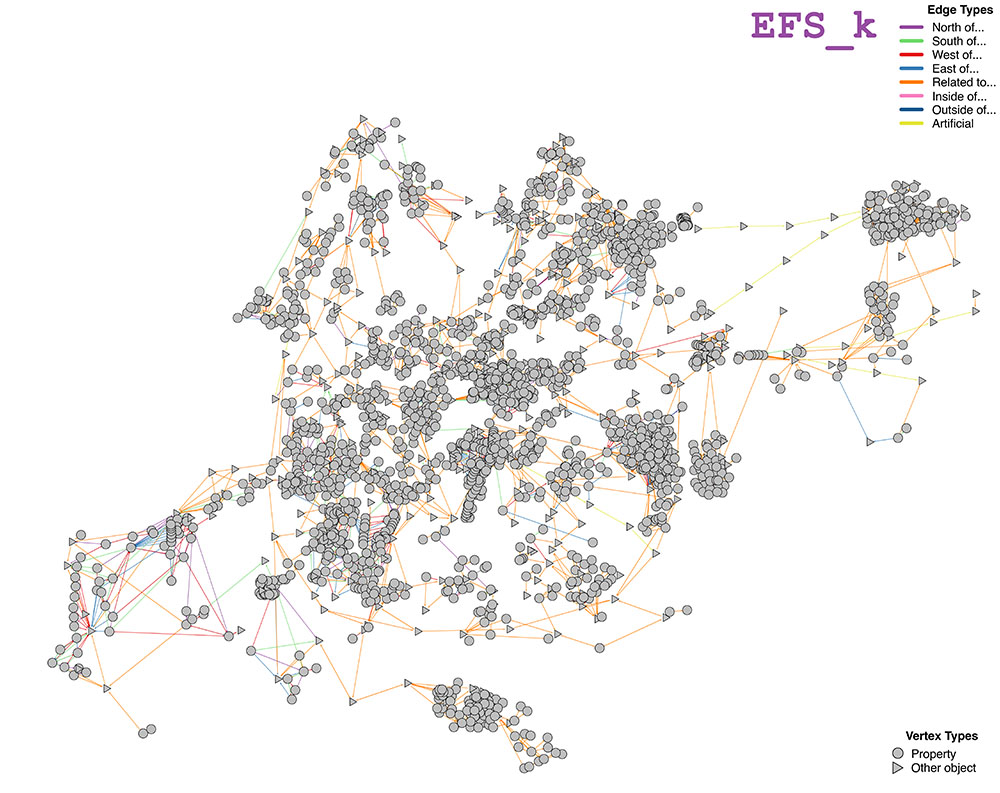}
    \hfill~\\
    \caption{Graphs extracted using the split vertex approaches (\texttt{S}). Vertices are positioned depending on their geographic location. Those whose spatial position is unknown are not shown. The left-hand column shows the graphs based on the raw data (\texttt{R}), whereas the right-hand one contains the extended versions that leverage additional relationships (\texttt{E}). The two top rows show the graphs that focus on \texttt{all} types of vertices, with both flat and hierarchical relations (\texttt{H}) for the first and only flat edges for the second (\texttt{F}). All non-punctual objects are split. The two bottom rows also show flat graphs, but without the non-punctual objects except the streets (\texttt{street}) and the shorter streets (\texttt{k}), respectively, which are split. Figure available at \href{http://doi.org/10.5281/zenodo.14175830}{10.5281/zenodo.14175830} under CC-BY license.}
    \label{fig:GraphsSplitLambert}
\end{figure}

Figures~\ref{fig:GraphsWholeLambert} and~\ref{fig:GraphsSplitLambert} show the graphs extracted through all the optional steps considered in the article. The layout corresponds to the spatial position of the vertices: exact when it is known, and estimated when it is unknown.

\section{Extraction Statistics}
\label{sec:ApdxStats}

\subsection{Long Streets}
\label{sec:ApdxStatsStreetProc}
\begin{figure}[!htbp]
    \centering
    \includegraphics[width=0.49\linewidth]{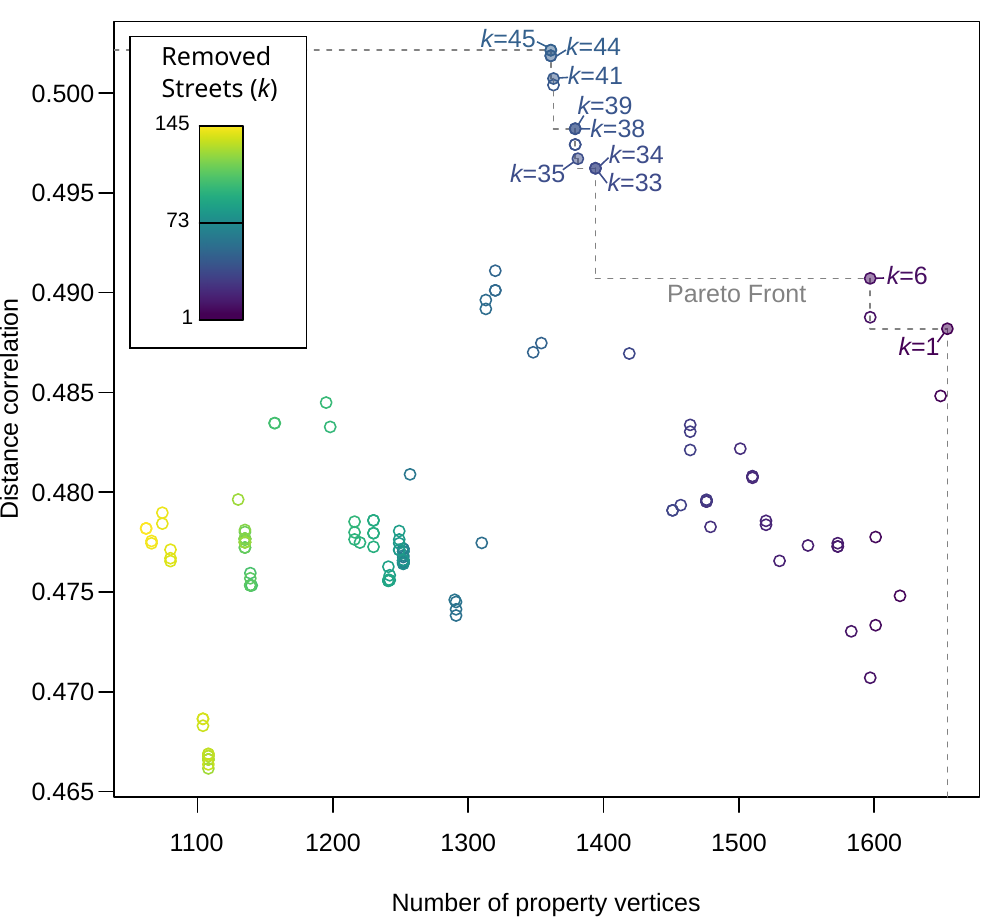}
    \hfill
    \includegraphics[width=0.49\linewidth]{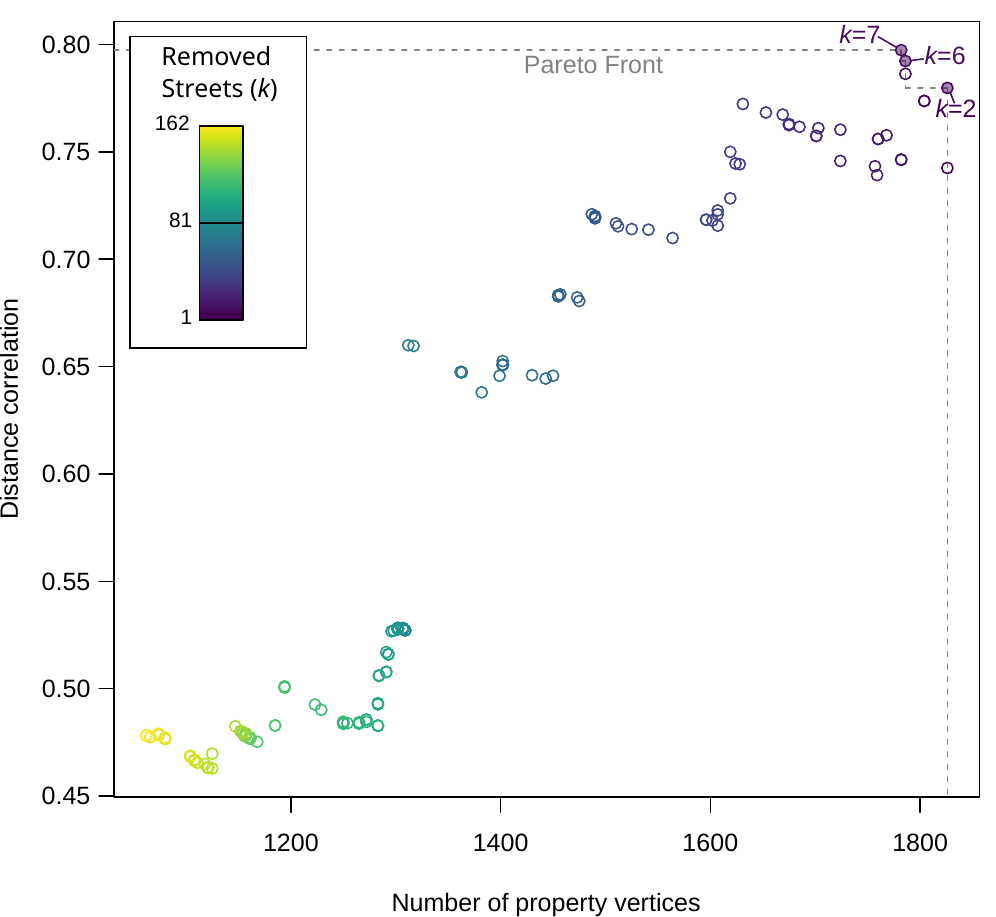}
    \caption{Selection of the optimal value of parameter $k$ for the \texttt{RFW\_k} (left) and \texttt{EFW\_k} (right) extraction methods (see Section~\ref{sec:MethCompStreets}). Figure available at \href{http://doi.org/10.5281/zenodo.14175830}{10.5281/zenodo.14175830} under CC-BY license.}
    \label{fig:ParetoK}
\end{figure}

Figure~\ref{fig:ParetoK} shows the plots used to select the optimal value of parameter $k$ during the extraction process of graphs \texttt{RFW\_k} (\textit{raw} data, flat relations, whole vertices, top $k$ streets removed) and \texttt{EFW\_k} (same thing, but with \textit{extended} data). This parameter corresponds to the number of streets removed during the extraction process. The removal starts with the longest streets (cf. Section~\ref{sec:ExtractionOptNonpunct}). The $y$ axis corresponds to Spearman's correlation between the graph and spatial distances. The $x$ axis shows the number of vertices that model properties and are still present in the graph after removing its longest streets. Each point shows both metrics for a specific value of $k$, indicated by its color. The dotted line materializes the Pareto front, which includes the solid dots.

\begin{figure}[!htbp]
    \centering
    \includegraphics[width=0.49\linewidth]{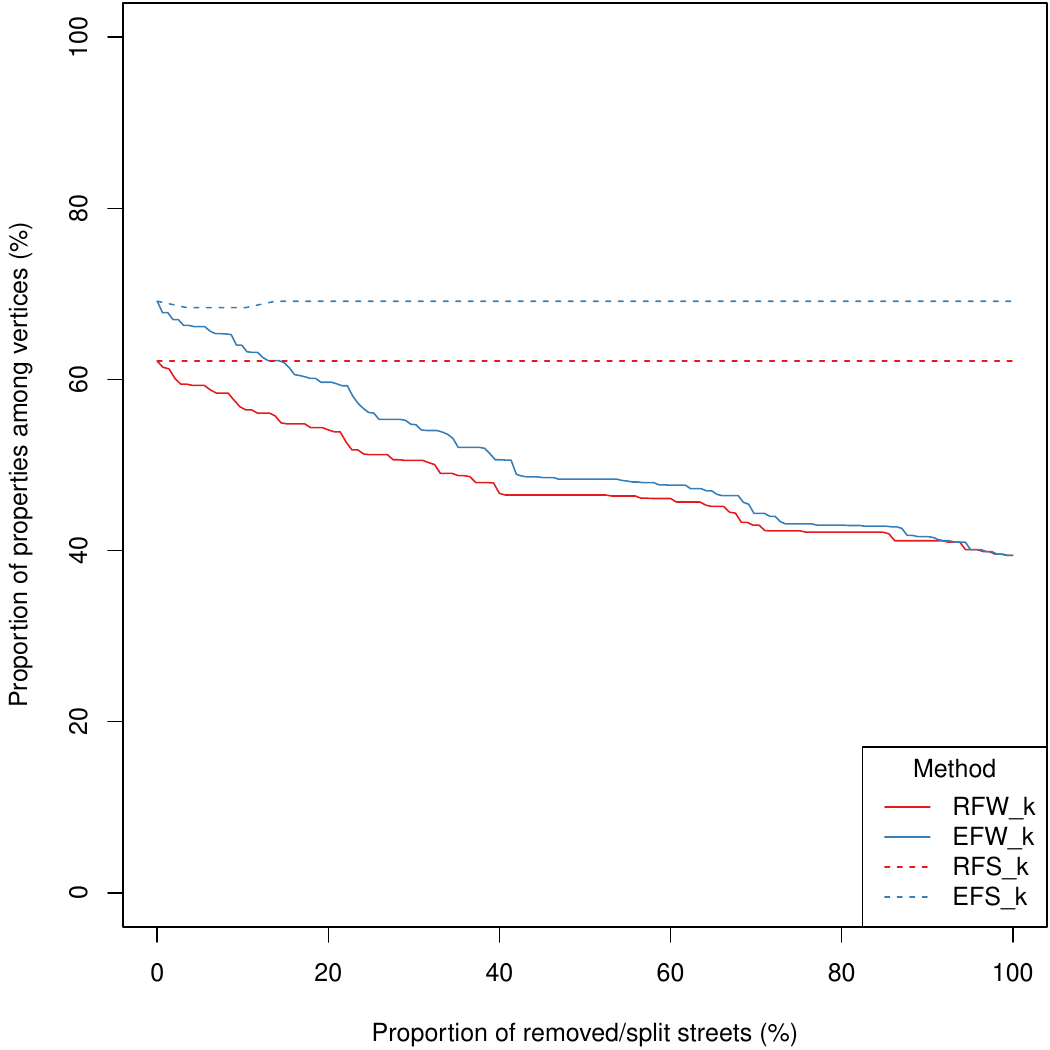}
    \hfill
    \includegraphics[width=0.49\linewidth]{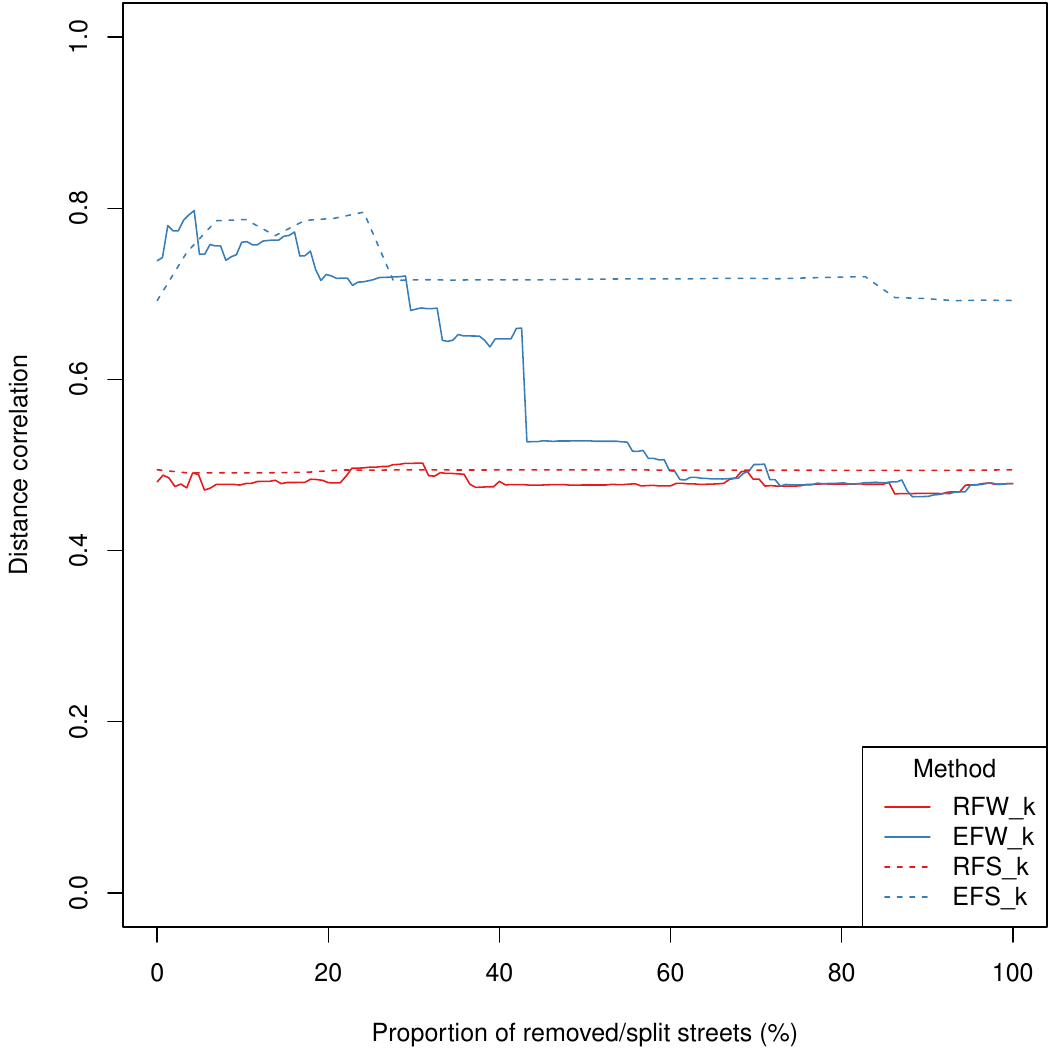}
    \caption{Evolution of the coverage (left) and distance correlation (right) as functions of the proportion of longest streets considered for removal or splitting in extraction methods \texttt{$\cdot$F$\cdot$\_k} (see Section~\ref{sec:MethCompSplit}). Figure available at \href{http://doi.org/10.5281/zenodo.14175830}{10.5281/zenodo.14175830} under CC-BY license.}
    \label{fig:EvolutionK}
\end{figure}

Figure~\ref{fig:EvolutionK} shows the evolution of our two criteria, as a function of the proportion of longest streets that undergo the removal from methods \texttt{$\cdot$FW\_k} (solid lines) or the splitting from methods \texttt{$\cdot$FS\_k} (dotted lines): on the left, the proportion of properties left in the produced graph, and on the right, Spearman's correlation between spatial and graph distances. Red lines show the results obtained on the raw data, whereas blue lines represent the extended data. Increasing $k$ when splitting has little effect on coverage, hence the horizontal dotted lines in the left plot. Street removal (solid lines) causes the deletion of all the properties that happen to be connected only to the deleted streets, hence the decrease in coverage. For distance correlation (right plot), increasing $k$ has not much effect when processing the raw data (red lines), be it through street removal or splitting. On the contrary, there is a clear effect when using the extended data: removing too many streets tend to decrease the distance correlation. 

\subsection{Distance Comparison}
\label{sec:ApdxStatsDistComp}
\begin{figure}[!htbp]
    \centering
    \hfill
    \includegraphics[width=0.24\linewidth]{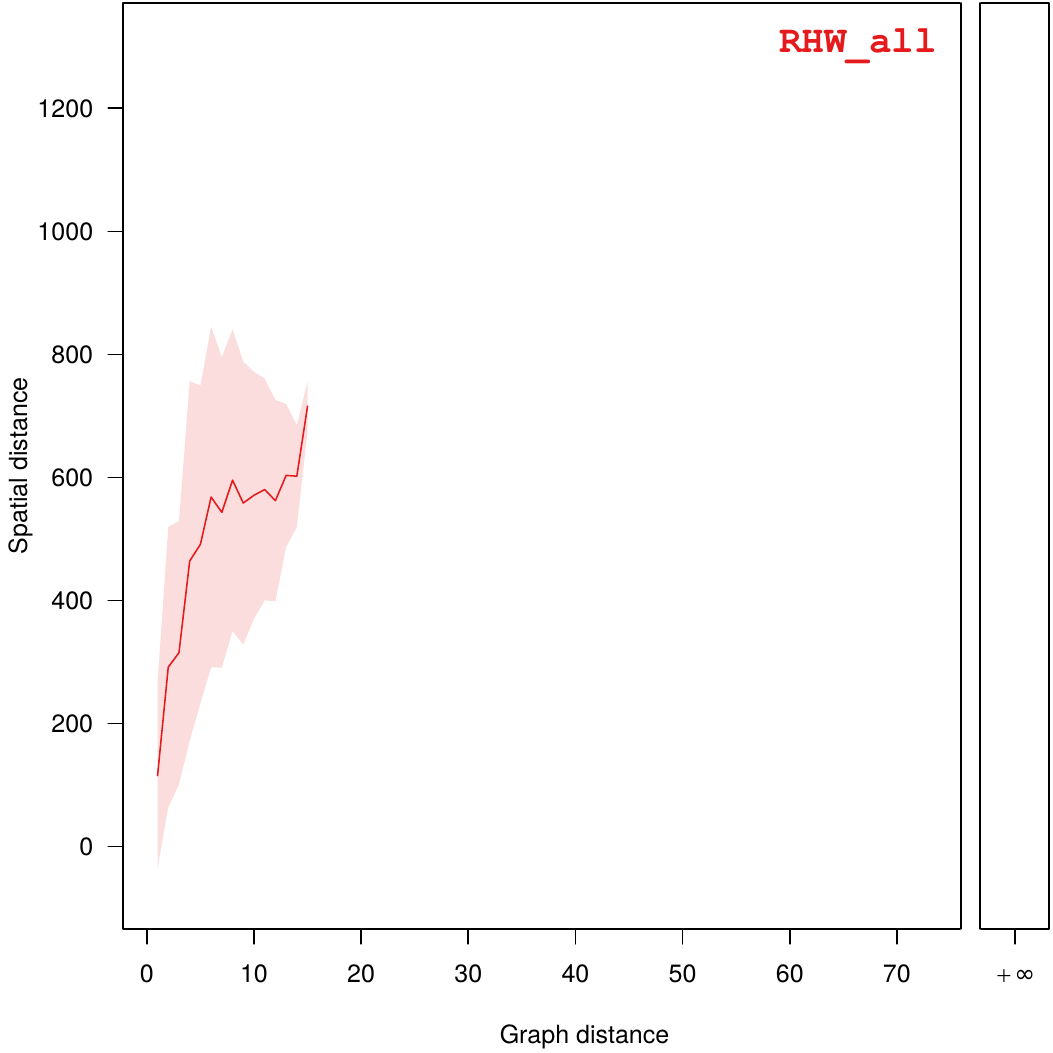}
    \hfill
    \includegraphics[width=0.24\linewidth]{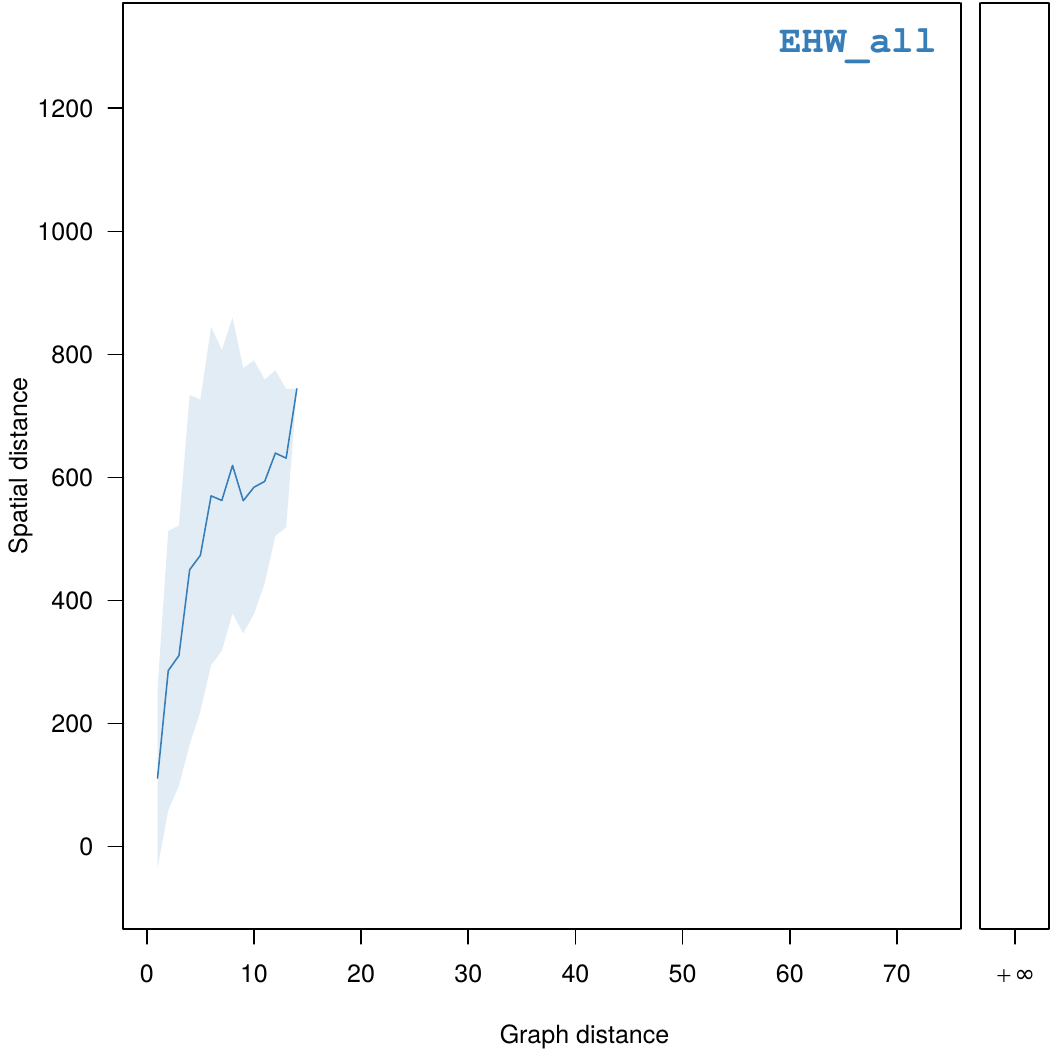}
    \hfill
    \includegraphics[width=0.24\linewidth]{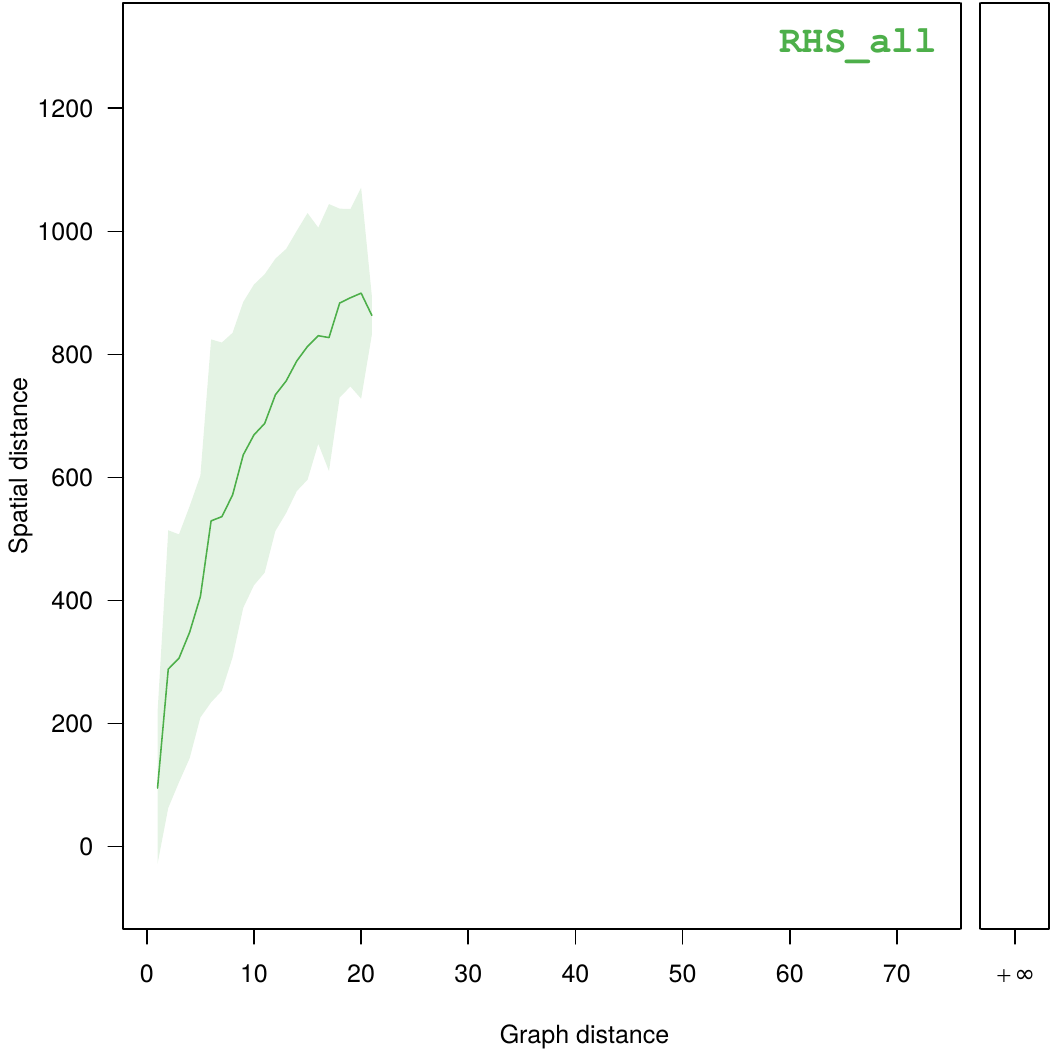}
    \hfill
    \includegraphics[width=0.24\linewidth]{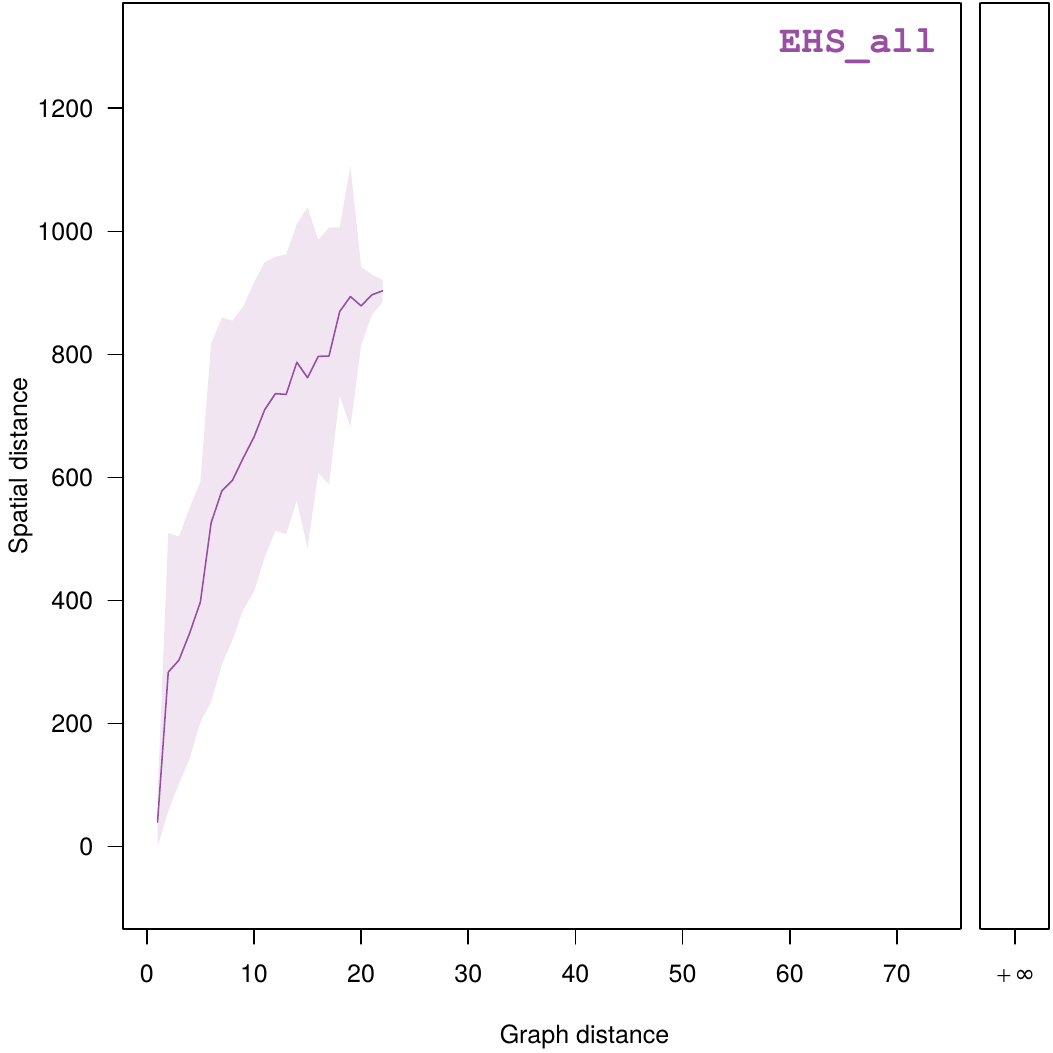}
    \hfill~\\[0.2cm]
    \hfill
    \includegraphics[width=0.24\linewidth]{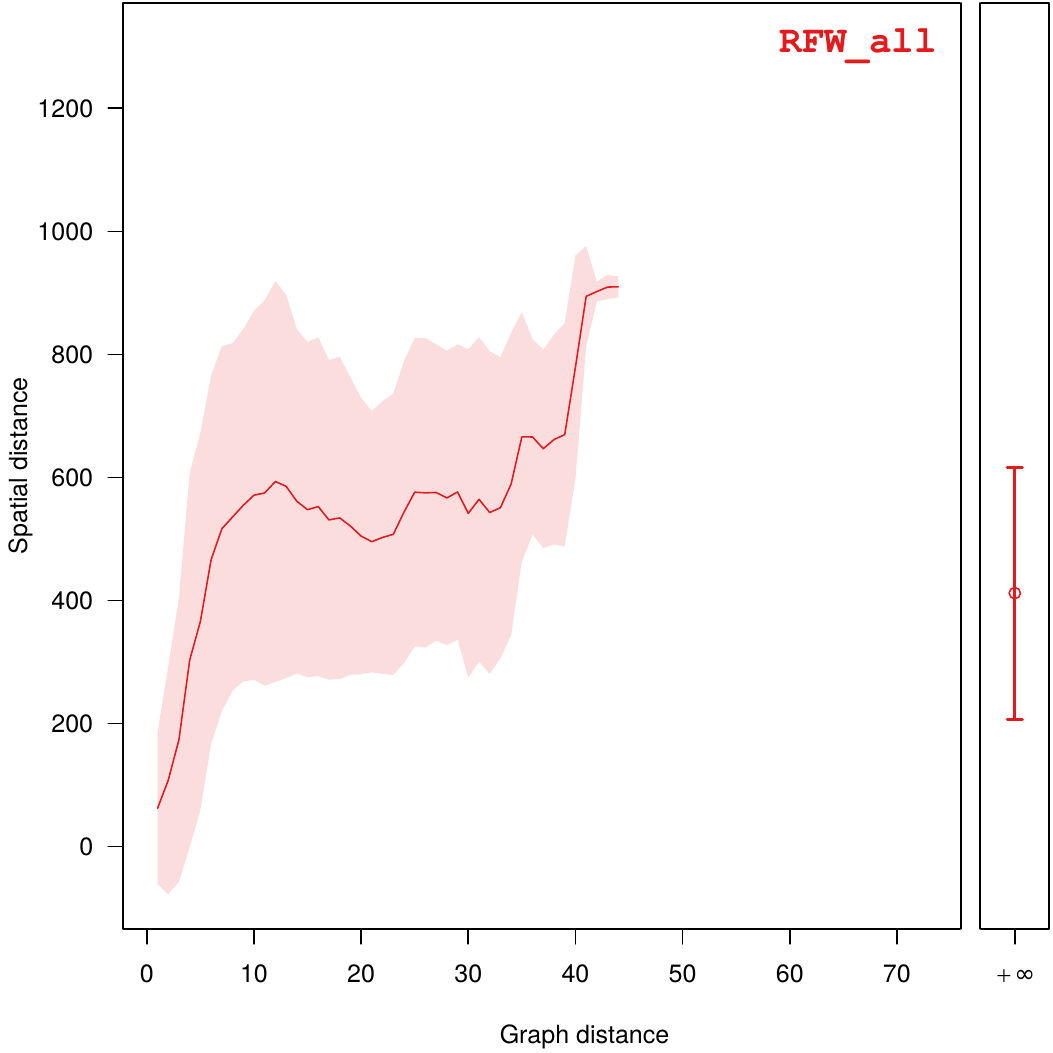}
    \hfill
    \includegraphics[width=0.24\linewidth]{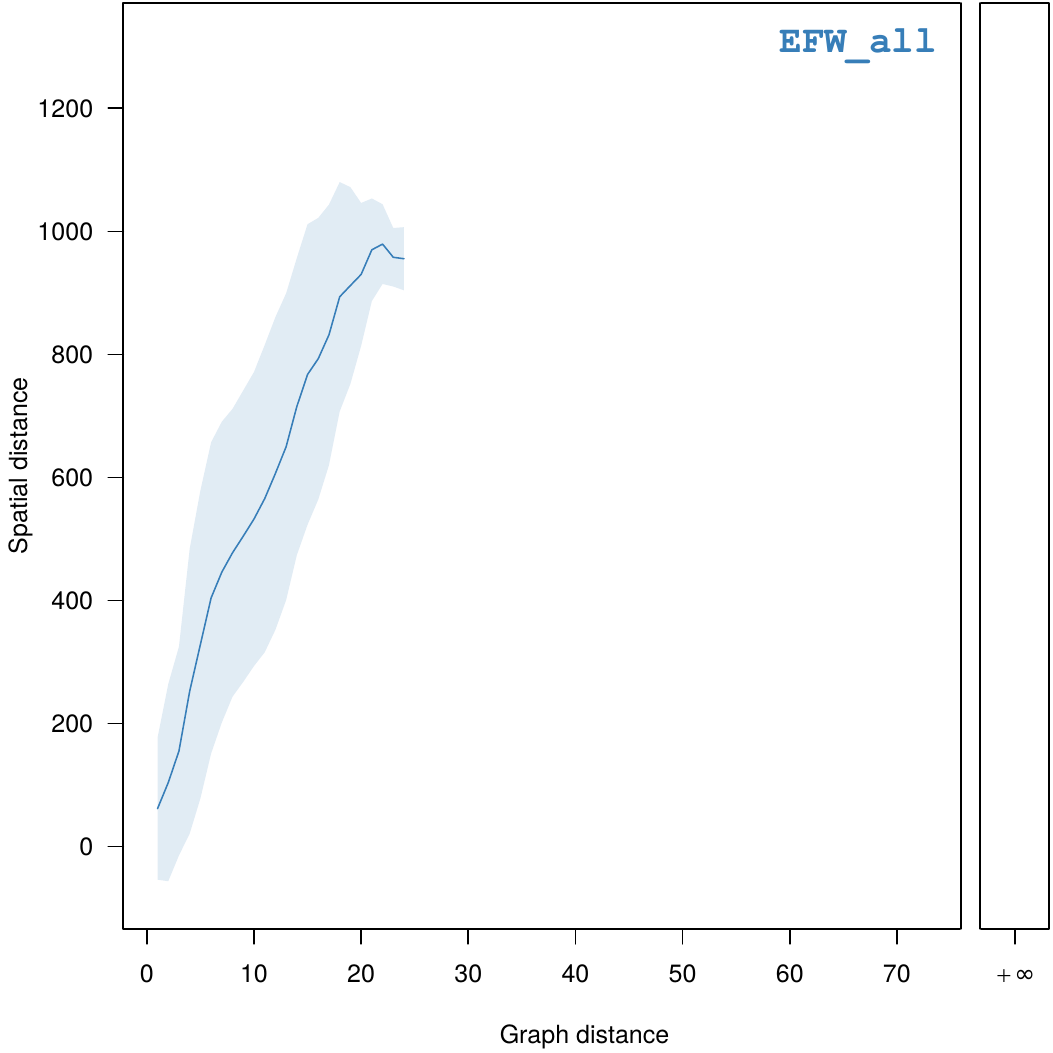}
    \hfill
    \includegraphics[width=0.24\linewidth]{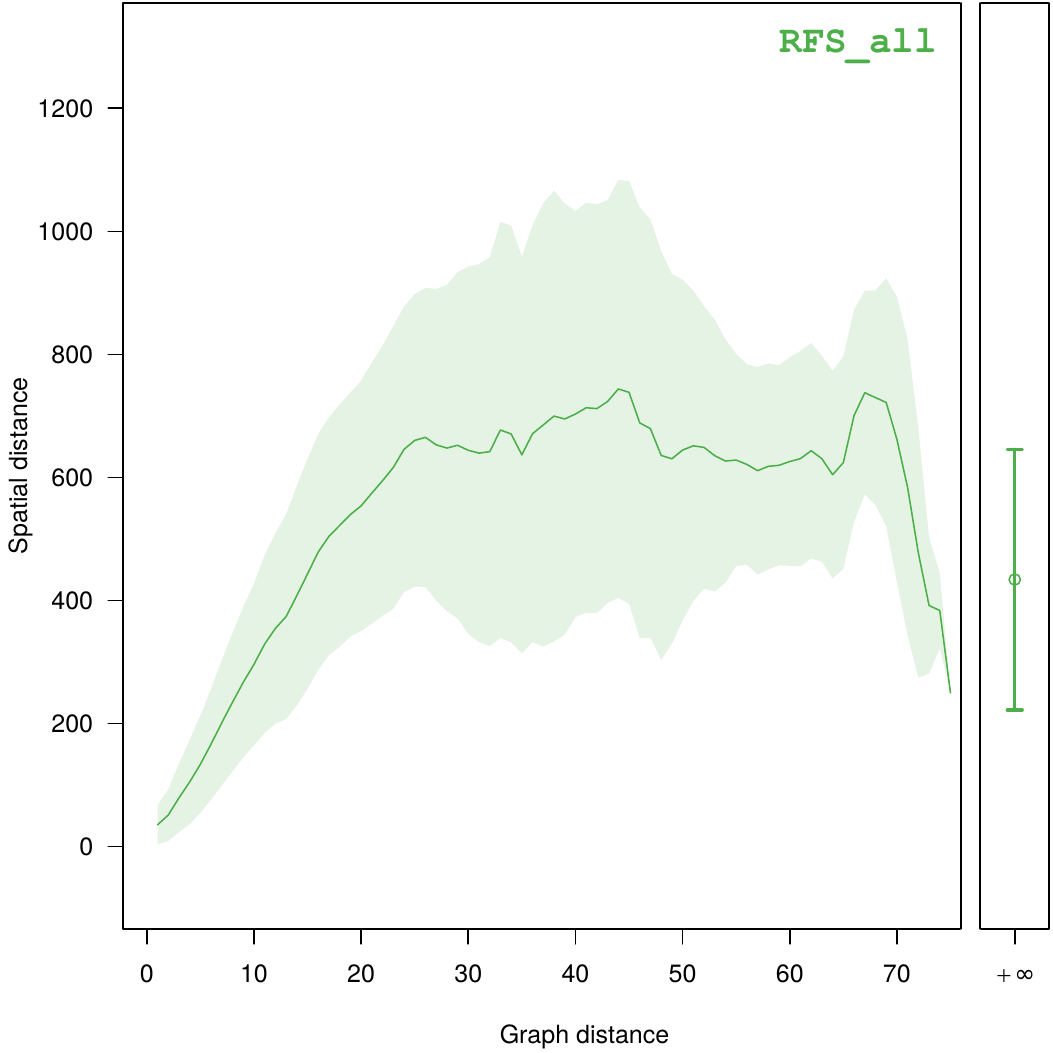}
    \hfill
    \includegraphics[width=0.24\linewidth]{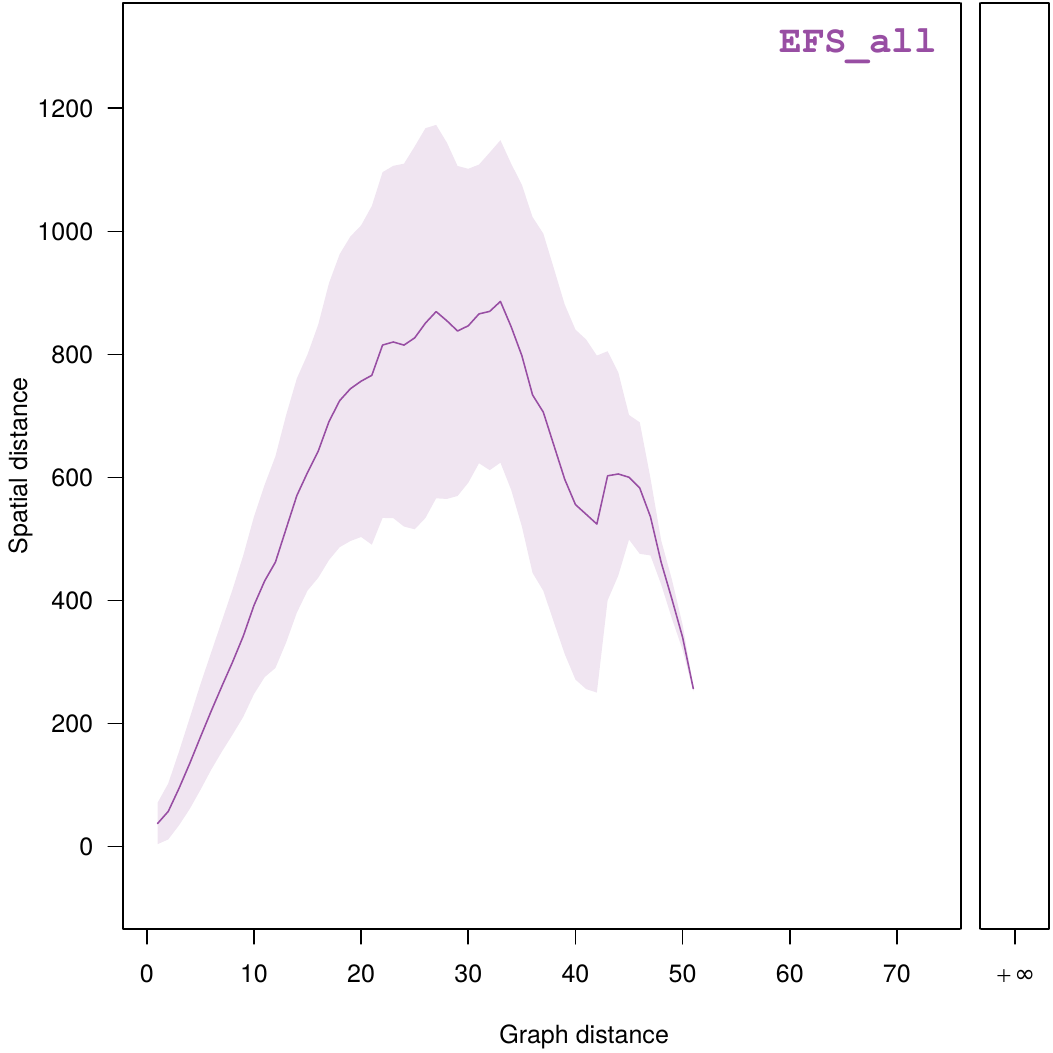}
    \hfill~\\[0.2cm]
    \hfill
    \includegraphics[width=0.24\linewidth]{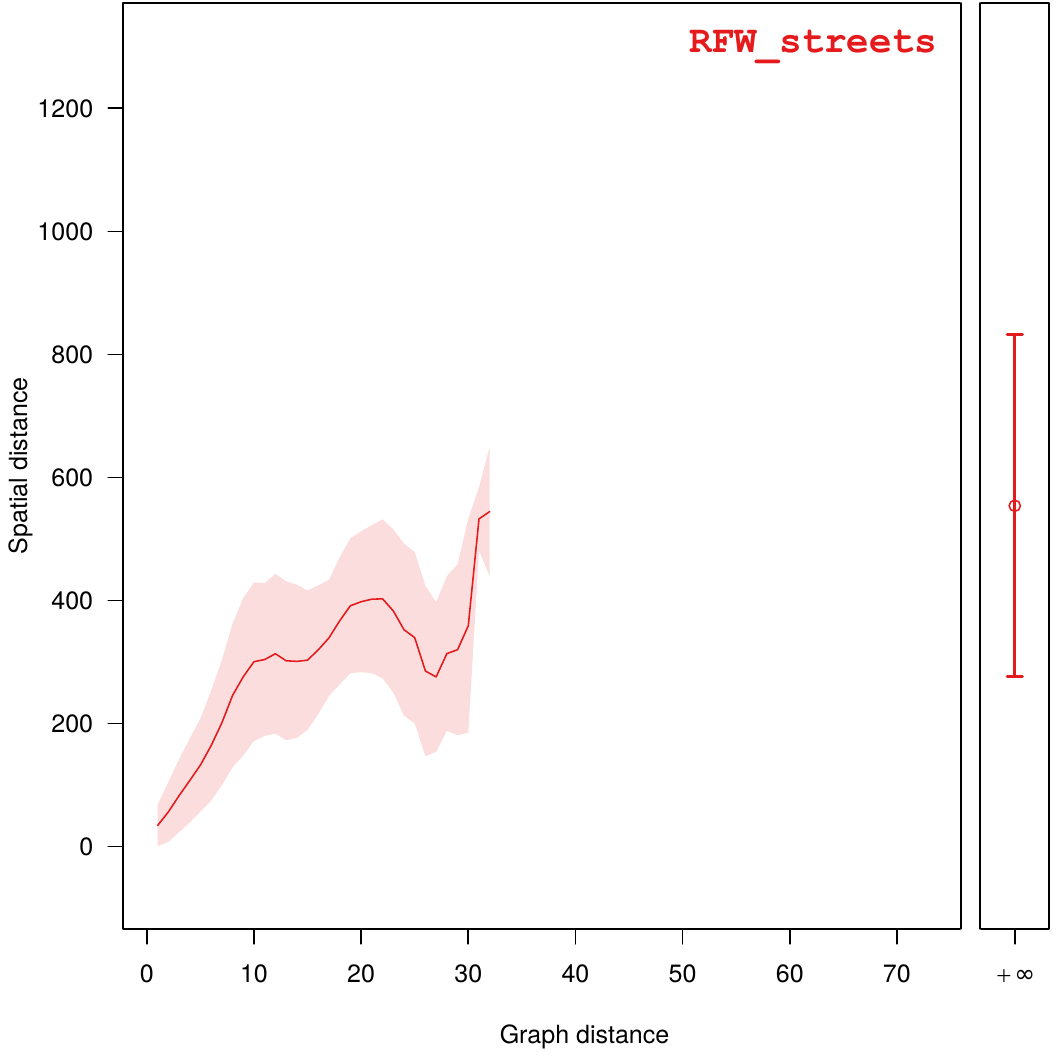}
    \hfill
    \includegraphics[width=0.24\linewidth]{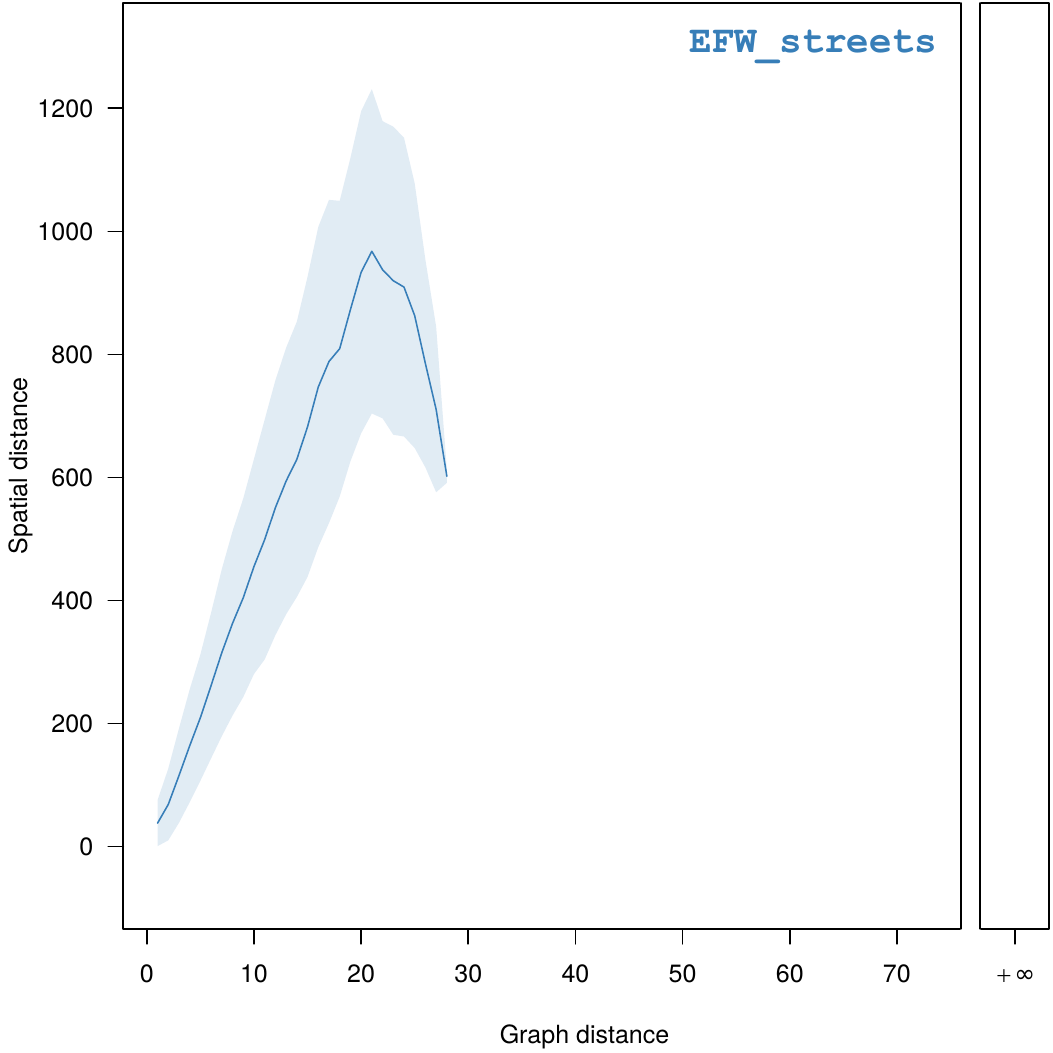}
    \hfill
    \includegraphics[width=0.24\linewidth]{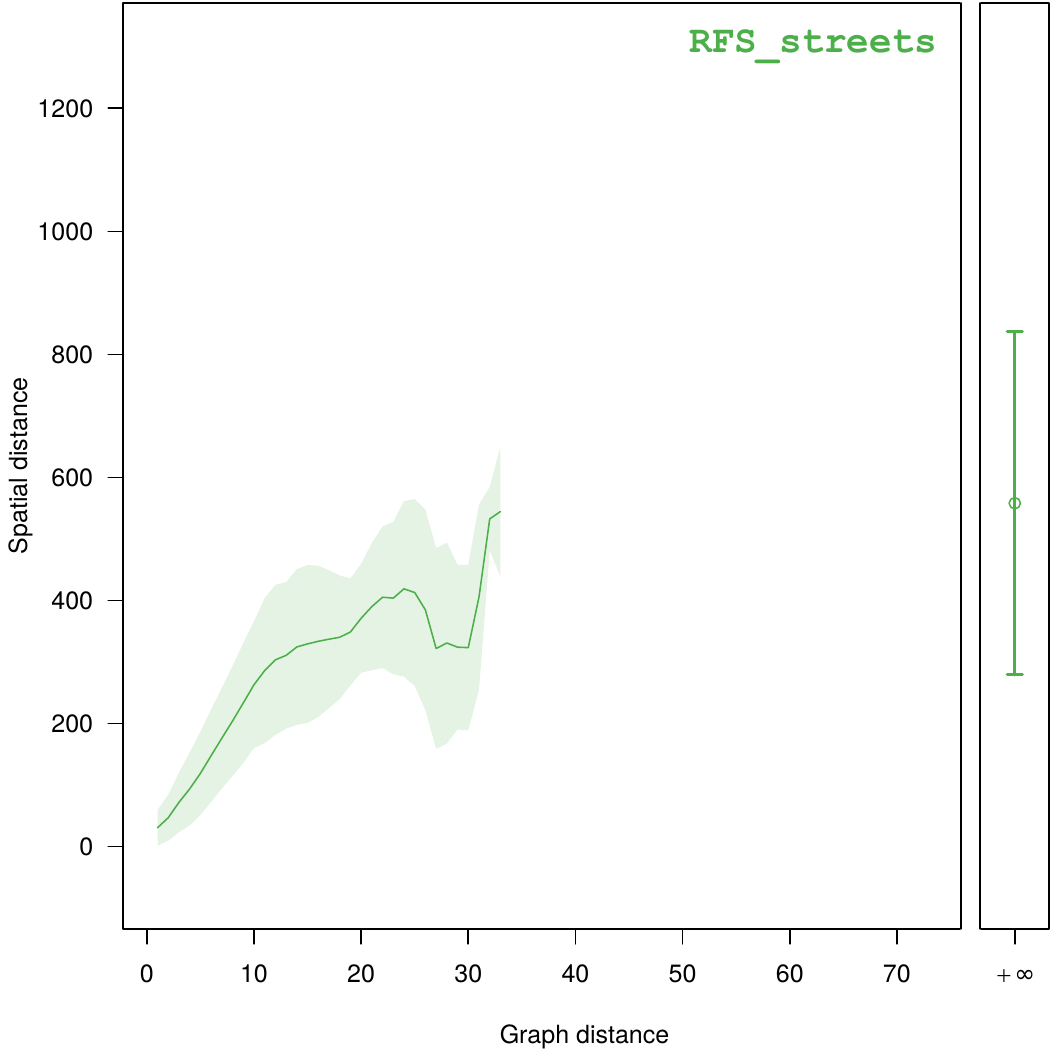}
    \hfill
    \includegraphics[width=0.24\linewidth]{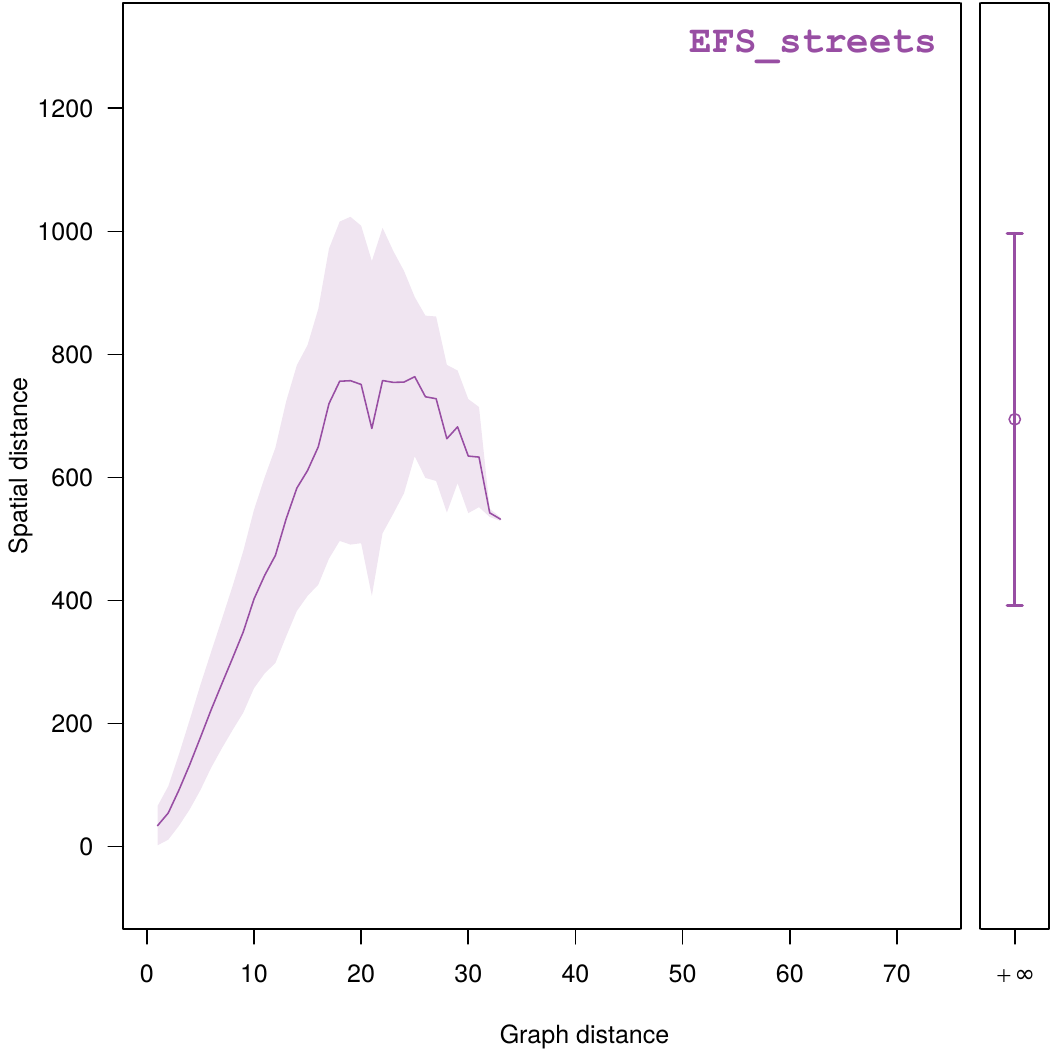}
    \hfill~\\[0.2cm]
    \hfill
    \includegraphics[width=0.24\linewidth]{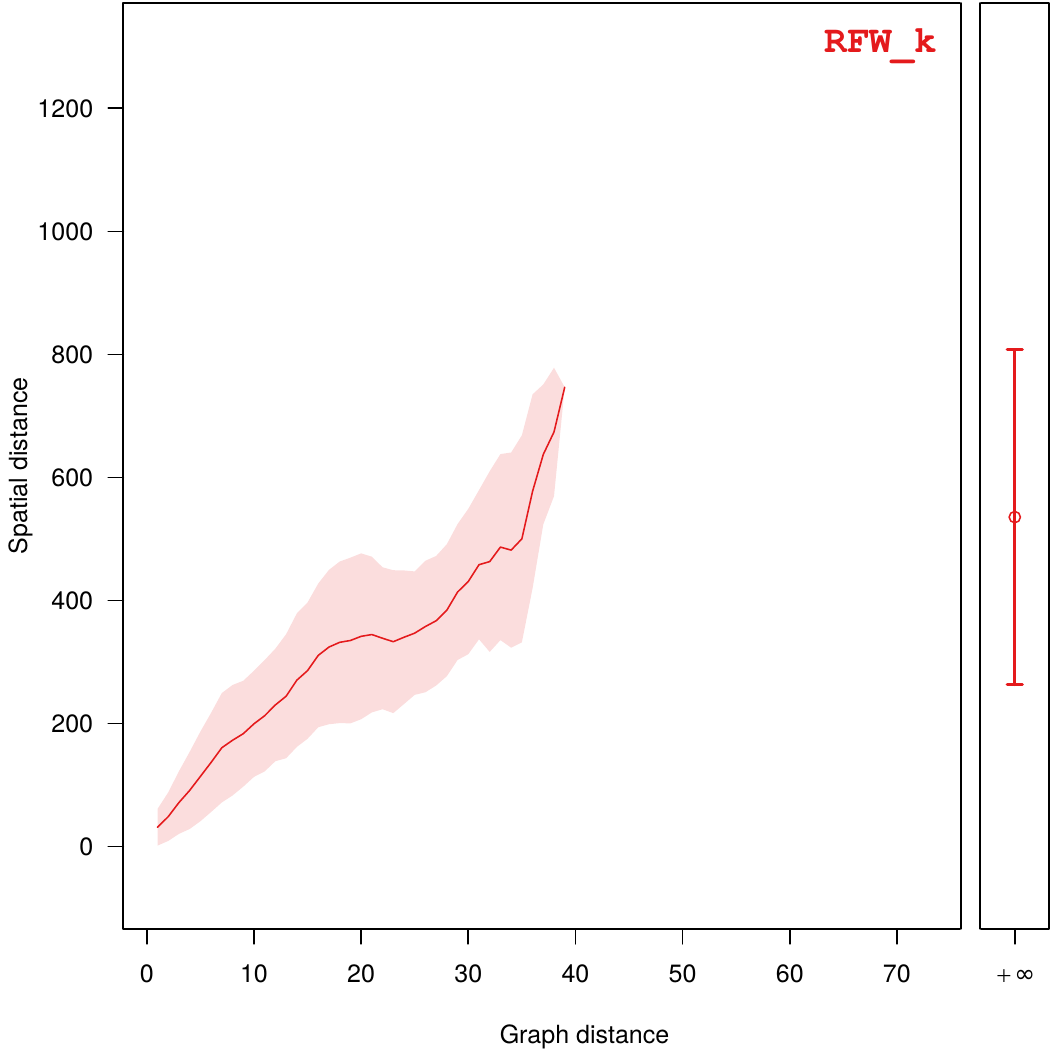}
    \hfill
    \includegraphics[width=0.24\linewidth]{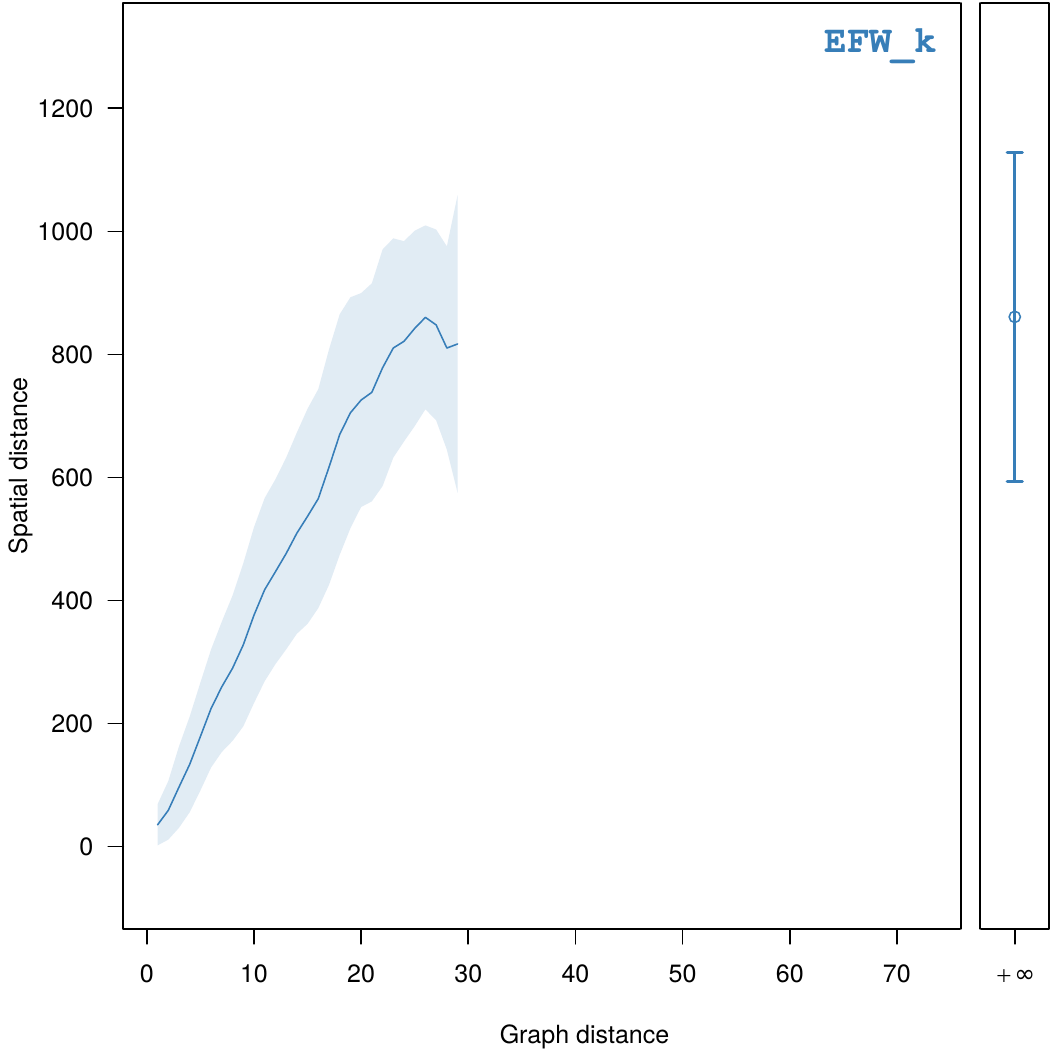}
    \hfill
    \includegraphics[width=0.24\linewidth]{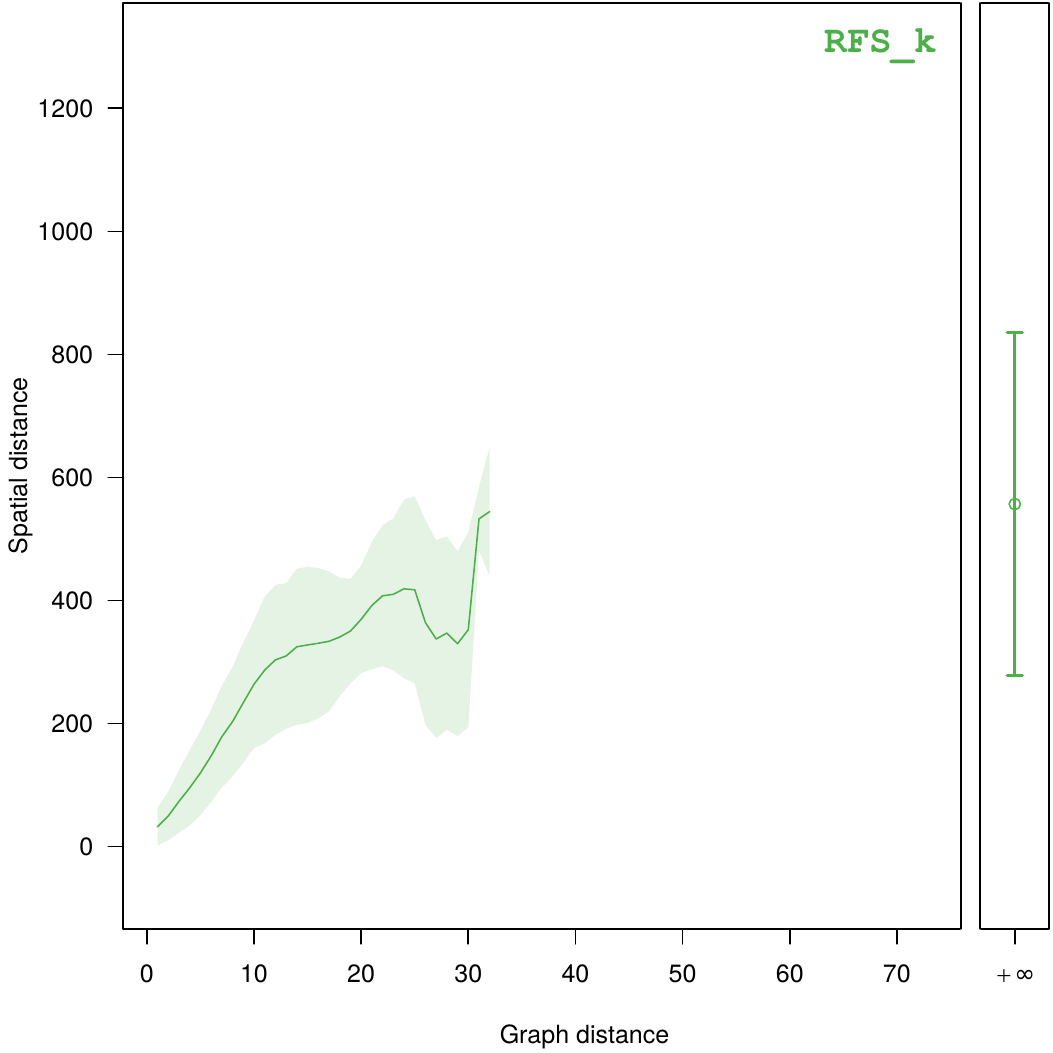}
    \hfill
    \includegraphics[width=0.24\linewidth]{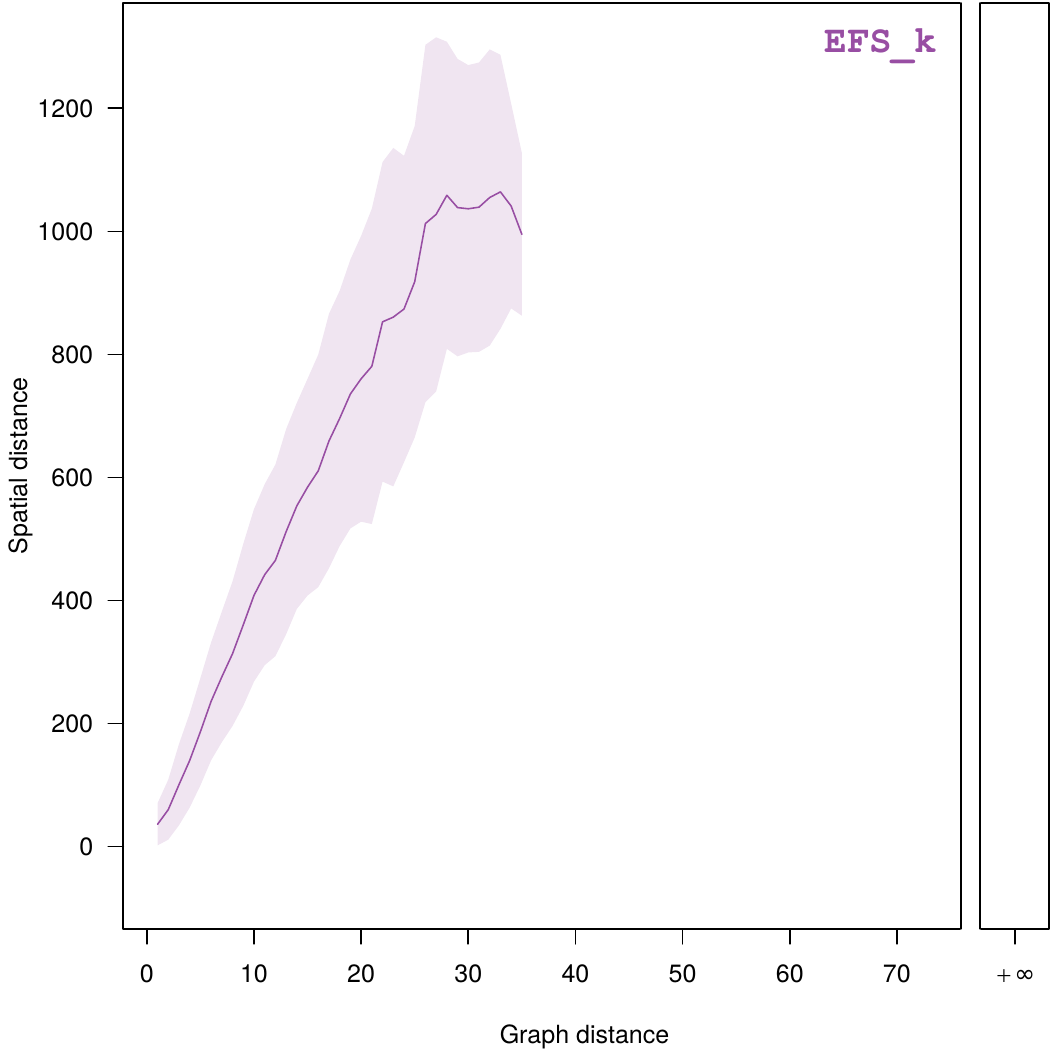}
    \hfill~\\
    \caption{Comparison of graph and spatial distances, for all graphs. Figure available at \href{http://doi.org/10.5281/zenodo.14175830}{10.5281/zenodo.14175830} under CC-BY license.}
    \label{fig:DistComp}
\end{figure}

Figure~\ref{fig:DistComp} shows the comparison of the graph and spatial distances, for each graph extracted using one of the methods considered in the article. The colors are the same as in Figure~\ref{fig:CompPareto}: red for \texttt{R$\dotp$W\_$\dotp$}, blue for \texttt{E$\dotp$W\_$\dotp$}, green for \texttt{R$\dotp$S\_$\dotp$}, and purple for \texttt{E$\dotp$S\_$\dotp$}. The lines represent the means and the areas are the standard deviations. 

The separated dot that appears on the right side of certain plots (ex. \texttt{RFW\_all}), reveals the existence of separated components in the corresponding graph, as the graph distance between vertices located in different components is infinite. Ideally, there should be a single component, and if there are several ones, then the spatial distances matching an infinite graph distance should be the largest of all.

Visually, a good result corresponds to a relatively straight line (ex. \texttt{EFW\_k}), or at least a monotonous function (ex. \texttt{RFW\_k}), with low dispersion. Some plots exhibit an increasing then decreasing trend (ex. \texttt{RFS\_all}), which means that large graph distances are associated with small spatial distances. This means that the graph is too linear: some shortcuts are missing to correctly encode the spatial distance.

Spanning a wide range of graph distances is also a good thing (ex. \texttt{EFS\_k}), as it suggests a better ``distance resolution'' on the graph (i.e. a better chance that two distinct spatial distances are not associated to the same graph distance). Graphs that include too many shortcut vertices tend to exhibit a very narrow range (ex. \texttt{RHW\_all}).

\section{Community Structure}
\label{sec:ApdxComs}
\begin{figure}[!htbp]
    \centering
    \includegraphics[width=0.88\linewidth]{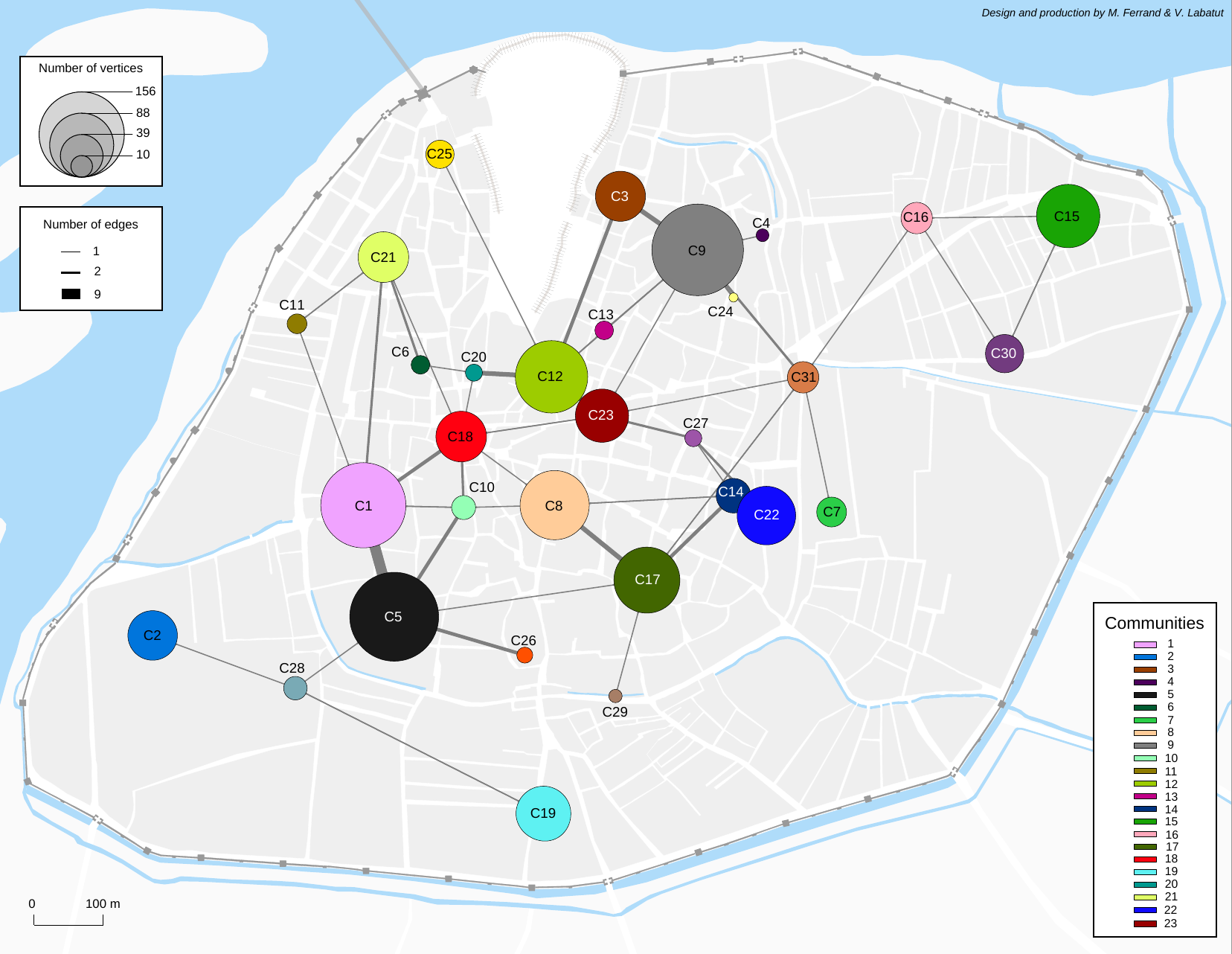}
    \\[3mm]
    \includegraphics[width=0.88\linewidth]{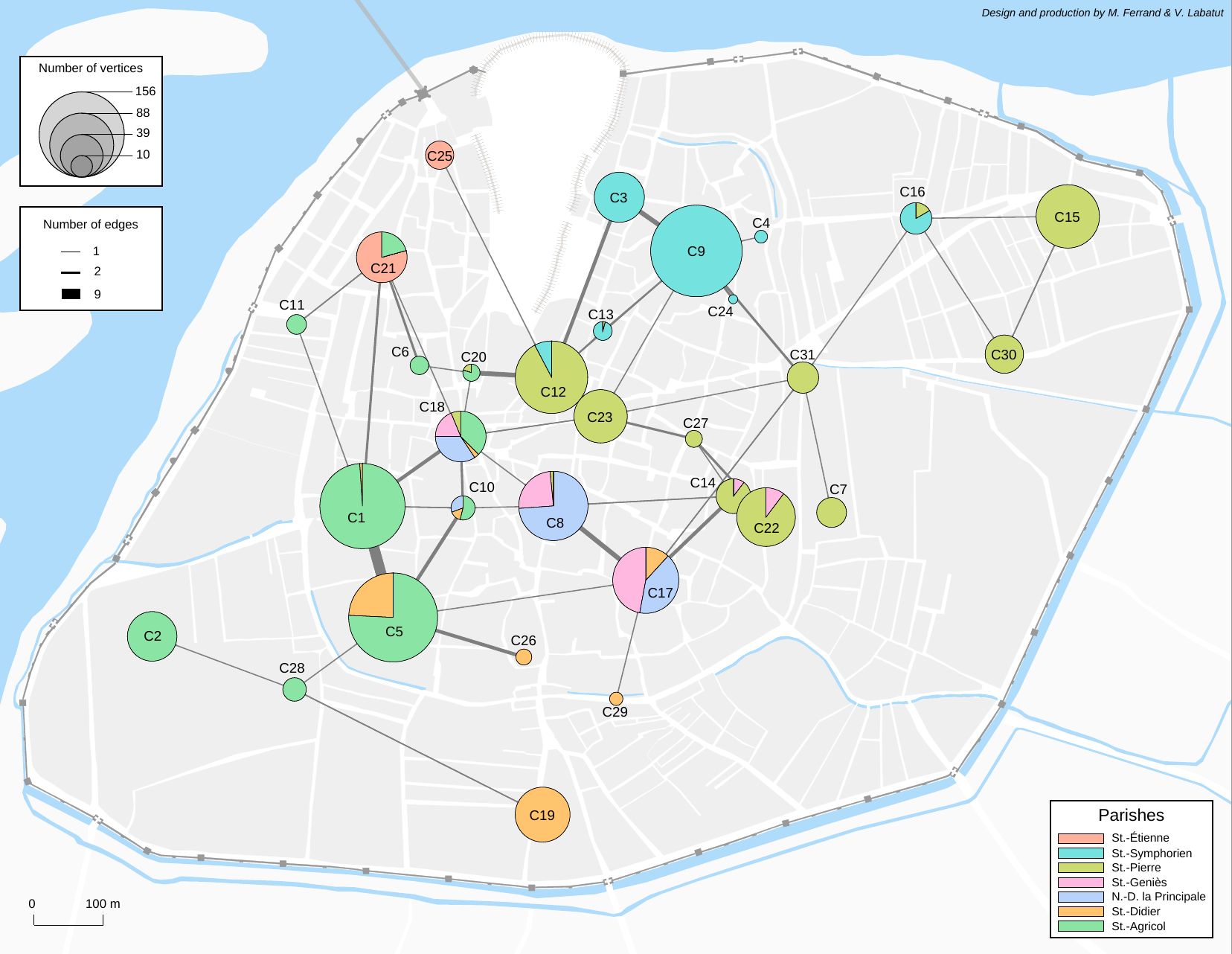}
    \caption{Two versions of the community network: each node represents a community from Section~\ref{sec:ComStruct}, its size is proportional to the number of vertices in the community, and the thickness of a link connecting two nodes corresponds to the number of edges between the matching communities in the original graph. \textbf{Top:} each community is represented using a unique color, as in Figure~\ref{fig:ComParish}. \textbf{Bottom:} each community is represented by a pie chart showing how the parochial membership of its constituting properties is distributed. Figure available at \href{http://doi.org/10.5281/zenodo.14175830}{10.5281/zenodo.14175830} under CC-BY license.}
    \label{fig:ComParishPies}
\end{figure}

Figure~\ref{fig:ComParishPies} provides additional visualization of the community structures, in addition to those presented in Section~\ref{sec:ComStruct} of the main text. It shows two so-called community networks, representing the community structure in a simplified way, where each node represents a whole community in the original graph. The left plot uses the same vertex colors as in Figure~\ref{fig:ComParish} to show the relative position and importance of the communities. The right plot uses the colors of the parishes, as in Figures~\ref{fig:ComParish} and~\ref{fig:ComHisto}, to represent the distribution of properties over parishes, for each community. 

\end{document}